\documentclass[prd,10pt,aps,superscriptaddress,twocolumn,floatfix,preprintnumbers]{revtex4-1}
\usepackage[centertags]{amsmath}
\usepackage{amsthm,amssymb,bbm,graphicx,hyperref,slashed,ifthen}
\graphicspath{{fig/}}
\allowdisplaybreaks
 
\DeclareMathOperator{\SU}{SU}
\DeclareMathOperator{\1}{\mathbbm{1}} 

\newcommand{\shownorm}[3]{#2 && #3}
\newcommand{\norm}[1]{\vert\vert #1 \vert\vert}
\renewcommand{\O}{{\mathcal{O}}}
\newcommand{\Qp}{Q_p}
\newcommand{\E}{F}

\newcommand{\msbar}{{\overline{\rm MS}}}
\newcommand{\ri}{{\rm RI}}

\newcommand{\rimom}{{\rm RI/MOM}}
\newcommand{\bfrimom}{{\bf RI/MOM}}
\newcommand{\ripmom}{\rm RI^{\prime}/MOM}
\newcommand{\rismom}{{\rm RI/SMOM}}
\newcommand{\bfrismom}{{\bf RI/SMOM}}
\newcommand{\rismomb}{{\rismom_{\gamma_{\mu}}}}
\newcommand{\sub}{{\rm sub}}
\newcommand{\ndr}{{\rm NDR}}

\newcommand{\NDR}{\msbar{\rm [NDR]}}
\newcommand{\bfNDR}{\bfmsbar{\bf [NDR]}}
\newcommand{\tree}{{\rm tree}}
\newcommand{\bfmsbar}{{\overline {\bf MS}}}
\newcommand{\reduced}{chiral}
\newcommand{\GAMMA}{\Upsilon}
\newcommand{\braket}[1]{\left\langle #1 \right\rangle}

\newcommand{\dIleft}{\bigl}
\newcommand{\dIright}{\bigr}
\newcommand{\dIbsp}{\!\!}
\newcommand{\dataINL}{} 
\newcommand{\dataIcuti}{100}
\newlength{\dataIla}
\newlength{\dataIlb}
\newlength{\dataIlc}
\newlength{\dataIld}
\newcount\datacnti 
\newcount\datacntj
\newcommand{\dataIab}[4]{\ifthenelse{ #1 < \dataIcuti }{\dataINL
    $(#1,#2)$ & & \parbox[t]{\dataIla}{\raggedright $#3$} &
    & \parbox[t]{\dataIlb}{\raggedright $#4$}
    \renewcommand{\dataINL}{\\}}{}}

\newcommand{\dataIabN}[4]{\ifthenelse{ #1 < \dataIcuti }{\dataINL
    $(#1,#2)$ & & \parbox[t]{\dataIla}{\raggedright $#3$} &
    & \parbox[t]{\dataIlb}{\raggedleft $#4$}
    \renewcommand{\dataINL}{\\}}{}}

\newcommand{\dataPg}[2]{\datacnti #1\relax\ifthenelse{ #1 > 3 }{\advance\datacnti 1\relax}{}
\edef\OldX{\OldX \OldS $\GAMMA^{G_1}_{\the\datacnti}$ } \edef\OldY{\OldY \OldS $#2$ }
  \edef\OldS{ & } }

\newcommand{\dataPab}[4]{\ifthenelse{ #1 < \dataIcuti
  }{\dataINL\datacnti #1\relax\datacntj #2\relax\ifthenelse{ #1 > 3
    }{\advance\datacnti 1\relax}{}\ifthenelse{ #2 > 3
    }{\advance\datacntj 1\relax}{}$(\the\datacnti,\the\datacntj)$ &
    & \parbox[t]{\dataIla}{\raggedright $#3$} &
    & \parbox[t]{\dataIlb}{\raggedleft $#4$}
    \renewcommand{\dataINL}{\\}}{}}

\newcommand{\dataPa}[3]{\ifthenelse{ #1 < \dataIcuti
  }{\dataINL\datacnti #1\relax\datacntj #2\relax\ifthenelse{ #1 > 3
    }{\advance\datacnti 1\relax}{}\ifthenelse{ #2 > 3
    }{\advance\datacntj 1\relax}{}$(\the\datacnti,\the\datacntj)$ &
    & \parbox[t]{7.25cm}{\raggedright $#3$}
    \renewcommand{\dataINL}{\\}}{}}

\newcommand{\dataPgc}[2]{\ifthenelse{ #1 < \dataIcuti
  }{\dataINL\datacnti #1\relax\ifthenelse{ #1 > 3
    }{\advance\datacnti 1\relax}{}$(\the\datacnti)$ &
    & \parbox[t]{7.35cm}{\raggedright $#2$}
    \renewcommand{\dataINL}{\\}}{}}

\newcommand{\dataPupsilon}[6]{\ifthenelse{ #1 < \dataIcuti
  }{\dataINL\datacnti #1\relax\datacntj #2\relax\ifthenelse{ #1 > 3
    }{\advance\datacnti 1\relax}{}\ifthenelse{ #2 > 3
    }{\advance\datacntj 1\relax}{}$(\the\datacnti,\the\datacntj)$ 
    & & \parbox[t]{\dataIla}{\raggedright $#3$} 
    & & \parbox[t]{\dataIlb}{\raggedright $#4$}
    & & \parbox[t]{\dataIlc}{\raggedright $#5$}
    & & \parbox[t]{\dataIld}{\raggedright $#6$}
    \renewcommand{\dataINL}{\\}}{}}

\begin{document}
\title{\boldmath Matching factors for $\Delta S=1$ four-quark
  operators in $\bfrismom$ schemes}

\author{Christoph Lehner} \affiliation{RIKEN/BNL Research Center,
  Brookhaven National Laboratory, Upton, NY-11973, USA}
\author{Christian Sturm} \affiliation{Max-Planck-Institut f\"ur
  Physik, F\"ohringer Ring 6, 80805 M\"unchen, Germany}

\date{April 25, 2011}

\begin{abstract}
  The non-perturbative renormalization of four-quark operators plays a
  significant role in lattice studies of flavor physics.  For this
  purpose, we define regularization-independent symmetric
  momentum-subtraction ($\rismom$) schemes for $\Delta S=1$
  flavor-changing four-quark operators and provide one-loop matching
  factors to the $\msbar$ scheme in naive dimensional regularization.
  The mixing of two-quark operators is discussed in terms of two
  different classes of schemes.  We provide a compact expression for
  the finite one-loop amplitudes which allows for a straightforward
  definition of further $\rismom$ schemes.
\end{abstract}

\preprint{BNL-95075-2011-JA, MPP-2011-48, RBRC 897}

\maketitle

\section{Introduction} 
The study of physical processes which change the strangeness by one
unit ($\Delta S=1$), such as the decay of a kaon into two pions, is
important for the understanding of CP violation within the Standard
Model (SM) and its possible extensions.  Such processes can be used to
measure the parameter of direct CP violation
$\epsilon^\prime/\epsilon$, to study the $\Delta I=1/2$ rule, and to
calculate long-distance contributions to $K_0-\overline{K}_0$ mixing
and the parameter of indirect CP violation $\epsilon$
\cite{Noaki:2001un,Blum:2001xb,Buras:2003zz,Buras:2010pza,Christ:2010gi}.
The resulting constraints for the Cabibbo-Kobayashi-Maskawa (CKM)
matrix elements allow for a precise test of the SM.  The weak
interaction which mediates these processes with change in the
strangeness can be described by local four-fermion operators at low
energy scales, where the character of the vector boson interaction is
essentially point like, see
Refs.~\cite{Vainshtein:1975sv,Witten:1975bh,Shifman:1975tn,Witten:1976kx,
  Gilman:1979bc} and
Refs.~\cite{Buchalla:1995vs,Buras:1998raa,Buras:2011we} for
reviews. Matrix elements which describe, e.g., two-pion decays of
kaons can then be computed with the help of lattice simulations.

In order to perform the renormalization of relevant operators in the
lattice computation one can adopt a renormalization scheme which is
independent of the regulator.  Such a scheme can then be implemented
in both non-perturbative lattice calculations and continuum
perturbation theory.  This allows for a conversion of lattice results
to the modified minimal subtraction ($\msbar$) scheme which is not
directly applicable in lattice simulations.  In
Ref.~\cite{Martinelli:1994ty} the non-perturbative renormalization
(NPR) technique and regularization independent (RI)
momentum-subtraction schemes were defined for this purpose.

In the context of light up, down, and strange quark-mass
determinations quark bilinear operators need to be studied for the NPR
procedure.  The required matching factors which convert the quark
masses and fields from the RI schemes to the $\msbar$ scheme are known
up to three-loop
order~\cite{Martinelli:1994ty,Franco:1998bm,Chetyrkin:1999pq,Gracey:2003yr}
in perturbative Quantum Chromodynamics (QCD).  The renormalization
constants in these regularization independent momentum-subtraction
($\rimom$) schemes are determined at an exceptional momentum point of
the considered amplitude.  In the case of the quark bilinear operators
the exceptional momentum configuration is distinguished by the fact
that no momentum leaves the operator.  However, a lattice simulation
with an exceptional momentum configuration for the renormalization
constants is more disposed to effects of chiral symmetry breaking
\cite{Aoki:2007xm}.  Furthermore, in the $\rimom$ scheme unwanted
infrared effects exist and the matching factors show a poor
convergence behavior.  For these reasons a non-exceptional momentum
configuration was proposed in Ref.~\cite{Aoki:2007xm} and the
framework and concepts of new $\rismom$ schemes with a symmetric
subtraction point were worked out in Ref.~\cite{Sturm:2009kb}.  The
symmetric subtraction point is characterized by the fact that a
momentum leaves the inserted operator.

The matching factors for the conversion of quark masses from the
schemes with a symmetric subtraction point to the $\msbar$ scheme were
computed to two-loop
order~\cite{Sturm:2009kb,Gorbahn:2010bf,Almeida:2010ns,Gracey:2011fb}
by considering the amputated Green's functions with insertion of the
scalar or pseudo-scalar operator.  These schemes exhibit a better
infrared behavior, and also the coefficients of the perturbative
expansion of the matching factors are smaller.  Therefore their use
led to a significant reduction of the systematic uncertainties in the
light quark mass determinations following this
approach~\cite{Aoki:2010dy} compared to previous studies, see
Ref.~\cite{Allton:2008pn}.

For the insertion of any multi-quark operator into an amputated
Green's function the fermion field for each external leg needs to be
renormalized. In Ref.~\cite{Sturm:2009kb} two schemes, the $\rismom$
and $\rismomb$ scheme, were suggested for the renormalization with a
symmetric subtraction point. It was also shown that the former is
equivalent to the $\ripmom$ scheme and is thus known to three-loop
order~\cite{Martinelli:1994ty,Chetyrkin:1999pq,Gracey:2003yr}, whereas
the latter is known to two-loop
order~\cite{Sturm:2009kb,Almeida:2010ns,Gorbahn:2010bf,Gracey:2011fb}.
The corresponding calculation requires the computation of amputated
Green's functions with insertion of the vector or axial-vector
operator and the utilization of Ward-Takahashi-identities.

We would like to mention that also the tensor operator has
applications in lattice simulations, see, e.g.,
Ref.~\cite{Donnellan:2007xr} and a scheme with a symmetric subtraction
point has been introduced in Ref.~\cite{Sturm:2009kb}. Its matching
factor and anomalous dimension is known to two-loop order
\cite{Sturm:2009kb,Almeida:2010ns,Gracey:2011fb}.  Also moments of
twist-2 operators used in deep inelastic scattering have been studied
in a $\rismom$ scheme in Refs.~\cite{Gracey:2010ci,Gracey:2011zn}.

In view of these successes and advantages, the $\rismom$ definition
has been extended to $\Delta S=2$ flavor-changing four-quark operators
in Ref.~\cite{Aoki:2010pe}, where the one-loop QCD corrections to
different matching factors have been computed, and also the anomalous
dimensions were provided.  These results were then used for the
determination of the $B_K$ parameter which is needed to parametrize
the hadronic matrix element for the theoretical description of
$K_0-\overline{K}_0$ mixing.  For the case of an exceptional
subtraction point the matching factors were determined at
next-to-leading order in Refs.~\cite{Ciuchini:1997bw,Buras:2000if}
based on Ref.~\cite{Donini:1995xj}.  Similarly the matching for
$\Delta S=1$ flavor-changing four-quark operators with an exceptional
subtraction point was determined in
Refs.~\cite{Ciuchini:1995cd,Buras:2000if}.  The purpose of this paper
is to introduce $\rismom$ schemes with a non-exceptional subtraction
point for $\Delta S=1$ flavor-changing four-quark operators as well as
to provide the corresponding matching factors for the conversion from
these $\rismom$ schemes to the $\msbar$ scheme in naive dimensional
regularization ($\ndr$).  To this end we first present the framework
needed to properly take into account the mixing with two-quark
operators which was not needed in the $\Delta S=2$ case.  We then
study the insertion of the $\Delta S=1$ operators into amputated
Green's functions in perturbative QCD at one-loop order to determine
the renormalization constants.

The outline of this work is as follows.  In Sec.~\ref{sec:conventions}
we define the set of $\Delta S=1$ four-quark operators used in this
work.  In Sec.~\ref{sec:renorm} we discuss some generalities of the
renormalization of the $\Delta S=1$ operators in the $\NDR$ scheme as
well as in a general $\ri$ scheme and introduce our notation.  In
Sec.~\ref{sec:results} we provide a classification of projectors used to
define the $\ri$ schemes and present our results for the finite one-loop
amplitude as well as conversion factors from different $\rismom$ schemes
to the $\NDR$ scheme.  Finally we close with a summary and conclusions
in Sec.~\ref{sec:conclusion}.

\section[The bases of $\Delta S=1$ operators]{\boldmath The bases of
  $\Delta S=1$ operators\label{sec:conventions}}
In this section we define operator bases of the effective $\Delta S=1$
Hamiltonian of electroweak interactions, where we closely follow the
notation of Ref.~\cite{Buras:1993dy}. We work in an effective
three-flavor theory including the up, down, and strange quark. This
effective theory is valid for energies below the charm quark mass.
The effective Hamiltonian reads
\begin{align}
\label{eq:Hamilton}
\mathcal{H}^{\Delta S=1}_{\rm eff}={G_{F}\over\sqrt{2}}\sum_{i}C^{x}_{i}(\mu)O^{x}_{i}(\mu)
\end{align}
with Fermi coupling constant $G_F$ and renormalization scale
$\mu$. The symbols $C_{i}(\mu)$ denote Wilson coefficients and
$O_{i}(\mu)$ are four-quark operators, which we will discuss in terms
of ``physical'' and ``chiral'' operator bases in different schemes
labeled $x$.  In the case of the physical operator bases the effective
Hamiltonian is expressed in terms of ten operators which are grouped
into current-current operators, QCD penguin operators, and electroweak
penguin operators.  The physical operator bases are classified by the
physical origin of their respective operators, whereas the chiral
operator basis is classified by irreducible representations of the
$\SU(3)_L \otimes \SU(3)_R$ chiral symmetry.

\subsection{The physical bases}
Let us start with the traditional physical operator basis of
Refs.~\cite{Gilman:1979bc,Gilman:1982ap,Buras:1992tc,Ciuchini:1993vr}.
The current-current operators are defined by
\begin{align}\label{eqn:q1FTdef}
  Q_1 &= (\bar s_a u_b)_{V-A} (\bar u_b d_a)_{V-A}\,,\notag\\
  Q_2 &= (\bar s_a u_a)_{V-A} (\bar u_b
  d_b)_{V-A}\,,
\end{align}
the QCD penguin operators are defined by
\begin{align}
  Q_3 &= (\bar s_a d_a)_{V-A} \sum_{q=u,d,s}  (\bar q_b q_b)_{V-A} \,, \notag\\
  Q_4 &= (\bar s_a d_b)_{V-A} \sum_{q=u,d,s}  (\bar q_b q_a)_{V-A} \,, \notag\\
  Q_5 &= (\bar s_a d_a)_{V-A} \sum_{q=u,d,s}  (\bar q_b q_b)_{V+A} \,, \notag\\
  Q_6 &= (\bar s_a d_b)_{V-A} \sum_{q=u,d,s} (\bar
  q_b q_a)_{V+A}\,,
\end{align}
the electroweak penguin operators are defined by
\begin{align}
  Q_7 &= \frac32 (\bar s_a d_a)_{V-A} \sum_{q=u,d,s} e_q (\bar q_b q_b)_{V+A} \,, \notag\\
  Q_8 &= \frac32 (\bar s_a d_b)_{V-A} \sum_{q=u,d,s} e_q (\bar q_b q_a)_{V+A}
\end{align}
with $e_u=2/3$, $e_d=e_s=-1/3$, and
\begin{align}
  Q_9 &= \frac32 (\bar s_a d_a)_{V-A} \sum_{q=u,d,s} e_q (\bar q_b q_b)_{V-A} \,, \notag\\\label{eqn:q10def}
  Q_{10} &= \frac32 (\bar s_a d_b)_{V-A} \sum_{q=u,d,s} e_q (\bar q_b q_a)_{V-A}\,,
\end{align}
where $(\bar q q)_{V \pm A}$ refers to the spinor structure $\bar q
\gamma_\mu (1 \pm \gamma_5) q$, $a$ and $b$ are color indices, and
$u$, $d$, and $s$ are the fields of the up, down, and strange quarks.
This basis of operators $\{O_i\}=\{Q_1,\ldots,Q_{10}\}$ is referred to
as ``basis I'' in the following.  Alternatively we can Fierz transform
$Q_1$ and $Q_2$ to
\begin{align}\label{eqn:q1def}
  \tilde Q_1 &= (\bar s_a d_a)_{V-A} (\bar u_b u_b)_{V-A}\,, \notag\\
  \tilde Q_2 &= (\bar s_a d_b)_{V-A} (\bar u_b
  u_a)_{V-A}\,.
\end{align}
The basis of operators $\{O_i\}=\{\tilde Q_1,\tilde
Q_2,Q_3,\ldots,Q_{10}\}$ is called ``basis II'' in the following.
Operators with color contractions as in $\tilde Q_1$ are called color
diagonal, operators with color contractions as in $\tilde Q_2$ are
called color mixed.  It will also be useful to define the Fierz
transformation of $Q_3$, i.e.,
\begin{align}
  \tilde Q_3 = \sum_{q=u,d,s} (\bar s_a q_b)_{V-A} (\bar
  q_b d_a)_{V-A}\,.
\end{align}
In an explicit four-dimensional regularization scheme, such as lattice
regularization, $Q_i$ and $\tilde Q_i$ can be used interchangeably.
In dimensional regularization, however, the contribution of evanescent operators
such as
\begin{align}\label{eqn:defevanqqtilde}
  E_{1i} = Q_i - \tilde Q_i
\end{align}
has to be included. The evanescent operators vanish at tree level if
the regulator is removed, see, e.g.,
Refs.~\cite{Buras:1989xd,Dugan:1990df,Buras:1992tc,Herrlich:1994kh}.

\subsection{The chiral basis}
The operators $Q_1,\ldots,Q_{10}$ are not linearly independent, i.e.,
one can eliminate three operators by expressing them as linear
combinations of the remaining ones.  In a regularization which breaks
Fierz transformations also evanescent operators enter these relations.
The reduced operator basis of linearly independent operators can then
be classified according to irreducible representations of $\SU(3)$ and
$\SU(2)$ flavor symmetries \cite{Vainshtein:1975sv,Shifman:1975tn},
and the resulting operator basis will be referred to as the
``\reduced{} basis'' in the following.  The linear independence of its
elements will become important later for the non-perturbative
definition of RI schemes with the help of projectors.

We proceed along the lines of Ref.~\cite{Blum:2001xb} amending their
discussion by the contributions of evanescent operators since we work
in dimensional regularization.  We first eliminate the operators
$Q_4$, $Q_9$, and $Q_{10}$ using
\begin{align}\label{eqn:diffeqndim}
  Q_4 &= Q_2 + Q_3 - Q_1 - E_{12} - E_{13}\,, \notag\\
  Q_9 &= \frac32 Q_1 - \frac12 Q_3 - \frac32 E_{11} \,, \notag\\
  Q_{10} &= \frac12 (Q_1-Q_3) + Q_2 + \frac12 E_{13} - E_{12}.
\end{align}

The remaining seven operators can then be recombined according to
irreducible representations of the chiral flavor-symmetry group
$\SU(3)_L \otimes \SU(3)_R$.  The details of this decomposition
including evanescent operators are given in
App.~\ref{app:flavordecompose}.  The \reduced{} operator basis is thus
given by
\begin{align}\label{eqn:qprimeqe}
  (27,1) && Q'_1 &= 3 Q_1 + 2 Q_2 - Q_3 - 3 E_{11} \,, \notag\\
  (8,1) && Q'_2 &= \frac15(2 Q_1 - 2 Q_2 + Q_3) - \frac25 E_{11}\,, \notag\\
  (8,1) &&Q'_3 &= \frac15(-3 Q_1 + 3 Q_2 + Q_3) + \frac35 E_{11}\,, \notag\\
  (8,1) &&Q'_{5,6} &= Q_{5,6}\,,  \notag\\
  (8,8) &&Q'_{7,8} &= Q_{7,8}\,, 
\end{align}
where $(L,R)$ denotes the respective irreducible representation of
$\SU(3)_L \otimes \SU(3)_R$.  Instead of expressing
Eqs.~\eqref{eqn:diffeqndim} and \eqref{eqn:qprimeqe} in terms of the
operators of basis I and evanescent operators we could have also
eliminated the latter by introducing the operators $\tilde Q_1$,
$\tilde Q_2$ of basis II and $\tilde Q_3$ with the help of
Eq.~\eqref{eqn:defevanqqtilde}.

\section{Renormalization\label{sec:renorm}}
In this section we discuss the renormalization of the four-quark
operators $O_i$ starting with some generalities concerning operator
renormalization at fixed gauge.  We then describe the one-loop
off-shell renormalization of the operators in the different bases in
the $\NDR$ scheme in detail and provide a discussion of
renormalization in a general $\ri$ scheme.  A brief discussion of some
details concerning the Wilson coefficients of the \reduced{} basis
concludes this section.

\subsection{Renormalization at fixed gauge}
The renormalization schemes described in this work are defined at a
fixed covariant gauge with gauge fixing parameter $\xi$, where $\xi=0$
($\xi=1$) corresponds to the Landau (Feynman) gauge.  The gauge-fixing
procedure explicitly breaks the gauge symmetry, and it can be shown
that mixing with three classes of operators can occur
\cite{KlubergStern:1975hc,Joglekar:1975nu,Deans:1978wn,Collins:1984xc,Buras:1992tc,Dawson:1997ic}:
(i) gauge-invariant operators which do not vanish using the equations
of motion, (ii) gauge-invariant operators which vanish using the
equations of motion, and (iii) gauge non-invariant operators which are
either BRST invariant or vanish using the equations of motion.

In a general renormalization scheme $x$, a renormalized four-quark
operator $O_i^x$ can be written as
\begin{align}\label{eqn:genren}
  O_i^x = Z^x_{ij} O_j + b^x_{ik} \E_k + c^x_{il} G_l + d^x_{im} N_m\,,
\end{align}
where $O_j$ are the bare four-quark operators, $\E_k$ are evanescent
operators, $G_l$ ($N_m$) are gauge-invariant (gauge non-invariant)
operators involving only two quark fields. The symbols $Z^x_{ij}$,
$b^x_{ik}$, $c^x_{il}$, and $d^x_{im}$ denote renormalization
constants.  The sum over the respective operator basis for $O_j$,
$\E_k$, $G_l$, and $N_m$ is implied.  The operators $O_j$ belong to
class (i), the operators $G_l$ belong to class (i) or (ii), and the
operators $N_m$ belong to class (iii).  Operators of class (ii) and
(iii) do not contribute to physical amplitudes
\cite{KlubergStern:1975hc,Joglekar:1975nu,Deans:1978wn,Collins:1984xc}.
The mixing of operators $N_m$ can be avoided by using the background
field gauge \cite{Buras:1992tc,Ciuchini:1993vr,Abbott:1980hw}.

\subsection[The $\NDR$ scheme]{The $\bfNDR$ scheme\label{sec:MSbar}}
In the following we discuss the off-shell renormalization of the
operators $O_i$ in the $\msbar$ scheme with massless quarks at
one-loop order in perturbative QCD.  We use naive dimensional
regularization ($\ndr$) in $d=4-2\varepsilon$ space-time dimensions
with a naive anti-commutation definition of $\gamma_5$.  For
multi-loop calculations the operator basis of
Ref.~\cite{Chetyrkin:1997gb} is more convenient and allows for a
straightforward treatment of $\gamma_5$.  Since we restrict ourselves
in this work to the one-loop order, we adhere to the traditional bases
in order to connect with previous works in this field.  The
$\msbar$-renormalization of the four-quark operators $Q_i$ with a
focus on on-shell renormalization is discussed, e.g., in
Refs.~\cite{Gilman:1979bc,Wise:1980qb} at one loop and in
Refs.~\cite{Buras:1992tc,Ciuchini:1993vr} at two-loop order.

The four-quark operators $O_i$ are of mass-dimension $6$, so that in
the massless limit mixing can only occur with operators $\E_k$, $G_l$,
and $N_m$ of mass-dimension $6$.  For off-shell external states the
operator
\begin{align}
  G_1 = \frac4 {ig^2} \bar s \gamma_\nu (1-\gamma_5)
  [D_\mu,[D^\mu,D^\nu]] d
\end{align}
mixes under renormalization with the operators $O_i$ at the one-loop
level \cite{Buras:1992tc,Wise:1979at}.  This operator is of class (i),
i.e., it is nonzero in the limit of on-shell external states.  In this
limit, however, the operator $G_1$ becomes linearly dependent on the
four-quark operators $O_i$, and one finds
\cite{Gilman:1979bc,Wise:1979at,Buras:1992tc}
\begin{align}\label{eqn:onshelllimitgone}
  G_1 \stackrel{\mbox{\tiny{on-shell}}}{\longrightarrow} \Qp
\end{align}
with
\begin{align}\label{eqn:defqpbasisI}
 \Qp = Q_4 + Q_6 - \frac1{N_c}(Q_3 + Q_5)\,.
\end{align}

\begin{figure}[t]
  \begin{tabular}{lcccr}
    \includegraphics[width=3.0cm]{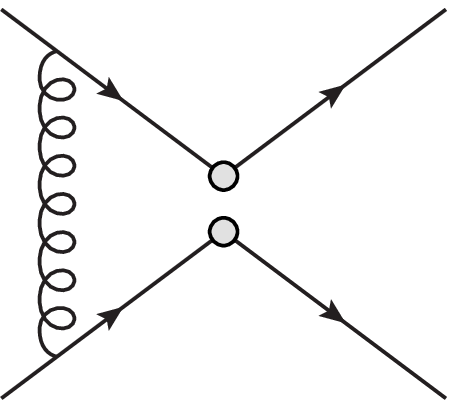}&~\hspace{0.1cm}~&
    \includegraphics[width=3.0cm]{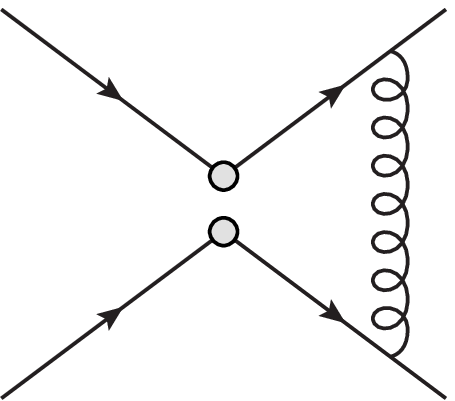}\\
    \includegraphics[width=3.0cm]{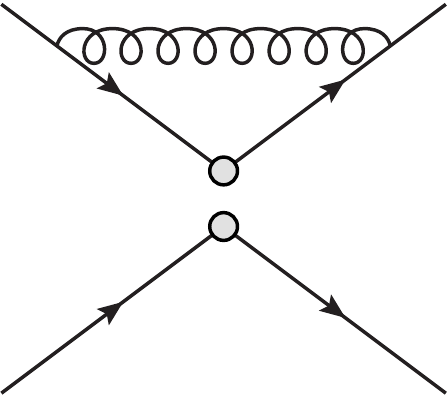}&~\hspace{0.1cm}~&
    \includegraphics[width=3.0cm]{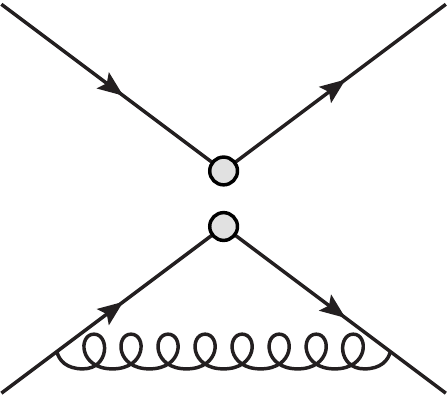}\\
    \includegraphics[width=3.0cm]{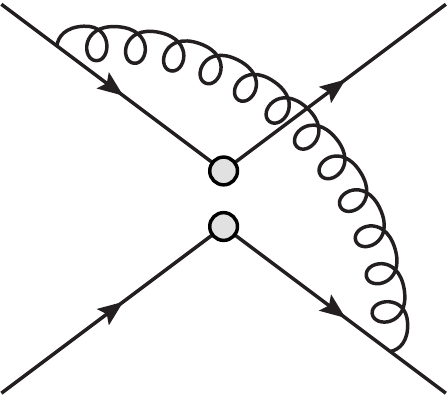}&~\hspace{0.1cm}~&
    \includegraphics[width=3.0cm]{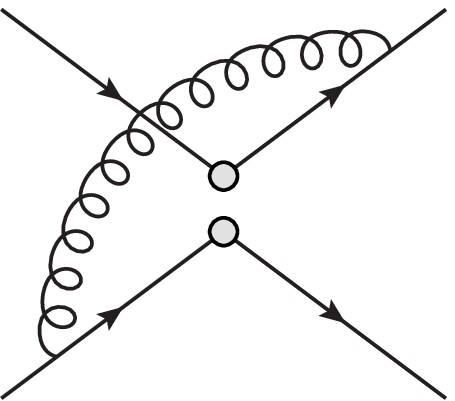}
  \end{tabular}
  \caption{Current-current contributions at one-loop order.}
  \label{fig:ccone}
\end{figure}

\begin{figure}[t]
  \includegraphics[height=2.5cm]{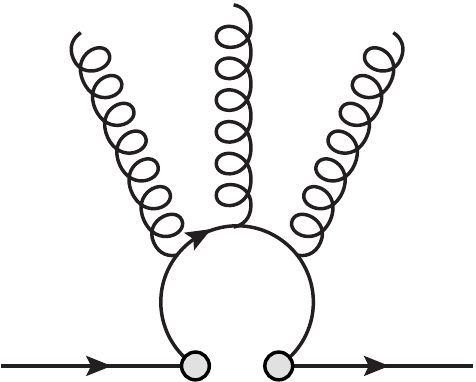}
  \includegraphics[height=2.5cm]{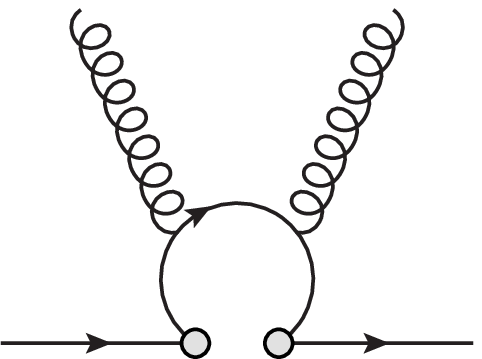}
  \includegraphics[height=2.5cm]{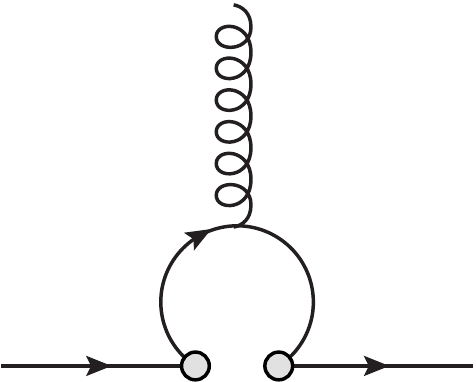}
  \caption{Penguin-type diagrams with up to three external gluons.
    Three analog penguin diagrams, where the gluons attach to a closed
    fermion loop, also need to be considered but are not shown here.}
  \label{fig:peng}
\end{figure}

At one-loop order the renormalized operator in the $\msbar$ scheme is
given by
\begin{align}\label{eqn:renorm}
  O_i^\msbar = O_i + a^\msbar_{ij} O_j + b^\msbar_{ik} \E_k +
  c^\msbar_i G_1
\end{align}
with
\begin{align}
 a^\msbar = Z^\msbar - \1\,.
\end{align}
There are two different types of diagrams that need to be considered:
current-current diagrams, where each fermion line involving a quark
field of the operator $O_i$ extends to the external quark fields, and
penguin diagrams, where one fermion line involving quark fields of the
operator $O_i$ begins and ends at the operator.  We depict the
corresponding one-loop diagrams in Fig.~\ref{fig:ccone} and
Fig.~\ref{fig:peng}. The one-loop current-current diagrams determine
the mixing of the four-quark operators with themselves (given in
$a^\msbar$), while the penguin diagrams determine the mixing of $G_1$
with the four-quark operators (given in $c^\msbar$).

We first consider the off-shell renormalization of operator basis I
and II.  The current-current contributions in $a^{\msbar}$ separate in
$2\times2$ blocks given
by~\cite{Gilman:1979bc,Wise:1980qb,Buras:1992tc,Ciuchini:1993vr}
\begin{align}
  (a^{\msbar}_{1,2}) &= (a^{\msbar}_{3,4}) 
  = (a^{\msbar}_{9,10})
  \notag\\ &=\frac{\alpha_s}{4\pi\epsilon}
  \begin{pmatrix}
    -3/N_c & 3 \\ 3 & -3/N_c
  \end{pmatrix},
\end{align}
and
\begin{align}
  (a^{\msbar}_{5,6}) &= (a^{\msbar}_{7,8})
  =\frac{\alpha_s}{4\pi\epsilon}
  \begin{pmatrix}
    3/N_c & -3 \\ 0 & 3/N_c-3N_c
  \end{pmatrix},
\end{align}
where the subscripts $i$, $j$ of $(a^{\msbar}_{i,j})$ denote that the
matrix acts on the space of operators $O_i$ and $O_j$.  This notation
is also used for other block-diagonal matrices in the remainder of
this section.  The matrix $a^\msbar$ is identical for basis I and II.

The penguin contributions in $c^\msbar$ are given by
\begin{align}
  c^\msbar_2 &= -c^\msbar_9 = \frac{\alpha_s}{4\pi\varepsilon}
  \frac13\,,
  & c^\msbar_3 &= \frac{\alpha_s}{4\pi\varepsilon} \frac23\,, \notag\\
  c^\msbar_4 &= c^\msbar_6 = \frac{\alpha_s}{4\pi\varepsilon}\,, &
  c^\msbar_{1,5,7,8,10}&=0
\end{align}
for basis I and II.  For on-shell matrix elements we use
Eq.~\eqref{eqn:onshelllimitgone} in Eq.~\eqref{eqn:renorm} and thus
reproduce the results for the anomalous dimensions of
Refs.~\cite{Gilman:1979bc,Wise:1980qb,Buras:1992tc,Ciuchini:1993vr}.

The set of evanescent operators $\{\E_k\}$ used in
Eq.~\eqref{eqn:renorm} to define the $\msbar$ scheme consists of
operators
\begin{align}\label{eqn:evanstart}
  E_{1} &= (\bar s_a \gamma_\mu\gamma_\nu\gamma^L_\rho u_b
  ) (\bar u_b \gamma^\mu\gamma^\nu{\gamma^\rho}^L d_a)
  - (16-4\varepsilon) Q_1\,, \notag\\
  E_{2} &= (\bar s_a \gamma_\mu\gamma_\nu\gamma^L_\rho u_a )
  (\bar u_b \gamma^\mu\gamma^\nu{\gamma^\rho}^L
  d_b)\notag\\&\quad
  -(16-4\varepsilon) Q_2\,,
\end{align}
and
\begin{align}
  \tilde E_{1} &= (\bar s_a \gamma_\mu\gamma_\nu\gamma^L_\rho d_a
  ) (\bar u_b \gamma^\mu\gamma^\nu{\gamma^\rho}^L u_b)
  - (16-4\varepsilon)\tilde Q_1\,, \notag\\
  \tilde E_{2} &= (\bar s_a \gamma_\mu\gamma_\nu\gamma^L_\rho d_b )
  (\bar u_b \gamma^\mu\gamma^\nu{\gamma^\rho}^L
  u_a) \notag\\&\quad-(16-4\varepsilon)\tilde Q_2\,,
\end{align}
as well as
\begin{align}
  E_3 &= (\bar s_a \gamma_\mu\gamma_\nu\gamma^L_\rho d_a
  ) \sum_{q=u,d,s} (\bar q_b \gamma^\mu\gamma^\nu{\gamma^\rho}^L q_b) \notag\\&\quad
  - (16-4\varepsilon) Q_3\,, \notag\\
  E_4 &= (\bar s_a \gamma_\mu\gamma_\nu\gamma^L_\rho d_b )
  \sum_{q=u,d,s} (\bar q_b \gamma^\mu\gamma^\nu{\gamma^\rho}^L
  q_a) \notag\\&\quad-(16-4\varepsilon) Q_4\,, \notag\\
  E_5 &= (\bar s_a \gamma_\mu\gamma_\nu\gamma^L_\rho d_a
  ) \sum_{q=u,d,s} (\bar q_b \gamma^\mu\gamma^\nu{\gamma^\rho}^R q_b) \notag\\&\quad
  - (4+4\varepsilon) Q_5\,, \notag\\
  E_6 &= (\bar s_a \gamma_\mu\gamma_\nu\gamma^L_\rho d_b )
  \sum_{q=u,d,s} (\bar q_b \gamma^\mu\gamma^\nu{\gamma^\rho}^R
  q_a) \notag\\&\quad-(4+4\varepsilon) Q_6\,,
\end{align}
and
\begin{align}
  E_7 &= \frac32 (\bar s_a \gamma_\mu\gamma_\nu\gamma^L_\rho d_a
  ) \sum_{q=u,d,s} e_q (\bar q_b \gamma^\mu\gamma^\nu{\gamma^\rho}^R q_b) \notag\\&\quad
  - (4+4\varepsilon) Q_7\,, \notag\\
  E_8 &= \frac32 (\bar s_a \gamma_\mu\gamma_\nu\gamma^L_\rho d_b )
  \sum_{q=u,d,s} e_q (\bar q_b \gamma^\mu\gamma^\nu{\gamma^\rho}^R
  q_a) \notag\\&\quad-(4+4\varepsilon) Q_8\,,
\end{align}
and finally
\begin{align}
  E_9 &= \frac32 (\bar s_a
  \gamma_\mu\gamma_\nu\gamma^L_\rho d_a ) \sum_{q=u,d,s} e_q (\bar q_b
  \gamma^\mu\gamma^\nu{\gamma^\rho}^L q_b) \notag\\&\quad
  - (16-4\varepsilon) Q_9\,, \notag\\
  E_{10} &= \frac32 (\bar s_a
  \gamma_\mu\gamma_\nu\gamma^L_\rho d_b ) \sum_{q=u,d,s} e_q (\bar q_b
  \gamma^\mu\gamma^\nu{\gamma^\rho}^L q_a)
  \notag\\&\quad-(16-4\varepsilon) Q_{10}\,, \label{eqn:evanend}
\end{align}
where $\gamma^{L/R}_\rho=\gamma_\rho(1\mp\gamma_5)$, and $a$ and $b$
are color indices.  The explicit contributions of order $\varepsilon$
are determined by the ``Greek'' method \cite{Tracas:1982gp} in
accordance with two-loop calculations such as
Ref.~\cite{Buras:2000if}.  For operator basis I we have $\{\E_k\} =
\{E_1,\ldots,E_{10}\}$.  The renormalization coefficients of the
evanescent operators in Eq.~\eqref{eqn:renorm} decompose in $2\times
2$ blocks and are given by
\begin{align}
  (b^{\msbar}_{1,2}) &= \frac{\alpha_s}{4\pi\varepsilon}
  \begin{pmatrix}
     N_c/4 - 1/(2N_c) & 1/4 \\
    1/2 & - 1/(2N_c)
  \end{pmatrix}, \notag\\
  (b^{\msbar}_{3,4}) &= (b^{\msbar}_{5,6}) = (b^{\msbar}_{7,8}) = (b^{\msbar}_{9,10}) \notag\\
  &=\frac{\alpha_s}{4\pi\varepsilon}\begin{pmatrix}
    - 1/(2N_c) & 1/2 \\
     1/4 & N_c/4 - 1/(2N_c)
   \end{pmatrix}
\end{align}
for basis I.  Similarly for basis II we have $\{\E_k\} = \{\tilde
E_1,\tilde E_2,E_3,\ldots,E_{10}\}$, and one obtains
\begin{align}
  (b^{\msbar}_{1,2}) &= (b^{\msbar}_{3,4}) = (b^{\msbar}_{5,6}) = (b^{\msbar}_{7,8}) = (b^{\msbar}_{9,10}) \notag\\
  &=\frac{\alpha_s}{4\pi\varepsilon}\begin{pmatrix}
    - 1/(2N_c) & 1/2 \\
    1/4 & N_c/4 - 1/(2N_c)
  \end{pmatrix}.
\end{align}

If operators transform in an irreducible representation of a given
symmetry, they only mix with other operators transforming in the same
irreducible representation.  In the case of the \reduced{} basis the
decomposition of operators according to irreducible representations of
$\SU(3)_L\otimes\SU(3)_R$ is given in Eqs.~\eqref{eqn:qprimeqe}.  The
set of operators used in Eq.~\eqref{eqn:renorm} for the \reduced{}
basis is given by $\{O_i\} = \{Q'_1,Q'_2,Q'_3,Q'_5,\ldots,Q'_8\}$ and
$\{\E_k\}=\{E_1,\ldots,E_{13}\}$, where the operators $E_{11}$,
$E_{12}$, and $E_{13}$ are defined in Eq.~\eqref{eqn:defevanqqtilde}.
The matrix ${a}^\msbar$ is then again block-diagonal with
\begin{align}
  ({a}^\msbar_1) &= \frac{\alpha_s}{4\pi\varepsilon} (3 -
  3/N_c)\,, \notag\\
  ({a}^\msbar_{2,3}) &= \frac{\alpha_s}{4\pi\varepsilon}
\begin{pmatrix}
  -3/N_c & 3  \\
  3 & -3/N_c
\end{pmatrix}, \notag\\
({a}^\msbar_{5,6}) &= ({a}^\msbar_{7,8}) =  \frac{\alpha_s}{4\pi\varepsilon}
\begin{pmatrix}
3/N_c & -3 \\
0 & 3/N_c - 3N_c
\end{pmatrix},
\end{align}
and
\begin{align}
  {c}^\msbar =\frac{\alpha_s}{4\pi\varepsilon}
  \begin{pmatrix}
    0 & 0 &\frac13 & 0 & 1 & 0 & 0
  \end{pmatrix}^T.
\end{align}
The nonzero elements of ${b}^\msbar$ are given in
Tab.~\ref{tab:bprimed}.  In the \reduced{} basis the on-shell limit of
$G_1$ is given by
\begin{align}\label{eqn:defqpreducedbasis}
  \Qp &= (2-3/N_c) Q'_2 + (3-2/N_c) Q'_3 - Q'_5/N_c + Q'_6
  \notag\\&\quad - E_{11} - E_{12} - E_{13}\,.
\end{align}
\begin{table}[t]
  \centering
  \begin{tabular}{lcl}\hline
    $(i,j)$ & & ${b}^\msbar_{ij}/\frac{\alpha_s}{4\pi\varepsilon}$ \\\hline\hline
    $(1,1)$ & & $1$\\ 
    $(1,2)$ & & $-1/N_c$\\
    $(1,9)$ & & $-1/N_c$\\ 
    $(1,10)$ & & $1$\\ 
    $(1,11)$ & & $9$\\ 
    $(1,12)$ & & $-6$\\ 
    $(1,13)$ & & $3$\\ 
    $(2,1)$ & & $-1/5$\\ 
    $(2,2)$ & & $1/(5N_c)$\\ 
    $(2,3)$ & & $-1/(6N_c)$\\ 
    $(2,4)$ & & $1/6$\\ 
    $(2,9)$ & & $-2/(15N_c)$\\ 
    $(2,10)$ & & $2/15$\\ 
    $(2,11)$ & & $-9/5$\\ 
    $(2,12)$ & & $-9/5$\\ 
    $(2,13)$ & & $-3/5$\\ 
    \hline
  \end{tabular}
  \hspace{0.25cm}
  \begin{tabular}{lcl}\hline
    $(i,j)$ & & ${b}^\msbar_{ij}/\frac{\alpha_s}{4\pi\varepsilon}$ \\\hline\hline
    $(3,1)$ & & $3/10$\\
    $(3,2)$ & & $-3/(10N_c)$\\
    $(3,9)$ & & $1/(5N_c)$\\ 
    $(3,10)$ & & $-1/5$\\ 
    $(3,11)$ & & $6/5$\\ 
    $(3,12)$ & & $6/5$\\ 
    $(3,13)$ & & $-3/5$\\
    $(5,5)$ & & $-1/(2N_c)$\\ 
    $(5,6)$ & & $1/2$\\ 
    $(6,5)$ & & $1/4$\\ 
    $(6,6)$ & & $N_c/4-1/(2N_c)$\\ 
    $(7,7)$ & & $-1/(2N_c)$\\ 
    $(7,8)$ & & $1/2$\\ 
    $(8,7)$ & & $1/4$\\ 
    $(8,8)$ & & $N_c/4-1/(2N_c)$\\\\
    \hline
  \end{tabular}
  \caption{The nonzero elements of ${b}^\msbar$ for the \reduced{} basis.}
  \label{tab:bprimed}
\end{table}

\subsection{Regularization-independent schemes - The mixing of
  four-quark operators\label{sec:rifour}}
In the following we define $\ri$ schemes for the $\Delta S=1$
four-quark operator bases.  The $\ri$ schemes are defined
non-perturbatively, so that they can be used in lattice simulations as
well as in continuum perturbation theory.  In lattice calculations the
$\ri$ schemes serve as intermediate schemes and allow for a
straightforward conversion of the studied quantity to the $\msbar$
scheme.  In this subsection we focus on the mixing of four-quark
operators $O_i$ among themselves.  In terms of Eq.~\eqref{eqn:genren}
this means that we provide the $\ri$ renormalization conditions to
determine the renormalization matrix $Z^\ri$.  The renormalization
conditions which determine the mixing of two-quark operators $G_l$ and
$N_m$ with the four-quark operators $O_i$, i.e., $c^\ri$ and $d^\ri$,
are discussed in subsection \ref{sec:ritwo}.  While the content of
this subsection thus suffices to define the $\ri$ schemes for the
$(27,1)$ and $(8,8)$ operators of the \reduced{} basis, the discussion
of subsection \ref{sec:ritwo} is necessary to complete the $\ri$
schemes for the $(8,1)$ operators.

Let us consider a set of bare operators $\{\O_i\}$ that is closed
under renormalization and contains the set of four-quark operators
$\{O_i\}$, i.e., $\{O_i\} \subset \{\O_i\}$.  The renormalized
operators $\O^x_i$ in the $\msbar$ or an $\ri$ scheme can be expressed
in terms of the bare operators by
\begin{align}
  \O^x_i = Z^x_{ij} \O_j,\quad x\in\{\msbar,\ri\}.
\end{align}
The renormalization conditions of the $\ri$ schemes are formulated in
terms of renormalized amputated Green's functions $\Gamma^y_n(\O^x_i)$
with a single insertion of such an operator $\O^x_i$, where $y$
denotes the wave function renormalization scheme and $n$ indicates the
external states of the Green's function.  In this subsection we only
consider Green's functions with four external quarks which we denote
by $n=4$.

The renormalized amputated Green's function $\Gamma^y_4$ is related to
the bare amputated Green's function $\Gamma_4$ by
\begin{align}
\label{eq:green}
  \Gamma^y_4(\O^x_i) = \frac1{{(Z^{y}_q)}^2} Z^x_{ij} \Gamma_4(\O_j)\,,
\end{align}
where $Z^y_q$ is the quark wave function renormalization constant,
which relates the bare quark field $f$ to the renormalized quark field
$f^y={(Z^y_q)}^{1/2}f$ in the scheme $y$.

Various $\rismom$ wave function renormalization schemes have been
proposed in Ref.~\cite{Sturm:2009kb} and will be used later.  In order
to convert the quark fields from the $\ri$ scheme to the $\msbar$
scheme matching factors
\begin{align}
\label{eq:cq}
C^y_q =\frac{Z^\msbar_q}{Z^y_q}
\end{align}
have been computed with $f^\msbar={(C^y_q)}^{1/2}f^y$.

The operators which are renormalized in an $\ri$ scheme can be
converted to the $\msbar$ scheme
\begin{align}\label{eqn:genconv}
  \O^\msbar_i = S^{\ri \to \msbar}_{ij} \O^\ri_j
\end{align}
using the conversion matrix
\begin{align}
\label{eq:co}
   S^{\ri \to \msbar} = Z^\msbar (Z^\ri)^{-1}\,.
\end{align} 
The renormalization matrix $Z^\msbar$ has been discussed in the
previous Sec.~\ref{sec:MSbar}, whereas the matrix $Z^\ri$ is
determined by the renormalization conditions of the $\ri$ scheme.
Using Eqs.~\eqref{eq:cq} and \eqref{eq:co} one can express
Eq.~\eqref{eq:green} completely in terms of matching factors and
renormalized Green's functions
\begin{align}\label{eqn:gammaritomsbar}
  \Gamma^y_4(\O^\ri_i) &= {(C^y_q)}^2 {(S^{\ri \to \msbar})}^{-1}_{ij} \Gamma_4^\msbar(\O^\msbar_j)\,.
\end{align}

In our case the set of operators $\{\O_i\}$ is given by
$\{O_i,G_l,N_m,\E_k\}$, where $\{O_i\}$ are the four-quark operators
of basis I, II, or the \reduced{} basis and $\{\E_k\}$ are the
corresponding evanescent operators used to define the $\msbar$ scheme
in the previous section.  Since the evanescent operators are an
artifact of dimensional regularization and escape a
regularization-independent definition, their contribution to the
right-hand side of Eq.~\eqref{eqn:genconv} should be avoided.  In
order to achieve this, one defines new subtracted operators
$\{\O^\sub_i\}$ from the set of bare operators
$\{\O_i\}=\{O_i,G_l,N_m\}$.

In general for any regulator one first has to subtract all
contributions specific to the regularization from a given bare
operator.  Hence, in dimensional regularization we perform modified
minimal subtraction of the evanescent operators
\cite{Ciuchini:1995cd,Donini:1995xj}, i.e., the subtracted operator is
defined as
\begin{align}\label{eqn:defsubQE}
  \O^{\sub}_i = \O_i + s^\msbar_{ik} \E_k \,,
\end{align}
where $s^\msbar$ is chosen such that it cancels all contributions of
$\E_k$ proportional to a pole in $\varepsilon$.  In principle the
choice of evanescent operators $\{\E_k\}$ used on the right-hand side
of Eq.~\eqref{eqn:defsubQE} is not unique.  A useful choice is the
basis $\{\E_k\}$ given in the previous section to define the $\msbar$
scheme for operator basis I, II, and the \reduced{} basis, and for
which we therefore have
\begin{align}
  \O^{\sub,\msbar}_i = \O^{\msbar}_i\,.
\end{align}
For convenience we adopt this definition of the subtracted operator in
the following.  In a lattice regularization one has to perform a
similar subtraction of lower-dimensional two-quark operators
\cite{Blum:2001xb} which do not occur in dimensional regularization.
From now on we only consider such subtracted operators and therefore
drop the explicit notation of the superscript ``$\sub$''.

In the $\ri$ schemes one imposes the renormalization condition
\cite{Martinelli:1994ty} that amputated Green's functions with given
off-shell external states at a given momentum point and in a fixed
gauge coincide with their tree-level value.  This condition is made
explicit by choosing a certain set of projectors $\{P_{4j}\}$ in
spinor, color, and flavor space that is applied to the four-quark
amputated Green's function and imposing
\begin{align}\label{eqn:rischemeA}
  P_{4j} \Gamma^y_4(O^\ri_i)\big\vert_{\mbox{\tiny{mom. conf.}}} &=
  P_{4j}\Gamma^\tree_4(O_i)\,,
\end{align}
where $\Gamma^\tree_4(O_i)$ denotes the insertion of the operator
$O_i$ in the amputated Green's function $\Gamma_4$ at tree level.
Since we are only interested in $\ri$-renormalized operators
$O^\ri_i$, we do not provide $\ri$ conditions for the operators $G_l$
and $N_m$ in this work.  For a set of $n$ operators $O_i \in
\{O_1,\ldots,O_n\}$ we will provide $n$ projectors $P_{4j} \in \{
P_{41},\ldots,P_{4n} \}$ to determine the $n\times n$ elements of the
renormalization matrix $Z^\ri$.  If no two-quark operators mix with
the four-quark operators $O_i$, i.e., $c^\ri=0$ and $d^\ri=0$ in
Eq.~\eqref{eqn:genren}, the renormalization matrix is given by
\cite{Blum:2001xb}
\begin{align}\label{eqn:defzria}
  Z^\ri  &= (Z^y_q)^2 F_4 [M_4(O)]^{-1} \,,
\end{align}
where
\begin{align}
  [M_4(O)]_{ij} &= P_{4j} \Gamma_4(O_i)\big\vert_{\mbox{\tiny{mom. conf.}}} \,.
\end{align}
and $F_4 = \lim_{\alpha_s\to 0} M_4(O)$.  If two-quark operators mix
with the operators $O_i$, Eq.~\eqref{eqn:defzria} has to be modified
slightly, which is discussed in subsection \ref{sec:ritwo}.

The matrix $Z^\ri$ given in Eq.~\eqref{eqn:defzria} depends on the
regulator used to define the $\ri$ scheme, i.e., the matrix $Z^\ri$
obtained using a lattice regulator is different from the matrix
$Z^\ri$ obtained in dimensional regularization.  The $\ri$ condition
of Eq.~\eqref{eqn:rischemeA}, however, fixes the physical amplitudes
of the $\ri$-renormalized operators at a certain off-shell momentum
point to its tree-level value, which is independent of the choice of
the regulator.  Therefore, the physical amplitudes of the
$\ri$-renormalized operators agree for all choices of the regulator.

We define the $\ri$ scheme in the limit of vanishing quark masses.
The choice of projectors $\{P_{4j}\}$, the gauge fixing, and the
momentum configuration of the off-shell amputated Green's functions
$\Gamma^y_4$ defines the scheme up to mixing with two-quark operators.
The explicit form of the projectors will be discussed later in
Sec.~\ref{sec:results}.  Note that Eq.~\eqref{eqn:rischemeA} matches
the amputated Green's function of a certain physical process with
insertion of $\ri$ operators at a certain off-shell momentum point.

\begin{figure}[t]
  \includegraphics[scale=0.7]{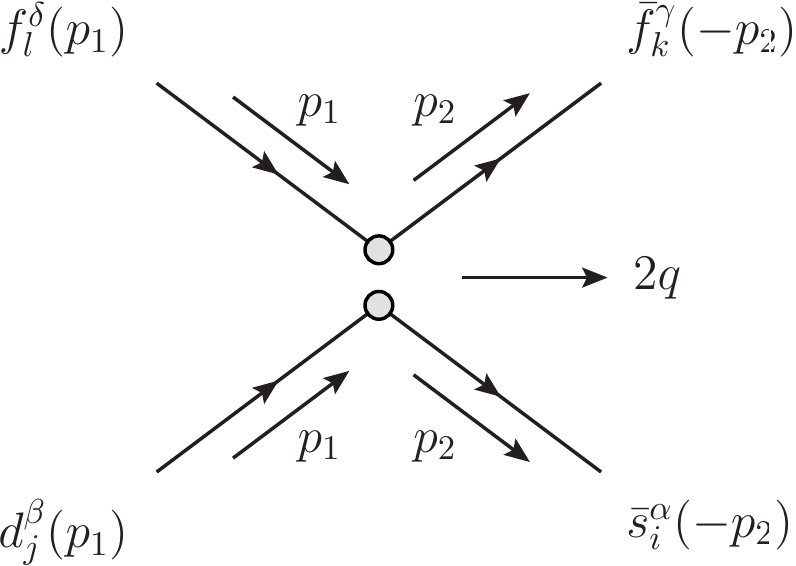}
  \caption{Momentum flow and open indices for the four-quark Green's
    functions $\Gamma_4$.  We explicitly write the quark fields $\bar
    s^\alpha_i$, $d^\beta_j$, $\bar f^\gamma_k$, and $f^\delta_l$ with
    spinor indices $\alpha$, $\beta$, $\gamma$, and $\delta$ and color
    indices $i$, $j$, $k$, and $l$.  The momenta of the quark fields
    are given in brackets, which are counted as incoming, i.e., they
    flow towards the four-quark operator. An additional momentum of
    $2q$ leaves the operator as indicated by the arrow.}
  \label{fig:cctree}
\end{figure}

\begin{figure}[t]
  \begin{tabular}{cc}
    \includegraphics[scale=0.7]{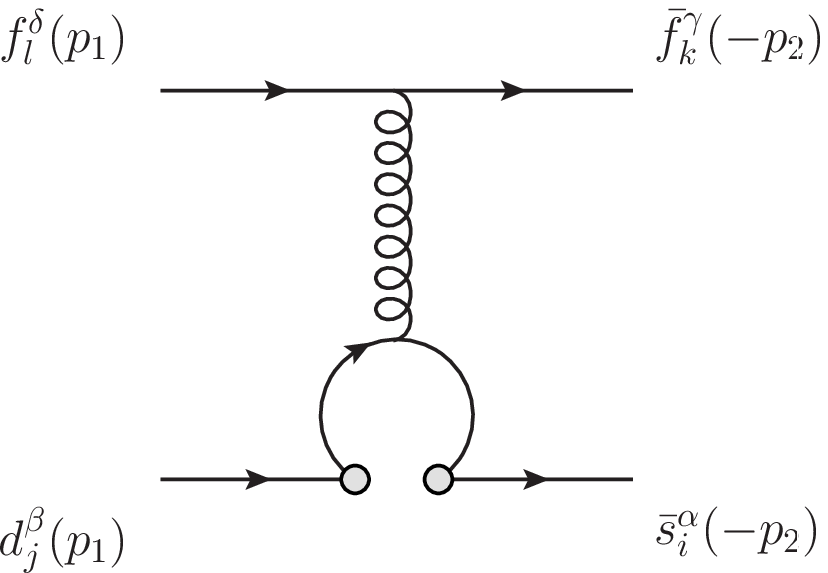}\\
    \multicolumn1c{$(a)$} \\ \\
    \includegraphics[scale=0.7]{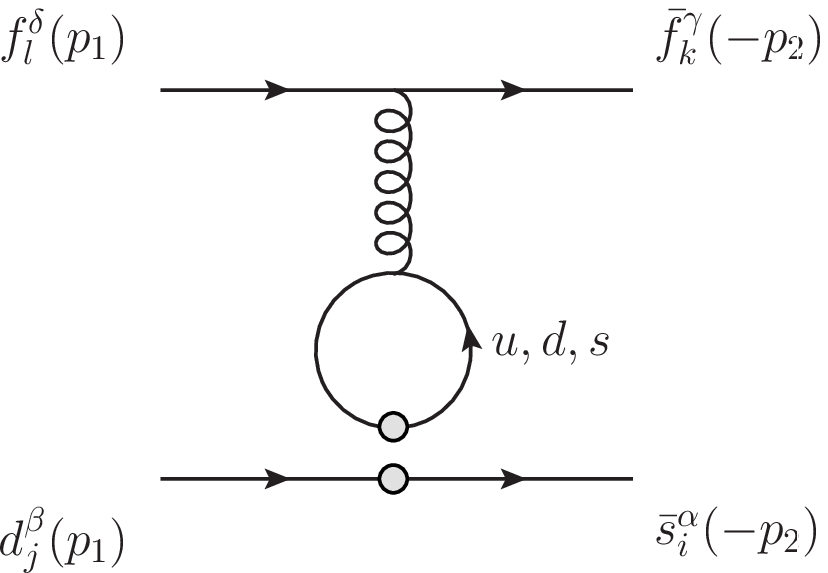}\\
    \multicolumn1c{$(b)$}
  \end{tabular}
  \caption{Penguin contributions at one loop and the respective
    momentum configuration.  We explicitly write the fields and
    momentum configuration as in Fig.~\ref{fig:cctree}.  We do not
    explicitly show the flow of the additional momentum $2q$ here.}
  \label{fig:pone}
\end{figure}

In particular for the $\Delta S=1$ operators we consider the off-shell
amputated Green's functions
$\Gamma_4^{\alpha\beta\gamma\delta;ijkl;f}$ of the process
\begin{align}
  d(p_1)\bar s(-p_2) \to \bar f(-p_3) f(p_4)
\end{align}
with quarks $f=u,d,s$, momenta $p_1$, $p_2$, $p_3$, $p_4$, color
indices $i,j,k,l$, and spinor indices $\alpha$, $\beta$, $\gamma$,
$\delta$.  Using crossing symmetry we could equally well consider the
scattering amplitude
\begin{align}
 d(p_1)f(p_3) \to s(p_2)f(p_4)\,.
\end{align}
The momentum configuration used in Eq.~\eqref{eqn:rischemeA} to define
the $\rismom$ scheme is then given by
\begin{align}
  p_3 =p_1 \,, \qquad p_4=p_2
\end{align}
with
\begin{align}\label{eqn:momconfne}
  p_1^2=p_2^2=q^2 = -\mu_s^2 \,, \qquad  q=p_1-p_2\,,
\end{align}
in Minkowski space, where $\mu_s$ is the subtraction scale.  This
momentum configuration and our convention for open indices for the
amputated Green's functions $\Gamma_4$ is shown in
Fig.~\ref{fig:cctree}.  In Fig.~\ref{fig:pone} we explicitly show the
corresponding penguin diagrams at one-loop order, where a momentum
transfer $2q$ leaves the operator.  This choice of momenta is
non-exceptional (no partial sum of incoming external momenta vanishes)
which has the advantage of suppressing unwanted infrared effects in
the lattice simulation.
\relax

In Ref.~\cite{Ciuchini:1995cd} a $\rimom$ scheme was defined which
uses exceptional kinematics and a different momentum point for
current-current and penguin diagrams, see Fig.~5 of
Ref.~\cite{Ciuchini:1995cd}.  The momentum configuration in the
$\rimom$ scheme is given by $p=p_1=p_2=p_3=p_4$ at $\mu_s^2=-p^2$ for
current-current diagrams and $p=p_1=p_4$ and $p'=p_2=p_3$ at
$\mu_s^2=-q^2$ for penguin diagrams.  We consider the $\rimom$ scheme
in the following for completeness, illustration, and check of our
calculation.  Another $\ri$ scheme with exceptional kinematics which
uses the same momentum point for current-current and penguin diagrams
is discussed in Ref.~\cite{Liu:2011zzz}.

\subsection{Regularization-independent schemes - The mixing of
  two-quark operators\label{sec:ritwo}}
In the following we discuss the mixing of the two-quark operators
$G_l$ and $N_m$ with the four-quark operators $O_i$ in the $\ri$
schemes.  Such mixing occurs, e.g., for the $(8,1)$ operators of the
\reduced{} basis.  The two-quark operators mix through the penguin
diagrams, and therefore their mixing should be determined from
amplitudes which only receive contributions from penguin-type
contractions.

\begin{figure}[t]
  \includegraphics[scale=0.7]{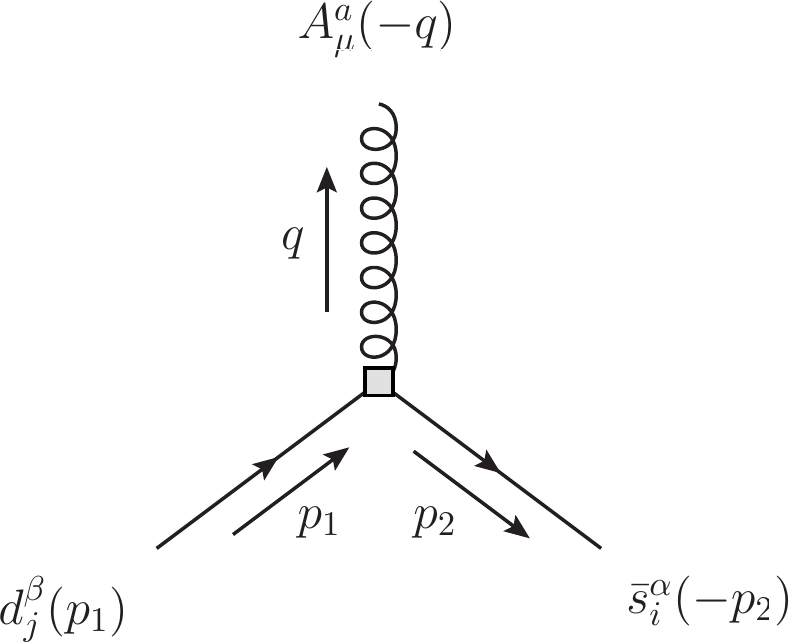}
  \caption{Momentum flow and open indices for the two-quark Green's
    functions $\Gamma_2$.  We explicitly write the quark fields $\bar
    s^\alpha_i$, $d^\beta_j$ and the gluon field $A^a_\mu$ with spinor
    indices $\alpha$, $\beta$, color indices $i$, $j$, and Lorentz
    index $\mu$.  The index $a=1,\ldots,N_c^2-1$ is contracted with
    the generators of $\SU(N_c)$.  The incoming momenta of the fields
    are also given in brackets.}
  \label{fig:tqtree}
\end{figure}

Two kinds of such amplitudes will be considered in the following: (a)
amputated Green's functions $\Gamma_2$ with two external quarks and
one external gluon and (b) amputated Green's functions $\Gamma_{4p}$
with four external quarks corresponding to the process $d f \to s f$,
where the quark flavor $f \notin \{ u,d,s \}$.  The momentum flow and
our convention for open indices in case (a) is shown in
Fig.~\ref{fig:tqtree}.  The corresponding momenta $p_1$, $p_2$, and
$q$ for the $\rismom$ scheme satisfy Eq.~\eqref{eqn:momconfne}.  This
momentum configuration is also non-exceptional.  In case (b) we adhere
to the momentum configuration and choice of indices shown in
Figs.~\ref{fig:cctree} and \ref{fig:pone}.  In the following we define
the $\ri$ schemes in such a way that both cases, (a) and (b), lead to
identical one-loop conversion factors from the $\ri$ scheme to the
$\NDR$ scheme.  At higher loops, however, the $\ri$ schemes defined by
(a) differ from the $\ri$ schemes defined by (b).  Case (b) can be
implemented in lattice simulations in a straightforward way
\cite{Blum:2001xb}, while case (a) requires an external gluonic state.
Nevertheless, the availability of both cases should be beneficial for
lattice simulations, in particular to estimate higher-loop effects
that are neglected in this work.
\relax

Case (a) is formulated in terms of renormalized amputated Green's
functions $\Gamma^y_2$ which are related to the bare amputated Green's
functions $\Gamma_2$ by
\begin{align}
\label{eq:greentwo}
  \Gamma^y_2(\O^x_i) = \frac1{Z^{y}_q (Z^y_A)^{1/2}} Z^x_{ij} \Gamma_2(\O_j)\,,
\end{align}
where $Z^y_A$ is the gluon wave function renormalization constant,
which relates the bare gluon field $A_\mu$ to the renormalized gluon
field $A^y_\mu=(Z^y_A)^{1/2} A_\mu$ in the scheme $y$.  Using
Eqs.~\eqref{eq:cq} and \eqref{eq:co} one can express
Eq.~\eqref{eq:greentwo} as
\begin{align}\label{eqn:gammatworitomsbar}
  \Gamma^y_2(\O^\ri_i) &= (Z^\msbar_A/Z^y_A)^{1/2} C^y_q \notag\\&\quad\times {(S^{\ri \to
      \msbar})}^{-1}_{ij} \Gamma^\msbar_2(\O^\msbar_j)\,.
\end{align}
We then determine the renormalization coefficients $c^\ri$ and $d^\ri$
by imposing
\begin{align}\label{eqn:rischemeB}
  P_{2k} \Gamma^y_2(O^\ri_i)\big\vert_{\mbox{\tiny{mom. conf.}}} &=
  P_{2k}\Gamma^\tree_2(O_i) = 0
\end{align}
for a certain set of projectors $\{P_{2k}\}$.  Since only tree-level
insertions of $G_l$ and $N_m$ need to be considered in this work, we
do not provide renormalization conditions to define $\ri$-renormalized
operators $G^\ri_l$ and $N^\ri_m$ .  The definition of $\ri$ schemes
for the two-quark operators $G_l$ and $N_m$ is beyond the scope of
this work.

The $\ri$ conditions given in Eqs.~\eqref{eqn:rischemeA} and
\eqref{eqn:rischemeB} then allow for a non-perturbative determination
of the renormalization matrices $Z^\ri$, $c^\ri$, and $d^\ri$ defined
in Eq.~\eqref{eqn:genren}.  Without loss of generality we set
$d^\ri=0$ in the following.  We find
\begin{align}\label{eqn:defzcdA}
  Z^\ri  &= (Z^y_q)^2 F_4 [M_4(O)]^{-1} (\1 - z^\ri)^{-1} \,, \notag\\
  c^\ri &= -Z^\ri M_2(O) [M_2(G)]^{-1}
\end{align}
with
\begin{align}\label{eqn:defzcdC}
  z^\ri & = M_2(O) [M_2(G)]^{-1} M_4(G) [M_4(O)]^{-1}
\end{align}
and
\begin{align}\label{eqn:defzcdB}
  [M_2(\O)]_{ik} &= P_{2k} \Gamma_2(\O_i)\big\vert_{\mbox{\tiny{mom. conf.}}}\,.
\end{align}

Case (b) is formulated in terms of the renormalized amputated Green's
function $\Gamma^y_{4p}$ corresponding to the process $ sf \to d f$
that is shown in Fig.~\ref{fig:pone} at one loop.  Note that the quark
flavor $f \notin \{ u,d,s\}$ for $\Gamma^y_{4p}$ as opposed to $f \in
\{ u,d,s\}$ for $\Gamma^y_4$ which was used in subsection
\ref{sec:rifour}.  We then determine the mixing of the two-quark
operators by imposing
\begin{align}\label{eqn:rischemeC}
  P_{4p,k} \Gamma^y_{4p}(O^\ri_i)\big\vert_{\mbox{\tiny{mom. conf.}}} &=
  P_{4p,k}\Gamma^\tree_{4p}(O_i) = 0
\end{align}
for a certain set of projectors $\{P_{4p,k}\}$.  The renormalization
coefficients for case (b) can be obtained from
Eqs.~\eqref{eqn:defzcdA} and \eqref{eqn:defzcdC} by replacing $M_2 \to
M_{4p}$ with
\begin{align}\label{eqn:defzcdD}
  [M_{4p}(\O)]_{ik} &= P_{4p,k} \Gamma_{4p}(\O_i)\big\vert_{\mbox{\tiny{mom. conf.}}}\,.
\end{align}
We provide the explicit projectors used to define the $\rismom$
schemes in Sec.~\ref{sec:results}.  We will define the set of
projectors $\{P_{2k}\}$ and $\{P_{4p,k}\}$ such that the resulting
$\rismom$ schemes agree at one loop.  From now on we refer to
$\rismom$ schemes defined using case (a) as $\rismom_2$ schemes, while
we do not write a subscript for $\rismom$ schemes defined using case
(b).

Since in the physical application of the $\ri$ scheme one is not
interested in the conversion of the operators $G_l$ and $N_m$ from the
$\ri$ scheme to the $\msbar$ scheme, it is useful to decompose the
matrix $S^{\ri\to\msbar}_{ij}$ of Eq.~\eqref{eq:co} into several
blocks. The conversion relation of Eq.~\eqref{eqn:genconv} for
subtracted operators is then given by
\begin{align}\label{eqn:generalritomsbarblock}
  O_i^{\msbar} &= O^{\ri}_i + \Delta a^{\ri\to\msbar}_{ij} O^{\ri}_j
  + \Delta c^{\ri\to\msbar}_{il} G^{\ri}_l
  \notag\\&\quad+ \Delta d^{\ri\to\msbar}_{im} N^{\ri}_m\,,
\end{align}
where the operators $O_i$ are either of basis I, II, or the chiral
basis.  The three blocks $\Delta a^{\ri\to\msbar}_{ij}$, $\Delta
c^{\ri\to\msbar}_{il}$, and $\Delta d^{\ri\to\msbar}_{im}$ are
obtained by imposing the renormalization condition of
Eqs.~\eqref{eqn:rischemeA} and \eqref{eqn:rischemeB}, i.e., we have to
evaluate $\Gamma^\msbar_n(O_i^\msbar)$ with off-shell external legs,
see Eq.~\eqref{eqn:gammaritomsbar}.

We denote on-shell matrix elements of the operators $O_i^\msbar$ with
a general external momentum setting by $\braket{O_i^\msbar}$ for which
one finds
\begin{align}\label{eqn:defr}
  \braket{O_i^\msbar} = R^{\ri\to\msbar}_{ij} \braket{O_j^\ri}
\end{align}
since the on-shell matrix elements of the two-quark operators $G_l$
and $N_m$ are related to the on-shell matrix elements of $O_i$.  At
the one-loop level only the two-quark operator $G_1$ contributes to
the right-hand side of Eq.~\eqref{eqn:generalritomsbarblock}, and one
has
\begin{align}
  \braket{G_1} = \braket{\Qp}\,.
\end{align}

In Sec.~\ref{sec:results} we compute $\Delta a^{\ri\to\msbar}$ and
$\Delta c^ {\ri\to\msbar}$ as well as the on-shell conversion factors
\begin{align}\label{eq:deltar}
  \Delta r^{\ri\to\msbar} = R^{\ri\to\msbar} - \1
\end{align}
at one loop for different $\ri$ schemes and for operator basis I, II,
and the \reduced{} basis.

\subsection{The Wilson coefficients of the \reduced{} basis\label{sec:wilson}}
For lattice calculations it is advantageous to consider the \reduced{}
operator basis that uses the classification of operators according to
irreducible representations of the chiral symmetry.  In this basis the
operators of each irreducible representation can be renormalized
independently.  Therefore the effective Hamiltonian should be
expressed in terms of the operators ${Q'}^\ri_i$.

On-shell matrix elements of the effective Hamiltonian defined in
Eq.~\eqref{eq:Hamilton} in terms of ${Q'}^\ri_i$ read
\begin{align}
  \braket{\mathcal{H}^{\Delta S=1}_{\rm eff}}&={G_{F}\over\sqrt{2}}\sum_{i}{C'}^{\msbar}_{i}(\mu) \braket{{Q'}^\msbar_i(\mu)} \notag\\
&={G_{F}\over\sqrt{2}}\sum_{i,j}{C'}^{\msbar}_{i}(\mu) {R'}^{\ri\to\msbar}_{ij}(\mu)\notag\\
&\qquad\qquad\times\braket{{Q'}^{\ri}_{j}(\mu)}\,,
\end{align}
where $\mu$ is the renormalization scale and ${R'}^{\ri\to\msbar}$ is
the conversion matrix of Eq.~\eqref{eqn:defr} for the \reduced{}
basis.  The Wilson coefficients are, however, typically given for the
traditional operator bases in
Refs.~\cite{Altarelli:1980fi,Buras:1991jm,Buras:1993dy,Ciuchini:1993vr}.
Therefore it remains to relate the above Wilson coefficients
${C'}^\msbar_i$ to the known Wilson coefficients $C^\msbar_i$
corresponding to the operators $O_i$ of basis I or II.

To this end we first note that Eqs.~\eqref{eqn:diffeqndim} and
\eqref{eqn:qprimeqe} hold for $\ri$-renormalized operators without
contributions of evanescent operators, i.e.,
\begin{align}\label{eqn:deft}
  O^\ri_i - T_{ij} {Q'}^\ri_j = 0
\end{align}
with
\begin{align}
  T &=\scriptsize
  \begin{pmatrix}
    1/5     & 1     & \cdot  & \cdot  & \cdot  & \cdot & \cdot \\
    1/5     & \cdot & 1      & \cdot  & \cdot  & \cdot & \cdot\\
    \cdot   & 3     & 2      & \cdot  & \cdot  & \cdot & \cdot\\
    \cdot   & 2     & 3      & \cdot  & \cdot  & \cdot & \cdot\\
    \cdot   & \cdot & \cdot  & 1      & \cdot  & \cdot & \cdot\\
    \cdot   & \cdot & \cdot  & \cdot  & 1      & \cdot & \cdot\\
    \cdot   & \cdot & \cdot  & \cdot  & \cdot  & 1     & \cdot\\
    \cdot   & \cdot & \cdot  & \cdot  & \cdot  & \cdot & 1\\
    3/10    & \cdot & -1     & \cdot  & \cdot  & \cdot & \cdot\\
    3/10    & -1    & \cdot  & \cdot  & \cdot  & \cdot & \cdot
  \end{pmatrix},
\end{align}
where the matrix $T$ can be read off from Eqs.~\eqref{eqn:diffeqndim}
and~\eqref{eqn:qprimeqe}.  This equation holds for the operators $O_i$
of basis I and II.  Note that $Q_5$--$Q_8$ are equal to
$Q'_5$--$Q'_8$, see Eqs.~\eqref{eqn:qprimeqe}, and therefore the
respective sub-block in $T$ is proportional to the unit matrix.  In
the $\msbar$ scheme, however, finite contributions of evanescent
operators modify Eq.~\eqref{eqn:deft} to
\begin{align}\label{eqn:deftms}
  O^\msbar_i - T_{ij} {Q'}^\msbar_j =  \Delta t_i G_1
\end{align}
at the one-loop level.  The coefficients $\Delta t_i$ are given by
\begin{align}
  \Delta t_4 = -\frac{\alpha_s}{4\pi}\,, \qquad \Delta t_i = 0 \quad \text{for} \quad i\ne 4
\end{align}
for operators $O_i$ of basis I and
\begin{align}
 \Delta t_2 &= -\frac13 \frac{\alpha_s}{4\pi}\,, \qquad  \Delta t_4 = -\frac{\alpha_s}{4\pi}\,, \notag\\
 \Delta t_i &= 0 \quad \text{for} \quad i\ne 2,4
\end{align}
for operators $O_i$ of basis II.

Since the Hamiltonian is independent of the choice of operator basis,
we have
\begin{align}
 \sum_{i=1}^7 {C'}^\msbar_i(\mu) \braket{ {Q'}^\msbar_i(\mu) } = \sum_{i=1}^{10} C^\msbar_i(\mu) \braket{ {O}^\msbar_i(\mu) }\,.
\end{align}
Therefore using Eqs.~\eqref{eqn:deftms} and $\braket{G_1} =
\braket{\Qp}$ we can determine $\Delta T$ in
\begin{align}\label{eqn:defdt}
  {C'}^\msbar_j(\mu) = C^\msbar_i(\mu) (T_{ij} + \Delta T_{ij})\,.
\end{align}
At one-loop order one finds \cite{Buras:1993dy}
\begin{align}
  \Delta T^\msbar_{\rm I} = \frac{\alpha_s}{4\pi} \scriptsize \begin{pmatrix}
    \cdot   & \cdot   & \cdot  & \cdot  & \cdot  & \cdot & \cdot \\
    \cdot   & \cdot   & \cdot  & \cdot  & \cdot  & \cdot & \cdot\\
    \cdot   & \cdot   & \cdot  & \cdot  & \cdot  & \cdot & \cdot\\
    \cdot   & \frac{3}{N_c}-2   & \frac{2}{N_c}-3  & \frac{1}{N_c} & -1  & \cdot & \cdot\\
    \cdot   & \cdot   & \cdot  & \cdot  & \cdot  & \cdot & \cdot\\
    \cdot   & \cdot   & \cdot  & \cdot  & \cdot  & \cdot & \cdot\\
    \cdot   & \cdot   & \cdot  & \cdot  & \cdot  & \cdot & \cdot\\
    \cdot   & \cdot   & \cdot  & \cdot  & \cdot  & \cdot & \cdot\\
    \cdot   & \cdot   & \cdot  & \cdot  & \cdot  & \cdot & \cdot\\
    \cdot   & \cdot   & \cdot  & \cdot  & \cdot  & \cdot & \cdot
  \end{pmatrix}
\end{align}
for the Wilson coefficients $C^\msbar_i$ corresponding to basis I
and
\begin{align}
  \Delta T^\msbar_{\rm II} = \frac{\alpha_s}{4\pi} \scriptsize \begin{pmatrix}
    \cdot   & \cdot   & \cdot  & \cdot  & \cdot  & \cdot & \cdot \\
    \cdot   & \frac{1}{N_c}-\frac{2}{3}   & \frac{2}{3 N_c}-1  & \frac{1}{3N_c}  & -\frac{1}{3} & \cdot & \cdot\\
    \cdot   & \cdot   & \cdot  & \cdot  & \cdot  & \cdot & \cdot\\
    \cdot   & \frac{3}{N_c}-2   & \frac{2}{N_c}-3  & \frac{1}{N_c} & -1  & \cdot & \cdot\\
    \cdot   & \cdot   & \cdot  & \cdot  & \cdot  & \cdot & \cdot\\
    \cdot   & \cdot   & \cdot  & \cdot  & \cdot  & \cdot & \cdot\\
    \cdot   & \cdot   & \cdot  & \cdot  & \cdot  & \cdot & \cdot\\
    \cdot   & \cdot   & \cdot  & \cdot  & \cdot  & \cdot & \cdot\\
    \cdot   & \cdot   & \cdot  & \cdot  & \cdot  & \cdot & \cdot\\
    \cdot   & \cdot   & \cdot  & \cdot  & \cdot  & \cdot & \cdot
  \end{pmatrix}
\end{align}
for the Wilson coefficients $C^\msbar_i$ corresponding to basis II.

In App.~\ref{app:rhut} we give an alternative way to determine $\Delta
T$ using our results for an $\ri$ scheme derived in the next section.

\section{Calculation and Results\label{sec:results}}
In the following we give the main results of this work.  We first
express the finite part of the $\NDR$ amplitudes defined in
Secs.~\ref{sec:rifour} and \ref{sec:ritwo} in a compact and
instructive way in Sec.~\ref{sec:ampgreen}.  We then discuss the
general matching procedure in Sec.~\ref{sec:matching} before giving
the conversion matrices for different $\rimom$ and $\rismom$ schemes
in Secs.~\ref{sec:convg} and \ref{sec:convq}.

\subsection{A compact expression for the amplitudes\label{sec:ampgreen}}
The $\ri$ schemes are defined by applying projectors $P_{nj}$ in
spinor, color, and flavor space to the renormalized amputated Green's
functions $\Gamma^y_n(O^\ri_i)$, see Eqs.~\eqref{eqn:rischemeA},
\eqref{eqn:rischemeB}, and \eqref{eqn:rischemeC}, and thus to
$\Gamma^\msbar_n(O^\msbar_i)$, see Eqs.~\eqref{eqn:gammaritomsbar} and
\eqref{eqn:gammatworitomsbar}.  In this section we calculate
$\Gamma^\msbar_2(O^\msbar_i)$, $\Gamma^\msbar_4(O^\msbar_i)$, and
$\Gamma^\msbar_{4p}(O^\msbar_i)$ as defined in Secs.~\ref{sec:rifour}
and \ref{sec:ritwo}.  The diagrams are generated with the program
QGRAF~\cite{Nogueira:1991ex}, and the symbolic manipulations are
carried out using FORM~\cite{Vermaseren:2000nd}.  We give results for
the respective exceptional momentum configuration of the $\rimom$
scheme as well as the non-exceptional momentum configuration of the
$\rismom$ schemes.  We consider the operators $O^\msbar_i$ of basis I,
II, and the \reduced{} basis.

We simplify the $\msbar$-renormalized amplitudes using the
decomposition \cite{Buras:2000if}
\begin{align}
  \gamma_\mu\gamma_\nu\gamma_\rho = g_{\mu\nu} \gamma_\rho +
  g_{\nu\rho} \gamma_\mu - g_{\mu\rho} \gamma_\nu + i
  \varepsilon_{\alpha\mu\nu\rho} \gamma^\alpha \gamma_5
\end{align}
and the Fierz identities
\begin{align}
  [\slashed{q} (1-\gamma_5)]_{ij} &[\slashed{q} (1-\gamma_5)]_{kl} =
  [\slashed{q} (1-\gamma_5)]_{il} [\slashed{q} (1-\gamma_5)]_{kj}
  \notag\\&
  - (q^2/2) [\gamma_\rho (1-\gamma_5)]_{il} [\gamma^\rho (1-\gamma_5)]_{kj}\,,\\
  [\gamma_\rho (1-\gamma_5)]_{ij} &[\gamma^\rho (1-\gamma_5)]_{kl} =
  \notag\\& - [\gamma_\rho (1-\gamma_5)]_{il} [\gamma^\rho
  (1-\gamma_5)]_{kj}\,.
\end{align}
The resulting one-loop expressions are written as
\begin{align}\label{eqn:msbargreenfullfour}
  \Gamma_4^\msbar(O^\msbar_i) &= \Gamma^\tree_4(O_i) +
  \frac{\alpha_s}{4\pi} \Bigl( \GAMMA^\gamma_{ij} \Gamma^\tree_4(O_j)
  \notag\\&\quad + \sum_{k=q,p_1,p_2}\GAMMA^{\slashed{k}}_{ij}
  \Gamma^\tree_4(X^{\slashed{k}}_j) \notag\\&\quad +
  \GAMMA^{\varepsilon}_{ij} \Gamma^\tree_4(Y_j) + \GAMMA^{G_1}_i
  \Gamma^\tree_4(G_1) \Bigr)\,, \\\label{eqn:msbargreenfulltwo}
  \Gamma_2^\msbar(O^\msbar_i) &= \frac{\alpha_s}{4\pi} \GAMMA^{G_1}_i
  \Gamma^\tree_2(G_1)\,, \\\label{eqn:msbargreenfullfourp}
  \Gamma_{4p}^\msbar(O^\msbar_i) &= \frac{\alpha_s}{4\pi}
  \GAMMA^{G_1}_i \Gamma^\tree_{4p}(G_1)\,,
\end{align}
where $\Gamma_n^\tree(O_j)$ denotes the insertion of the operator
$O_j$ of basis I, II or the \reduced{} basis at tree level and the sum
over repeated indices is implied.  The operators $X^{\slashed{k}}_j$
are obtained from the operators $O_j$ by replacing the gamma structure
$\gamma_\mu \otimes \gamma^\mu$ with $\slashed{k} \otimes \slashed{k}
/ k^2$, i.e.,
\begin{align}\label{eqn:x1FTdef}
  X^{\slashed{q}}_1 &= (\bar s_a \slashed{q}(1-\gamma_5) u_b) (\bar
  u_b \slashed{q}(1-\gamma_5) d_a) / q^2
\end{align}
for the case of $Q_1$ and $k=q$ and analog for all other $O_j$ and
$k\in\{q,p_1,p_2\}$.  Similarly the operators $Y_j$ are obtained from
the operators $O_j$ by replacing the gamma structure $\gamma_\mu
\otimes \gamma^\mu$ with $i\varepsilon_{\mu\nu\alpha\beta} \gamma^\mu
\otimes \gamma^\nu p_1^\alpha p_2^\beta / q^2$, i.e.,
\begin{align}\label{eqn:y1FTdef}
  Y_1 &= i\varepsilon_{\mu\nu\alpha\beta} (\bar
  s_a \gamma^\mu(1-\gamma_5) u_b) \notag\\&\quad\times (\bar u_b \gamma^\nu(1-\gamma_5)
  d_a) p_1^\alpha p_2^\beta / q^2
\end{align}
for the case of $Q_1$ and analog for all other $O_j$.  In
Tabs.~\ref{tab:gammaEPQ}--\ref{tab:gammaEPQG} we give the coefficients
$\GAMMA^z_{\ldots}$ with
$z\in\{\gamma,\slashed{q},\slashed{p}_1,\slashed{p}_2,\varepsilon,G_1\}$
at one loop for the \reduced{} basis and the exceptional and
non-exceptional momentum configuration.  The corresponding results for
basis I and II can be obtained from these tables using
Eq.~\eqref{eqn:deftms}.  The constant $C_0$ is defined as
\begin{align}
C_0 =(2/3)\Psi'(1/3)-(2\pi/3)^2\approx 2.34391\,,  
\end{align}
where $\Psi(x)$ is the digamma function \cite{abramowitz+stegun}.

\setlength{\dataIla}{9.0cm} \setlength{\dataIlb}{5.5cm}
\setlength{\dataIlc}{0.25cm} \setlength{\dataIld}{0.25cm}

  \renewcommand{\dataINL}{} 
  \begin{table*}[tpb]
    \centering
    \begin{tabular}{lclclclcl}
      \hline
      $(i,j)$ &  &  $\GAMMA^{\gamma}_{ij}$ & & $\GAMMA_{ij}^{\slashed{p}_1}$ & & $\GAMMA^{\slashed{q}}_{ij}$ & & $\GAMMA^{\varepsilon}_{ij}$ \\
      \hline\hline
      \dataPupsilon{1}{1}{\xi\dIleft(-\frac{4\log(2)}{ N_c }\dIbsp$ $+4\log(2)\dIbsp$ $+ N_c \dIbsp$ $+\frac{1}{ N_c }\dIbsp$ $-1\dIright)\dIbsp$ $-\frac{12\log(2)}{ N_c }\dIbsp$ $+12\log(2)\dIbsp$ $+\frac{7}{ N_c }\dIbsp$ $-7}{(-2 N_c \dIbsp$ $-2)\xi}{0}{0}
\dataPupsilon{2}{2}{\xi\dIleft(-\frac{4\log(2)}{ N_c }\dIbsp$ $+ N_c \dIbsp$ $+\frac{1}{ N_c }\dIright)\dIbsp$ $-\frac{12\log(2)}{ N_c }\dIbsp$ $+\frac{7}{ N_c }}{-2 N_c \xi}{0}{0}
\dataPupsilon{2}{3}{(4\log(2)\dIbsp$ $-1)\xi\dIbsp$ $+12\log(2)\dIbsp$ $-7}{-2\xi}{0}{0}
\dataPupsilon{3}{2}{(4\log(2)\dIbsp$ $-1)\xi\dIbsp$ $+12\log(2)\dIbsp$ $-7}{-2\xi}{0}{0}
\dataPupsilon{3}{3}{\xi\dIleft(-\frac{4\log(2)}{ N_c }\dIbsp$ $+ N_c \dIbsp$ $+\frac{1}{ N_c }\dIright)\dIbsp$ $-\frac{12\log(2)}{ N_c }\dIbsp$ $+\frac{7}{ N_c }}{-2 N_c \xi}{0}{0}
\dataPupsilon{4}{4}{\xi\dIleft(-\frac{4\log(2)}{3 N_c }\dIbsp$ $+ N_c \dIbsp$ $-\frac{5}{3 N_c }\dIright)\dIbsp$ $-\frac{4\log(2)}{3 N_c }\dIbsp$ $-\frac{5}{3 N_c }}{\xi\dIleft(-\frac{8\log(2)}{3 N_c }\dIbsp$ $-2 N_c \dIbsp$ $+\frac{8}{3 N_c }\dIright)\dIbsp$ $-\frac{8\log(2)}{3 N_c }\dIbsp$ $-\frac{4}{3 N_c }}{0}{0}
\dataPupsilon{4}{5}{\dIleft(\frac{4\log(2)}{3}\dIbsp$ $+\frac{2}{3}\dIright)\xi\dIbsp$ $+\frac{4\log(2)}{3}\dIbsp$ $+\frac{5}{3}}{\dIleft(\frac{8\log(2)}{3}\dIbsp$ $-\frac{2}{3}\dIright)\xi\dIbsp$ $+\frac{8\log(2)}{3}\dIbsp$ $+\frac{4}{3}}{0}{0}
\dataPupsilon{5}{4}{\dIleft(\frac{4\log(2)}{3}\dIbsp$ $-\frac{1}{3}\dIright)\xi\dIbsp$ $+\frac{4\log(2)}{3}\dIbsp$ $-\frac{7}{3}}{\dIleft(\frac{8\log(2)}{3}\dIbsp$ $-\frac{8}{3}\dIright)\xi\dIbsp$ $+\frac{8\log(2)}{3}\dIbsp$ $+\frac{4}{3}}{0}{0}
\dataPupsilon{5}{5}{\xi\dIleft(-\frac{4\log(2)}{3 N_c }\dIbsp$ $+2 N_c \dIbsp$ $-\frac{5}{3 N_c }\dIright)\dIbsp$ $-\frac{4\log(2)}{3 N_c }\dIbsp$ $+4 N_c \dIbsp$ $-\frac{5}{3 N_c }}{\xi\dIleft(\frac{8}{3 N_c }\dIbsp$ $-\frac{8\log(2)}{3 N_c }\dIright)\dIbsp$ $-\frac{8\log(2)}{3 N_c }\dIbsp$ $-\frac{4}{3 N_c }}{0}{0}
\\
      \hline
    \end{tabular}
    \caption{One-loop
  coefficients $\GAMMA^\gamma$, $\GAMMA^{\slashed{p}_1}$,
  $\GAMMA^{\slashed{q}}$, and $\GAMMA^{\varepsilon}$ for the
  exceptional momentum configuration and the \reduced{} basis.  The
  coefficient $\GAMMA_{ij}^{\slashed{p}_2}=0$ for the exceptional
  momentum configuration.  The coefficients for $Q_7$ and $Q_8$ are
  identical to the coefficients for $Q_5$ and $Q_6$.  The remaining
  matrix elements not given here are zero.}
    \label{tab:gammaEPQ}
  \end{table*}

\setlength{\dataIla}{5.5cm} \setlength{\dataIlb}{4.5cm}
\setlength{\dataIlc}{3.0cm} \setlength{\dataIld}{2.5cm}

  \renewcommand{\dataINL}{} 
  \begin{table*}[tpb]
    \centering
    \begin{tabular}{lclclclcl}
      \hline
      $(i,j)$ &  &  $\GAMMA^{\gamma}_{ij}$ & & $\GAMMA_{ij}^{\slashed{p}_1}$,
  $\GAMMA_{ij}^{\slashed{p}_2}$ & & $\GAMMA^{\slashed{q}}_{ij}$ & & $\GAMMA^{\varepsilon}_{ij}$ \\
      \hline\hline
      \dataPupsilon{1}{1}{\xi\dIleft(-\frac{2 C_0  N_c }{3}\dIbsp$ $+\frac{ C_0 }{ N_c }\dIbsp$ $-\frac{2 C_0 }{3}\dIbsp$ $-\frac{4\log(2)}{ N_c }\dIbsp$ $+4\log(2)\dIbsp$ $+2 N_c \dIbsp$ $-\frac{1}{ N_c }\dIright)\dIbsp$ $+\frac{ C_0  N_c }{3}\dIbsp$ $+\frac{ C_0 }{3}\dIbsp$ $-\frac{12\log(2)}{ N_c }\dIbsp$ $+12\log(2)\dIbsp$ $-2 N_c \dIbsp$ $+\frac{9}{ N_c }\dIbsp$ $-9}{-\frac{2 C_0  N_c }{3}\dIbsp$ $-\frac{2 C_0 }{3}\dIbsp$ $+\dIleft(-\frac{2 N_c }{3}\dIbsp$ $-\frac{2}{3}\dIright)\xi\dIbsp$ $+\frac{4 N_c }{3}\dIbsp$ $+\frac{4}{3}}{\xi\dIleft(\frac{2 C_0  N_c }{3}\dIbsp$ $+\frac{2 C_0 }{3}\dIbsp$ $-\frac{2 N_c }{3}\dIbsp$ $-\frac{2}{3}\dIright)\dIbsp$ $+\frac{4 N_c }{3}\dIbsp$ $+\frac{4}{3}}{0}
\dataPupsilon{2}{2}{\xi\dIleft(-\frac{2 C_0  N_c }{3}\dIbsp$ $+\frac{ C_0 }{ N_c }\dIbsp$ $-\frac{4\log(2)}{ N_c }\dIbsp$ $+2 N_c \dIbsp$ $-\frac{1}{ N_c }\dIright)\dIbsp$ $+\frac{ C_0  N_c }{3}\dIbsp$ $-\frac{12\log(2)}{ N_c }\dIbsp$ $-2 N_c \dIbsp$ $+\frac{9}{ N_c }}{-\frac{2 C_0  N_c }{3}\dIbsp$ $-\frac{2 N_c \xi}{3}\dIbsp$ $+\frac{4 N_c }{3}}{\xi\dIleft(\frac{2 C_0  N_c }{3}\dIbsp$ $-\frac{2 N_c }{3}\dIright)\dIbsp$ $+\frac{4 N_c }{3}}{0}
\dataPupsilon{2}{3}{\xi\dIleft(4\log(2)\dIbsp$ $-\frac{2 C_0 }{3}\dIright)\dIbsp$ $+\frac{ C_0 }{3}\dIbsp$ $+12\log(2)\dIbsp$ $-9}{-\frac{2 C_0 }{3}\dIbsp$ $-\frac{2\xi}{3}\dIbsp$ $+\frac{4}{3}}{\dIleft(\frac{2 C_0 }{3}\dIbsp$ $-\frac{2}{3}\dIright)\xi\dIbsp$ $+\frac{4}{3}}{0}
\dataPupsilon{3}{2}{\xi\dIleft(4\log(2)\dIbsp$ $-\frac{2 C_0 }{3}\dIright)\dIbsp$ $+\frac{ C_0 }{3}\dIbsp$ $+12\log(2)\dIbsp$ $-9}{-\frac{2 C_0 }{3}\dIbsp$ $-\frac{2\xi}{3}\dIbsp$ $+\frac{4}{3}}{\dIleft(\frac{2 C_0 }{3}\dIbsp$ $-\frac{2}{3}\dIright)\xi\dIbsp$ $+\frac{4}{3}}{0}
\dataPupsilon{3}{3}{\xi\dIleft(-\frac{2 C_0  N_c }{3}\dIbsp$ $+\frac{ C_0 }{ N_c }\dIbsp$ $-\frac{4\log(2)}{ N_c }\dIbsp$ $+2 N_c \dIbsp$ $-\frac{1}{ N_c }\dIright)\dIbsp$ $+\frac{ C_0  N_c }{3}\dIbsp$ $-\frac{12\log(2)}{ N_c }\dIbsp$ $-2 N_c \dIbsp$ $+\frac{9}{ N_c }}{-\frac{2 C_0  N_c }{3}\dIbsp$ $-\frac{2 N_c \xi}{3}\dIbsp$ $+\frac{4 N_c }{3}}{\xi\dIleft(\frac{2 C_0  N_c }{3}\dIbsp$ $-\frac{2 N_c }{3}\dIright)\dIbsp$ $+\frac{4 N_c }{3}}{0}
\dataPupsilon{4}{4}{\xi\dIleft(-\frac{2 C_0  N_c }{3}\dIbsp$ $+\frac{7 C_0 }{6 N_c }\dIbsp$ $-\frac{4\log(2)}{3 N_c }\dIbsp$ $+2 N_c \dIbsp$ $-\frac{8}{3 N_c }\dIright)\dIbsp$ $+\frac{ C_0  N_c }{3}\dIbsp$ $+\frac{7 C_0 }{6 N_c }\dIbsp$ $-\frac{4\log(2)}{3 N_c }\dIbsp$ $-2 N_c \dIbsp$ $+\frac{1}{3 N_c }}{-\frac{2 C_0  N_c }{3}\dIbsp$ $+\frac{2 C_0 }{3 N_c }\dIbsp$ $+\xi\dIleft(-\frac{4\log(2)}{3 N_c }\dIbsp$ $-\frac{2 N_c }{3}\dIbsp$ $+\frac{1}{ N_c }\dIright)\dIbsp$ $-\frac{4\log(2)}{3 N_c }\dIbsp$ $+\frac{4 N_c }{3}\dIbsp$ $-\frac{2}{ N_c }}{\xi\dIleft(\frac{2 C_0  N_c }{3}\dIbsp$ $-\frac{2 C_0 }{3 N_c }\dIbsp$ $-\frac{2 N_c }{3}\dIbsp$ $+\frac{2}{3 N_c }\dIright)\dIbsp$ $+\frac{4 N_c }{3}\dIbsp$ $-\frac{4}{3 N_c }}{-\frac{ C_0 \xi}{3 N_c }\dIbsp$ $-\frac{2 C_0  N_c }{3}\dIbsp$ $+\frac{ C_0 }{ N_c }}
\dataPupsilon{4}{5}{\xi\dIleft(-\frac{ C_0 }{2}\dIbsp$ $+\frac{4\log(2)}{3}\dIbsp$ $+\frac{2}{3}\dIright)\dIbsp$ $-\frac{3 C_0 }{2}\dIbsp$ $+\frac{4\log(2)}{3}\dIbsp$ $+\frac{5}{3}}{\dIleft(\frac{4\log(2)}{3}\dIbsp$ $-\frac{1}{3}\dIright)\xi\dIbsp$ $+\frac{4\log(2)}{3}\dIbsp$ $+\frac{2}{3}}{0}{\frac{ C_0 \xi}{3}\dIbsp$ $-\frac{ C_0 }{3}}
\dataPupsilon{5}{4}{\xi\dIleft(-\frac{2 C_0 }{3}\dIbsp$ $+\frac{4\log(2)}{3}\dIbsp$ $+\frac{2}{3}\dIright)\dIbsp$ $+\frac{ C_0 }{3}\dIbsp$ $+\frac{4\log(2)}{3}\dIbsp$ $-\frac{13}{3}}{-\frac{2 C_0 }{3}\dIbsp$ $+\dIleft(\frac{4\log(2)}{3}\dIbsp$ $-1\dIright)\xi\dIbsp$ $+\frac{4\log(2)}{3}\dIbsp$ $+2}{\dIleft(\frac{2 C_0 }{3}\dIbsp$ $-\frac{2}{3}\dIright)\xi\dIbsp$ $+\frac{4}{3}}{-\frac{2 C_0 }{3}}
\dataPupsilon{5}{5}{\xi\dIleft(-\frac{ C_0  N_c }{2}\dIbsp$ $+\frac{7 C_0 }{6 N_c }\dIbsp$ $-\frac{4\log(2)}{3 N_c }\dIbsp$ $+2 N_c \dIbsp$ $-\frac{8}{3 N_c }\dIright)\dIbsp$ $-\frac{3 C_0  N_c }{2}\dIbsp$ $+\frac{7 C_0 }{6 N_c }\dIbsp$ $-\frac{4\log(2)}{3 N_c }\dIbsp$ $+4 N_c \dIbsp$ $+\frac{1}{3 N_c }}{\frac{2 C_0 }{3 N_c }\dIbsp$ $+\xi\dIleft(\frac{1}{ N_c }\dIbsp$ $-\frac{4\log(2)}{3 N_c }\dIright)\dIbsp$ $-\frac{4\log(2)}{3 N_c }\dIbsp$ $-\frac{2}{ N_c }}{\xi\dIleft(\frac{2}{3 N_c }\dIbsp$ $-\frac{2 C_0 }{3 N_c }\dIright)\dIbsp$ $-\frac{4}{3 N_c }}{\xi\dIleft(\frac{ C_0  N_c }{3}\dIbsp$ $-\frac{ C_0 }{3 N_c }\dIright)\dIbsp$ $-\frac{ C_0  N_c }{3}\dIbsp$ $+\frac{ C_0 }{ N_c }}
\\
      \hline
    \end{tabular}
    \caption{One-loop coefficients
  $\GAMMA^\gamma$, $\GAMMA^{\slashed{p}_1}$, $\GAMMA^{\slashed{p}_2}$,
  $\GAMMA^{\slashed{q}}$, and $\GAMMA^{\varepsilon}$ for the
  non-exceptional momentum configuration and the \reduced{} basis.
  The coefficient
  $\GAMMA_{ij}^{\slashed{p}_1}=\GAMMA_{ij}^{\slashed{p}_2}$ for the
  exceptional momentum configuration.  The coefficients for $Q_7$ and
  $Q_8$ are identical to the coefficients for $Q_5$ and $Q_6$.  The
  remaining matrix elements not given here are zero.}
    \label{tab:gammaNEPQ}
  \end{table*}

  \renewcommand{\dataINL}{} 
  \def\OldS{}
  \def\OldX{}
  \def\OldY{}
  \dataPg{1}{0}\dataPg{2}{0}\dataPg{3}{-\frac{2}{9}}\dataPg{4}{0}\dataPg{5}{-\frac{5}{3}}\dataPg{6}{0}\dataPg{7}{0}
  \begin{table}[tp]
    \centering
    \begin{tabular}{cccccccccc}
      \hline
      \OldX \\
      \hline\hline
      \OldY \\
      \hline
    \end{tabular}
    \caption{One-loop coefficients $\GAMMA^{G_1}$ for the
  \reduced{} basis.  The results are identical for the exceptional and
  non-exceptional momentum configuration.}
    \label{tab:gammaEPQG}
  \end{table}

In order to determine the conversion factors from the $\ri$ schemes to
the $\NDR$ scheme we need to study projected Green's functions
\begin{align}
  \Lambda_{nst} = P_{nt} \Gamma_n^\msbar(O_s^\msbar)\,,
\end{align}
where $n \in \{2,4,4p\}$ and both $P_{nt}$ and $\Gamma_n^\msbar$ have
open spinor, color, and flavor indices.  The action of $P_{nt}$ on
$\Gamma_n^\msbar$ for $n \in \{4,4p\}$ is thus given by
\begin{align}
  \Lambda_{nst} = P_{nt}^{\alpha\beta\gamma\delta;ijkl;f}
  \Gamma^\msbar_{n;\alpha\beta\gamma\delta;ijkl;f}(O_s^\msbar)
\end{align}
with spinor indices $\alpha,\beta,\gamma,\delta$, color indices
$i,j,k,l$, and flavor index $f$, as shown in Fig.~\ref{fig:cctree}.
Similarly
\begin{align}
  \Lambda_{2st} = P_{2t}^{\alpha\beta;\mu;a;ij}
  \Gamma^\msbar_{2;\alpha\beta;\mu;a;ij}(O_s^\msbar)\,,
\end{align}
where $\mu$ is a Lorentz index and $a \in \{1,\ldots,N_c^2-1\}$
enumerates the generators of the $\SU(N_c)$ algebra, see
Fig.~\ref{fig:tqtree}.  The summation over repeated indices is
implied.  The resulting expression $\Lambda_{nst}$ naturally depends
on the choice of the projectors $P_{nt}$.  In the following we make
certain assumptions about the structure of the projectors, i.e., we
discuss different classes of projectors.  For each class of projectors
we can then give a very compact form of the amputated Green's function
that can, without loss of generality, be used to calculate all
projections $\Lambda$ of the respective class.

First we restrict the discussion to projectors which contain no
external momentum, except for at most the external momentum
\begin{align}\nonumber
  q = p_1 - p_2\,.
\end{align}
We also do not consider projectors that contain vectors such as
$\delta_{\mu1}$, $\delta_{\mu2}$, $\delta_{\mu3}$, or $\delta_{\mu4}$
which break Lorentz symmetry explicitly.  The corresponding class of
projectors shall be denoted by $P^{\slashed{q}}$.  We will also study
the sub-class $P^{\gamma_\mu} \subset P^{\slashed{q}}$ of projectors
that additionally do not contain the external momentum $q$.

Let us focus on the amputated Green's functions with four external
quarks and examine the projection of a spinor structure such as
$\Omega_1 \otimes \Omega_2$ under projectors of class
$P^{\slashed{q}}$, i.e.,
\begin{align}\nonumber
  P^{\slashed{q}} [  \Omega_1 \otimes \Omega_2]\,,
\end{align}
where the two factors belong to the two fermion lines, and the
combination $\Omega_1 \otimes \Omega_2$ contains strings of gamma
matrices without remaining open Lorentz indices.  Then the ansatz
\begin{align}\label{eqn:ptensoransatz}
 P^{\slashed{q}}[\gamma_\mu \Omega_1\otimes \gamma_\nu\Omega_2] = t_1 g_{\mu\nu} + t_2 q_\mu q_\nu / q^2
\end{align}
with Lorentz indices $\mu$ and $\nu$ is justified by the
Lorentz-transformation properties of both sides since $P^{\slashed{q}}$
does not contain any Lorentz vectors apart from the momentum $q$. 
The coefficients $t_1$ and $t_2$ can be determined by contraction with
$g_{\mu\nu}$ and $q_\mu q_\nu$, respectively.  We find
\begin{align}
  P^{\slashed{q}}&[\gamma_\mu \Omega_1\otimes \gamma_\nu\Omega_2] = \notag\\&
\!\!\!+ \frac{q_\mu q_\nu}{3q^2} \biggl( -
  P^{\slashed{q}}[\gamma_\rho\Omega_1 \otimes \gamma^\rho\Omega_2]
 +4 P^{\slashed{q}}[\slashed{q} \Omega_1
  \otimes \slashed{q}\Omega_2]/ q^2 \biggr)  \notag\\&
\!\!\!+ \frac{g_{\mu\nu}}{3} \biggl( P^{\slashed{q}}[\gamma_\rho\Omega_1
  \otimes \gamma^\rho\Omega_2] -
  P^{\slashed{q}}[\slashed{q} \Omega_1 \otimes \slashed{q}\Omega_2]/
  q^2 \biggr).
\end{align}
Therefore not all of the spinor structures that appear in the general
$\msbar$ amplitude of Eq.~\eqref{eqn:msbargreenfullfour} are
independent under projection with $P^{\slashed{q}}$, i.e.,
\begin{align}
  P^{\slashed{q}}[\Gamma^\tree_n(X_j^{\slashed{p}_1})] &=
  P^{\slashed{q}}[\Gamma^\tree_n(X_j^{\slashed{p}_2})]
  =\frac14 P^{\slashed{q}}[\Gamma^\tree_n(O_j)] \,, \notag\\
  P^{\slashed{q}}[\Gamma^\tree_n(Y_j)] &= 0
\end{align}
with $n \in \{4,4p\}$.  The resulting projected amputated Green's
functions for $n=4$ up to one-loop order can be written as
\begin{align}\label{eqn:msbargreen}
  P^{\slashed{q}}[&\Gamma_4^\msbar(O^\msbar_i)] =
  P^{\slashed{q}}[\Gamma^\tree_4(O_i)] \notag\\& +
  \frac{\alpha_s}{4\pi} \Bigl( (\GAMMA^\gamma_{ij} +
  \frac14\GAMMA^{\slashed{p}_1}_{ij} +
  \frac14\GAMMA^{\slashed{p}_2}_{ij})
  P^{\slashed{q}}[\Gamma^\tree_4(O_j)] \notag\\& +
  \GAMMA^{\slashed{q}}_{ij} P^{\slashed{q}}[\Gamma^\tree_4(X_j)] +
  \GAMMA^{G_1}_i P^{\slashed{q}}[\Gamma^\tree_4(G_1)] \Bigr)\,.
\end{align}

If we further restrict the discussion to projectors of the sub-class
$P^{\gamma_\mu} \subset P^{\slashed{q}}$, we find
\begin{align}\label{eqn:adhocfa}
  P^{\gamma_\mu}[\Gamma^\tree_n(X^{\slashed{q}}_j)] &= \frac14 P^{\gamma_\mu}[\Gamma^\tree_n(O_j)]\,, \notag\\
  P^{\gamma_\mu}[\Gamma^\tree_n(G_1)] &= \frac34
  P^{\gamma_\mu}[\Gamma^\tree_n(\Qp)]
\end{align}
with $n \in \{4,4p\}$ using the same arguments as in
Eq.~\eqref{eqn:ptensoransatz} without the term proportional to $t_2$.
Therefore, for projectors of class $P^{\gamma_\mu}$ and $n=4$ we can
write
\begin{align}\label{eqn:msbargreensimple}
  P^{\gamma_\mu}&[\Gamma_4^\msbar(O^\msbar_i)] =
  P^{\gamma_\mu}[\Gamma^\tree_4(O_i)] + \frac{\alpha_s}{4\pi}
  \Bigl( (\GAMMA^\gamma_{ij} + \frac14\GAMMA^{\slashed{p}_1}_{ij} \notag\\& +
  \frac14\GAMMA^{\slashed{p}_2}_{ij} +
  \frac14\GAMMA^{\slashed{q}}_{ij} + \frac34 \GAMMA^{G_1}_i \tau_j)
 P^{\gamma_\mu}[\Gamma^\tree_4(O_j)]
  \Bigr)
\end{align}
with $\Gamma^\tree_4(\Qp) = \tau_j \Gamma^\tree_4(O_j)$, where we can
read off the coefficients $\tau_j$ from Eq.~\eqref{eqn:defqpbasisI}
for basis I and II or from Eq.~\eqref{eqn:defqpreducedbasis} for the
\reduced{} basis.

\subsection{The matching procedure\label{sec:matching}}
In the following sections we give the one-loop conversion coefficients
from different $\ri$ schemes to the $\NDR$ scheme.  We first explain
the general procedure of calculating the conversion coefficients and
then give results for explicit $\rimom$ and $\rismom$ schemes.

At one loop the conversion from $\ri$ schemes to the $\NDR$ scheme in
terms of renormalized amputated Green's functions reads
\begin{align}\label{eqn:oneloopconvgreenb}
  C^y \Gamma_n^{\msbar}(O_i^{\msbar}) &= \Gamma_n^y(O^{\ri}_i) +
  \Delta a^{\ri\to\msbar}_{ij} \Gamma_n^y(O^{\ri}_j) \notag\\&\quad +
  \Delta c^{\ri\to\msbar}_i \Gamma_n^y(G^{\ri}_1)
\end{align}
with wave function conversion factor $C^y =(C^y_q)^2$ for $n \in
\{4,4p\}$ and $C^y = C^y_q (Z^\msbar_A/Z^y_A)^{1/2}$ for $n=2$, see
Eq.~\eqref{eqn:generalritomsbarblock}.  In order to define the $\ri$
scheme, and hence to determine the coefficients $\Delta
a_{ij}^{\ri\to\msbar}$ and $\Delta c_i^{\ri\to\msbar}$, we apply
projectors $P_{nl}$ to Eq.~\eqref{eqn:oneloopconvgreenb} and use the
$\ri$ conditions of Eqs.~\eqref{eqn:rischemeA} and
\eqref{eqn:rischemeB} for the $\rismom_2$ schemes and of
Eqs.~\eqref{eqn:rischemeA} and \eqref{eqn:rischemeC} for the $\rismom$
schemes, see Sec.~\ref{sec:ritwo}.

From Eqs.~\eqref{eqn:msbargreenfulltwo} and
\eqref{eqn:msbargreenfullfourp} and the $\ri$ conditions of
Eqs.~\eqref{eqn:rischemeB} and \eqref{eqn:rischemeC} we can already
conclude that
\begin{align}\label{eqn:detcri}
    \Delta c^{\ri\to\msbar} &= \frac{\alpha_s}{4\pi} \GAMMA^{G_1}
\end{align}
is the only possible result at one loop.  For convenience we provide a
specific projector to determine the mixing of the two-quark operator
$G_1$.  In an $\rismom_2$ scheme, we can use
\begin{align}
 P_{2,G_1} &= \lambda^a_{ji} (\gamma_\mu)_{\beta\alpha}\,,
\end{align}
where $\lambda^a$ is a generator of the $\SU(N_c)$ group algebra and
the convention of open indices is given in Fig.~\ref{fig:tqtree}.  In
an $\rismom$ scheme, we can use
\begin{align}\label{eq:PG1}
  P_{4p,G_1} &= \delta_{il}\delta_{kj}
  (\gamma^\mu)_{\beta\alpha}(\gamma_\mu)_{\delta\gamma} \,.
\end{align}
At higher loops the additional operators \cite{Buras:1992tc}
\begin{align}
  G_2 &= \frac4{ig^2}\bar s \{D_\mu D^\mu,D^\nu\} \gamma_\nu (1-\gamma_5) d \,, \\ \notag
  G_3 &= \frac4{ig^2}\bar s D_\mu D_\nu D_\lambda S^{\mu\nu\lambda} (1-\gamma_5) d \,, \\ \notag
  G_4 &= \frac4{ig^2}\bar s [D_\mu D^\mu,D^\nu] \gamma_\nu (1-\gamma_5) d \,,  
\end{align}
with $S^{\mu\nu\lambda}=\gamma^\mu\gamma^\nu\gamma^\lambda -
\gamma^\lambda\gamma^\nu\gamma^\mu$ and possibly other
non-gauge-invariant two-quark operators $N_m$ mix with the $(8,1)$
operators.  We therefore need to provide additional renormalization
conditions at higher loops in order to define the $\ri$ schemes for
the $(8,1)$ operators uniquely, i.e., we need to specify additional
projectors.  The specific choice of projectors to determine the mixing
with the operator $G_2$ and additional operators which occur at higher
loops is not relevant for the one-loop conversion factors given in
this work.  At one loop the conversion factors can thus be given
without specifying the details of higher-loop contributions.

In the following we use the $\ripmom$, $\rismom$, and $\rismomb$ wave
function renormalizations defined in
Refs.~\cite{Martinelli:1994ty,Chetyrkin:1999pq,Sturm:2009kb}. The
corresponding conversion factors up to one-loop order
\begin{align}
  C^y_q = 1+\frac{\alpha_s}{4\pi}\Delta^y_q
\end{align}
are given by
\begin{align}
 \Delta^{\ripmom}_q =  \Delta^{\rismom}_q = -\frac\xi2 \left( N_c - \frac1{N_c}\right)
\end{align}
for the $\ripmom$ and $\rismom$ wave function renormalization schemes as
well as
\begin{align}
  \Delta^\rismomb_q = -\frac12\left(N_c-\frac1{N_c} \right)\left(-1+\frac{\xi}2 (3-C_0)\right)
\end{align}
for the $\rismomb$ scheme.

\subsection{\boldmath The $\bfrimom$ and $\bfrismom(\gamma_\mu,y)$
  schemes\label{sec:convg}}

We first give the conversion coefficients for $\ri$ schemes that
either (i) use only projectors of class $P^{\gamma_\mu}$ or (ii) use
the exceptional momentum configuration and projectors of class
$P^{\slashed{q}}$.  The coefficients $\Delta c^{\ri\to\msbar}$ are
already determined in Eq.~\eqref{eqn:detcri}, and it remains to obtain
the coefficients $\Delta a^{\ri\to\msbar}$ by applying projectors
$P_{4k}$ to the amputated Green's function $\Gamma^\msbar_4$.

In case (i) the amputated Green's function $\Gamma^\msbar_4$
simplifies to
\begin{align}
  P^{\gamma_\mu}&[(C_q^y)^2\Gamma_4^\msbar(O^\msbar_i)] =
  P^{\gamma_\mu}[\Gamma^\tree_4(O_i)] + \frac{\alpha_s}{4\pi} \Bigl(
  (\GAMMA^\gamma_{ij} \notag\\&+ \frac14\GAMMA^{\slashed{p}_1}_{ij} +
  \frac14\GAMMA^{\slashed{p}_2}_{ij} +
  \frac14\GAMMA^{\slashed{q}}_{ij} + 2 \Delta_q^y \delta_{ij} +
  \frac34 \GAMMA^{G_1}_i \tau_j) \notag\\&\quad\times
  P^{\gamma_\mu}[\Gamma^\tree_4(O_j)] \Bigr)\,,
\end{align}
under projectors $P_{4k} \in P^{\gamma_\mu}$, see
Eq.~\eqref{eqn:msbargreensimple}.  This should be compared to
Eq.~\eqref{eqn:oneloopconvgreenb}, i.e.,
\begin{align}
  &P^{\gamma_\mu}[(C^y_q)^2\Gamma_4^{\msbar}(O_i^{\msbar})] =
  P^{\gamma_\mu}[\Gamma^y_4(O^{\ri}_i)] \notag\\&\quad + \Bigl(\Delta
  a^{\ri\to\msbar}_{ij} + \frac34\Delta c^{\ri\to\msbar}_i
  \tau_j\Bigr) P^{\gamma_\mu}[\Gamma^y_4(O^{\ri}_j)] \,,
\end{align}
where we used Eqs.~\eqref{eqn:adhocfa}.  If we impose the $\ri$
condition of Eq.~\eqref{eqn:rischemeA} and insert
Eq.~\eqref{eqn:detcri}, we find
\begin{align}\label{eqn:dadefdq}
  \Delta a^{\ri\to\msbar} &= \frac{\alpha_s}{4\pi}
  \Bigl(\GAMMA^\gamma+ \frac14(\GAMMA^{\slashed{p}_1}+
  \GAMMA^{\slashed{p}_2}+ \GAMMA^{\slashed{q}}) \notag\\
  &\quad\qquad + 2\Delta^y_q\1 \Bigr)
\end{align}
with identity matrix $\1$.

In case (ii) we have $\GAMMA^{\slashed{q}}=0$, and therefore
Eq.~\eqref{eqn:msbargreen} simplifies to
\begin{align}\label{eqn:msbargreensimpb}
  P^{\slashed{q}}&[\Gamma^\msbar_4(O^\msbar_i)] = 
  P^{\slashed{q}}[\Gamma^\tree_4(O_i)] 
+ \frac{\alpha_s}{4\pi} \Bigl( 
  (\GAMMA^\gamma_{ij} +\frac14\GAMMA^{\slashed{p}_1}_{ij} \notag\\&
+\frac14\GAMMA^{\slashed{p}_2}_{ij}) P^{\slashed{q}}[\Gamma^\tree_4(O_j)] 
+ \GAMMA^{G_1}_i P^{\slashed{q}}[\Gamma^\tree_4(G_1)] \Bigr)\,.
\end{align}
One can read off the conversion matrix $\Delta a^{\ri\to\msbar}$ by
comparing Eq.~\eqref{eqn:oneloopconvgreenb} to
Eq.~\eqref{eqn:msbargreensimpb} with
\begin{align}
  \Delta a^{\ri\to\msbar} &= \frac{\alpha_s}{4\pi} \Bigl(\GAMMA^\gamma
  + \frac14\GAMMA^{\slashed{p}_1}+ \frac14\GAMMA^{\slashed{p}_2} + 2\Delta^y_q\1 \Bigr) \,.
\end{align}

In both cases the conversion coefficients are unique up to the
definition of the $\ri$ quark wave function renormalization
$\Delta^y_q$.  In other words, in these cases the details of the
projectors do not matter, as long as one uses a sufficient number of
independent projectors to determine all elements of the conversion
matrices or $Z$-factors, respectively.

The choice of the $\ri$ wave function renormalization scheme only
affects the diagonal elements of $\Delta a^{\ri\to\msbar}$.  One can
also combine different $\ri$ wave function renormalization schemes
with different $\ri$ schemes for the operators $O_i$ in a
straightforward way using Eq.~\eqref{eqn:dadefdq}.

We call the $\ri$ scheme of case (i) $\rismom(\gamma_\mu,y)$ scheme,
where $y=\slashed{q}$ corresponds to the $\rismom$ wave function
renormalization scheme and $y=\gamma_\mu$ corresponds to the
$\rismomb$ wave function renormalization scheme.  The name
$\rismom(\gamma_\mu,y)$ reflects that we restrict the choice of
projectors to the class $P^{\gamma_\mu}$.

The scheme corresponding to case (ii) is the $\rimom$ scheme, and we
give results only for the $\ripmom$ wave function renormalization.

In Tabs.~\ref{tab:gammaCQETQ}--\ref{tab:gammaCQEPQ} we give the values
for $\Delta r^{\ri\to\msbar}$, which is defined in
Eq.~\eqref{eq:deltar}, for the $\rimom$ scheme with $\ripmom$ wave
function renormalization and operators of basis I, II, and the
\reduced{} basis.  In Tabs.~\ref{tab:gammaCQNEPQ} and
\ref{tab:gammaCGNEPQ} we present the results for $\Delta
r^{\ri\to\msbar}$ for the $\rismom(\gamma_\mu,\slashed{q})$ and the
$\rismom(\gamma_\mu,\gamma_\mu)$ scheme and operators of the
\reduced{} basis.

We would like to point out that the result of
Tab.~\ref{tab:gammaCQETQ} for $N_c=3$ and $\xi=0$ or $\xi=1$ agrees
with the result of Ref.~\cite{Ciuchini:1995cd} for the conversion from
the $\rimom$ scheme to the $\NDR$ scheme.  In Ref.~\cite{Buras:2000if}
the current-current contributions in the $\rimom$ scheme have been
considered.  Our result for these contributions agrees with the one of
Ref.~\cite{Buras:2000if}, which can be seen by using the results of
Eq.~\eqref{eqn:deftms} and Tab.~\ref{tab:gammaEPQ} and combining them
with the quark wave function renormalization constant of Eq.~(5.3) of
Ref.~\cite{Buras:2000if}.

Furthermore, the result for the $(27,1)$ operator $Q'_1$ in
Tabs.~\ref{tab:gammaCQNEPQ} and \ref{tab:gammaCGNEPQ} agrees with the
result of Ref.~\cite{Aoki:2010pe} for the conversion from the
$\rismom(\gamma_\mu,\slashed{q})$ and the
$\rismom(\gamma_\mu,\gamma_\mu)$ scheme to the $\NDR$ scheme for the
(VV+AA)--$(\Delta S=2)$ operator.

\setlength{\dataIla}{5.25cm} \setlength{\dataIlb}{1.5cm}

  \renewcommand{\dataINL}{} 
  \begin{table}[tp]
    \centering
    \begin{tabular}{lclcr}
      \hline
      $(i,j)$ &  &  $\Delta r^{\ri\to\msbar}_{ij}/\frac{\alpha_s}{4\pi}$ & & $\Delta r^{*}_{ij}/\frac{\alpha_s}{4\pi}$ \\
      \hline\hline
      \dataIabN{1}{1}{\xi\dIleft(-\frac{4\log(2)}{ N_c }\dIbsp$ $-\frac{ N_c }{2}\dIbsp$ $+\frac{2}{ N_c }\dIright)\dIbsp$ $-\frac{12\log(2)}{ N_c }\dIbsp$ $+\frac{7}{ N_c }}{-0.43926}
\dataIabN{1}{2}{\dIleft(4\log(2)\dIbsp$ $-\frac{3}{2}\dIright)\xi\dIbsp$ $+12\log(2)\dIbsp$ $-7}{1.31777}
\dataIabN{2}{1}{\dIleft(4\log(2)\dIbsp$ $-\frac{3}{2}\dIright)\xi\dIbsp$ $+12\log(2)\dIbsp$ $-7}{1.31777}
\dataIabN{2}{2}{\xi\dIleft(-\frac{4\log(2)}{ N_c }\dIbsp$ $-\frac{ N_c }{2}\dIbsp$ $+\frac{2}{ N_c }\dIright)\dIbsp$ $-\frac{12\log(2)}{ N_c }\dIbsp$ $+\frac{7}{ N_c }}{-0.43926}
\dataIabN{2}{3}{\frac{2}{9 N_c }}{0.07407}
\dataIabN{2}{4}{-\frac{2}{9}}{-0.22222}
\dataIabN{2}{5}{\frac{2}{9 N_c }}{0.07407}
\dataIabN{2}{6}{-\frac{2}{9}}{-0.22222}
\dataIabN{3}{3}{\xi\dIleft(-\frac{4\log(2)}{ N_c }\dIbsp$ $-\frac{ N_c }{2}\dIbsp$ $+\frac{2}{ N_c }\dIright)\dIbsp$ $-\frac{12\log(2)}{ N_c }\dIbsp$ $+\frac{67}{9 N_c }}{-0.29111}
\dataIabN{3}{4}{\dIleft(4\log(2)\dIbsp$ $-\frac{3}{2}\dIright)\xi\dIbsp$ $+12\log(2)\dIbsp$ $-\frac{67}{9}}{0.87332}
\dataIabN{3}{5}{\frac{4}{9 N_c }}{0.14815}
\dataIabN{3}{6}{-\frac{4}{9}}{-0.44444}
\dataIabN{4}{3}{\dIleft(4\log(2)\dIbsp$ $-\frac{3}{2}\dIright)\xi\dIbsp$ $+12\log(2)\dIbsp$ $+\frac{5}{3 N_c }\dIbsp$ $-7}{1.87332}
\dataIabN{4}{4}{\xi\dIleft(-\frac{4\log(2)}{ N_c }\dIbsp$ $-\frac{ N_c }{2}\dIbsp$ $+\frac{2}{ N_c }\dIright)\dIbsp$ $-\frac{12\log(2)}{ N_c }\dIbsp$ $+\frac{7}{ N_c }\dIbsp$ $-\frac{5}{3}}{-2.10592}
\dataIabN{4}{5}{\frac{5}{3 N_c }}{0.55556}
\dataIabN{4}{6}{-\frac{5}{3}}{-1.66667}
\dataIabN{5}{5}{\xi\dIleft(-\frac{2\log(2)}{ N_c }\dIbsp$ $-\frac{ N_c }{2}\dIright)\dIbsp$ $-\frac{2\log(2)}{ N_c }\dIbsp$ $-\frac{2}{ N_c }}{-1.12876}
\dataIabN{5}{6}{\dIleft(2\log(2)\dIbsp$ $+\frac{1}{2}\dIright)\xi\dIbsp$ $+2\log(2)\dIbsp$ $+2}{3.38629}
\dataIabN{6}{3}{\frac{5}{3 N_c }}{0.55556}
\dataIabN{6}{4}{-\frac{5}{3}}{-1.66667}
\dataIabN{6}{5}{(2\log(2)\dIbsp$ $-1)\xi\dIbsp$ $+2\log(2)\dIbsp$ $+\frac{5}{3 N_c }\dIbsp$ $-2}{-0.05815}
\dataIabN{6}{6}{\xi\dIleft( N_c \dIbsp$ $-\frac{2\log(2)}{ N_c }\dIright)\dIbsp$ $-\frac{2\log(2)}{ N_c }\dIbsp$ $+4 N_c \dIbsp$ $-\frac{2}{ N_c }\dIbsp$ $-\frac{5}{3}}{9.20457}
\dataIabN{7}{7}{\xi\dIleft(-\frac{2\log(2)}{ N_c }\dIbsp$ $-\frac{ N_c }{2}\dIright)\dIbsp$ $-\frac{2\log(2)}{ N_c }\dIbsp$ $-\frac{2}{ N_c }}{-1.12876}
\dataIabN{7}{8}{\dIleft(2\log(2)\dIbsp$ $+\frac{1}{2}\dIright)\xi\dIbsp$ $+2\log(2)\dIbsp$ $+2}{3.38629}
\dataIabN{8}{7}{(2\log(2)\dIbsp$ $-1)\xi\dIbsp$ $+2\log(2)\dIbsp$ $-2}{-0.61371}
\dataIabN{8}{8}{\xi\dIleft( N_c \dIbsp$ $-\frac{2\log(2)}{ N_c }\dIright)\dIbsp$ $-\frac{2\log(2)}{ N_c }\dIbsp$ $+4 N_c \dIbsp$ $-\frac{2}{ N_c }}{10.87124}
\dataIabN{9}{3}{-\frac{2}{9 N_c }}{-0.07407}
\dataIabN{9}{4}{\frac{2}{9}}{0.22222}
\dataIabN{9}{5}{-\frac{2}{9 N_c }}{-0.07407}
\dataIabN{9}{6}{\frac{2}{9}}{0.22222}
\dataIabN{9}{9}{\xi\dIleft(-\frac{4\log(2)}{ N_c }\dIbsp$ $-\frac{ N_c }{2}\dIbsp$ $+\frac{2}{ N_c }\dIright)\dIbsp$ $-\frac{12\log(2)}{ N_c }\dIbsp$ $+\frac{7}{ N_c }}{-0.43926}
\dataIabN{9}{10}{\dIleft(4\log(2)\dIbsp$ $-\frac{3}{2}\dIright)\xi\dIbsp$ $+12\log(2)\dIbsp$ $-7}{1.31777}
\dataIabN{10}{9}{\dIleft(4\log(2)\dIbsp$ $-\frac{3}{2}\dIright)\xi\dIbsp$ $+12\log(2)\dIbsp$ $-7}{1.31777}
\dataIabN{10}{10}{\xi\dIleft(-\frac{4\log(2)}{ N_c }\dIbsp$ $-\frac{ N_c }{2}\dIbsp$ $+\frac{2}{ N_c }\dIright)\dIbsp$ $-\frac{12\log(2)}{ N_c }\dIbsp$ $+\frac{7}{ N_c }}{-0.43926}
\\
      \hline
    \end{tabular}
    \caption{One-loop conversion matrix $\Delta
  r^{\ri\to\msbar}$ for basis I in the $\rimom$ scheme with $\ripmom$
  wave function renormalization.  We also give numerical values $\Delta
  r^{*}$ for $N_c=3$ and $\xi=0$.}
    \label{tab:gammaCQETQ}
  \end{table}

  \renewcommand{\dataINL}{} 
  \begin{table}[tp]
    \centering
    \begin{tabular}{lclcr}
      \hline
      $(i,j)$ &  &  $\Delta r^{\ri\to\msbar}_{ij}/\frac{\alpha_s}{4\pi}$ & & $\Delta r^{*}_{ij}/\frac{\alpha_s}{4\pi}$ \\
      \hline\hline
      \dataIabN{2}{3}{\frac{5}{9 N_c }}{0.18519}
\dataIabN{2}{4}{-\frac{5}{9}}{-0.55556}
\dataIabN{2}{5}{\frac{5}{9 N_c }}{0.18519}
\dataIabN{2}{6}{-\frac{5}{9}}{-0.55556}
\\
      \hline
    \end{tabular}
    \caption{One-loop conversion matrix
  $\Delta r^{\ri\to\msbar}$ for basis II in the $\rimom$ scheme with
  $\ripmom$ wave function renormalization.  All elements not given here
  are identical to Tab.~\ref{tab:gammaCQETQ}.}
    \label{tab:gammaCQEQ}
  \end{table}

  \renewcommand{\dataINL}{} 
  \begin{table}[tp]
    \centering
    \begin{tabular}{lclcr}
      \hline
      $(i,j)$ &  &  $\Delta r^{\ri\to\msbar}_{ij}/\frac{\alpha_s}{4\pi}$ & & $\Delta r^{*}_{ij}/\frac{\alpha_s}{4\pi}$ \\
      \hline\hline
      \dataPab{1}{1}{\xi\dIleft(-\frac{4\log(2)}{ N_c }\dIbsp$ $+4\log(2)\dIbsp$ $-\frac{ N_c }{2}\dIbsp$ $+\frac{2}{ N_c }\dIbsp$ $-\frac{3}{2}\dIright)\dIbsp$ $-\frac{12\log(2)}{ N_c }\dIbsp$ $+12\log(2)\dIbsp$ $+\frac{7}{ N_c }\dIbsp$ $-7}{0.87851}
\dataPab{2}{2}{\xi\dIleft(-\frac{4\log(2)}{ N_c }\dIbsp$ $-\frac{ N_c }{2}\dIbsp$ $+\frac{2}{ N_c }\dIright)\dIbsp$ $-\frac{12\log(2)}{ N_c }\dIbsp$ $+\frac{7}{ N_c }}{-0.43926}
\dataPab{2}{3}{\dIleft(4\log(2)\dIbsp$ $-\frac{3}{2}\dIright)\xi\dIbsp$ $+12\log(2)\dIbsp$ $-7}{1.31777}
\dataPab{3}{2}{\dIleft(4\log(2)\dIbsp$ $-\frac{3}{2}\dIright)\xi\dIbsp$ $+12\log(2)\dIbsp$ $+\frac{2}{3 N_c }\dIbsp$ $-\frac{67}{9}}{1.09554}
\dataPab{3}{3}{\xi\dIleft(-\frac{4\log(2)}{ N_c }\dIbsp$ $-\frac{ N_c }{2}\dIbsp$ $+\frac{2}{ N_c }\dIright)\dIbsp$ $-\frac{12\log(2)}{ N_c }\dIbsp$ $+\frac{67}{9 N_c }\dIbsp$ $-\frac{2}{3}}{-0.95777}
\dataPab{3}{4}{\frac{2}{9 N_c }}{0.07407}
\dataPab{3}{5}{-\frac{2}{9}}{-0.22222}
\dataPab{4}{4}{\xi\dIleft(-\frac{2\log(2)}{ N_c }\dIbsp$ $-\frac{ N_c }{2}\dIright)\dIbsp$ $-\frac{2\log(2)}{ N_c }\dIbsp$ $-\frac{2}{ N_c }}{-1.12876}
\dataPab{4}{5}{\dIleft(2\log(2)\dIbsp$ $+\frac{1}{2}\dIright)\xi\dIbsp$ $+2\log(2)\dIbsp$ $+2}{3.38629}
\dataPab{5}{2}{\frac{5}{ N_c }\dIbsp$ $-\frac{10}{3}}{-1.66667}
\dataPab{5}{3}{\frac{10}{3 N_c }\dIbsp$ $-5}{-3.88889}
\dataPab{5}{4}{(2\log(2)\dIbsp$ $-1)\xi\dIbsp$ $+2\log(2)\dIbsp$ $+\frac{5}{3 N_c }\dIbsp$ $-2}{-0.05815}
\dataPab{5}{5}{\xi\dIleft( N_c \dIbsp$ $-\frac{2\log(2)}{ N_c }\dIright)\dIbsp$ $-\frac{2\log(2)}{ N_c }\dIbsp$ $+4 N_c \dIbsp$ $-\frac{2}{ N_c }\dIbsp$ $-\frac{5}{3}}{9.20457}
\dataPab{6}{6}{\xi\dIleft(-\frac{2\log(2)}{ N_c }\dIbsp$ $-\frac{ N_c }{2}\dIright)\dIbsp$ $-\frac{2\log(2)}{ N_c }\dIbsp$ $-\frac{2}{ N_c }}{-1.12876}
\dataPab{6}{7}{\dIleft(2\log(2)\dIbsp$ $+\frac{1}{2}\dIright)\xi\dIbsp$ $+2\log(2)\dIbsp$ $+2}{3.38629}
\dataPab{7}{6}{(2\log(2)\dIbsp$ $-1)\xi\dIbsp$ $+2\log(2)\dIbsp$ $-2}{-0.61371}
\dataPab{7}{7}{\xi\dIleft( N_c \dIbsp$ $-\frac{2\log(2)}{ N_c }\dIright)\dIbsp$ $-\frac{2\log(2)}{ N_c }\dIbsp$ $+4 N_c \dIbsp$ $-\frac{2}{ N_c }}{10.87124}
\\
      \hline
    \end{tabular}
    \caption{One-loop conversion matrix $\Delta
  r^{\ri\to\msbar}$ for the \reduced{} basis in the $\rimom$ scheme with
  $\ripmom$ wave function renormalization.  We also give numerical
  values $\Delta r^{*}$ for $N_c=3$ and $\xi=0$.}
    \label{tab:gammaCQEPQ}
  \end{table}

  \renewcommand{\dataINL}{} 
  \begin{table}[tp]
    \centering
    \begin{tabular}{lclcr}
      \hline
      $(i,j)$ &  &  $\Delta r^{\ri\to\msbar}_{ij}/\frac{\alpha_s}{4\pi}$ & & $\Delta r^{*}_{ij}/\frac{\alpha_s}{4\pi}$ \\
      \hline\hline
      \dataPab{1}{1}{\xi\dIleft(-\frac{ C_0  N_c }{2}\dIbsp$ $+\frac{ C_0 }{ N_c }\dIbsp$ $-\frac{ C_0 }{2}\dIbsp$ $-\frac{4\log(2)}{ N_c }\dIbsp$ $+4\log(2)\dIbsp$ $+\frac{ N_c }{2}\dIbsp$ $-\frac{1}{2}\dIright)\dIbsp$ $-\frac{12\log(2)}{ N_c }\dIbsp$ $+12\log(2)\dIbsp$ $- N_c \dIbsp$ $+\frac{9}{ N_c }\dIbsp$ $-8}{-2.45482}
\dataPab{2}{2}{\xi\dIleft(-\frac{ C_0  N_c }{2}\dIbsp$ $+\frac{ C_0 }{ N_c }\dIbsp$ $-\frac{4\log(2)}{ N_c }\dIbsp$ $+\frac{ N_c }{2}\dIright)\dIbsp$ $-\frac{12\log(2)}{ N_c }\dIbsp$ $- N_c \dIbsp$ $+\frac{9}{ N_c }}{-2.77259}
\dataPab{2}{3}{\xi\dIleft(-\frac{ C_0 }{2}\dIbsp$ $+4\log(2)\dIbsp$ $-\frac{1}{2}\dIright)\dIbsp$ $+12\log(2)\dIbsp$ $-8}{0.31777}
\dataPab{3}{2}{\xi\dIleft(-\frac{ C_0 }{2}\dIbsp$ $+4\log(2)\dIbsp$ $-\frac{1}{2}\dIright)\dIbsp$ $+12\log(2)\dIbsp$ $+\frac{2}{3 N_c }\dIbsp$ $-\frac{76}{9}}{0.09554}
\dataPab{3}{3}{\xi\dIleft(-\frac{ C_0  N_c }{2}\dIbsp$ $+\frac{ C_0 }{ N_c }\dIbsp$ $-\frac{4\log(2)}{ N_c }\dIbsp$ $+\frac{ N_c }{2}\dIright)\dIbsp$ $-\frac{12\log(2)}{ N_c }\dIbsp$ $- N_c \dIbsp$ $+\frac{85}{9 N_c }\dIbsp$ $-\frac{2}{3}}{-3.29111}
\dataPab{3}{4}{\frac{2}{9 N_c }}{0.07407}
\dataPab{3}{5}{-\frac{2}{9}}{-0.22222}
\dataPab{4}{4}{\xi\dIleft(-\frac{ C_0  N_c }{2}\dIbsp$ $+\frac{ C_0 }{ N_c }\dIbsp$ $-\frac{2\log(2)}{ N_c }\dIbsp$ $+\frac{ N_c }{2}\dIbsp$ $-\frac{1}{ N_c }\dIright)\dIbsp$ $+\frac{3 C_0 }{2 N_c }\dIbsp$ $-\frac{2\log(2)}{ N_c }\dIbsp$ $- N_c \dIbsp$ $-\frac{1}{ N_c }}{-2.62348}
\dataPab{4}{5}{\xi\dIleft(-\frac{ C_0 }{2}\dIbsp$ $+2\log(2)\dIbsp$ $+\frac{1}{2}\dIright)\dIbsp$ $-\frac{3 C_0 }{2}\dIbsp$ $+2\log(2)\dIbsp$ $+2}{-0.12957}
\dataPab{5}{2}{\frac{5}{ N_c }\dIbsp$ $-\frac{10}{3}}{-1.66667}
\dataPab{5}{3}{\frac{10}{3 N_c }\dIbsp$ $-5}{-3.88889}
\dataPab{5}{4}{\xi\dIleft(2\log(2)\dIbsp$ $-\frac{ C_0 }{2}\dIright)\dIbsp$ $+2\log(2)\dIbsp$ $+\frac{5}{3 N_c }\dIbsp$ $-3}{-1.05815}
\dataPab{5}{5}{\xi\dIleft(-\frac{ C_0  N_c }{2}\dIbsp$ $+\frac{ C_0 }{ N_c }\dIbsp$ $-\frac{2\log(2)}{ N_c }\dIbsp$ $+ N_c \dIbsp$ $-\frac{1}{ N_c }\dIright)\dIbsp$ $-\frac{3 C_0  N_c }{2}\dIbsp$ $+\frac{3 C_0 }{2 N_c }\dIbsp$ $-\frac{2\log(2)}{ N_c }\dIbsp$ $+4 N_c \dIbsp$ $-\frac{1}{ N_c }\dIbsp$ $-\frac{5}{3}}{0.16227}
\dataPab{6}{6}{\xi\dIleft(-\frac{ C_0  N_c }{2}\dIbsp$ $+\frac{ C_0 }{ N_c }\dIbsp$ $-\frac{2\log(2)}{ N_c }\dIbsp$ $+\frac{ N_c }{2}\dIbsp$ $-\frac{1}{ N_c }\dIright)\dIbsp$ $+\frac{3 C_0 }{2 N_c }\dIbsp$ $-\frac{2\log(2)}{ N_c }\dIbsp$ $- N_c \dIbsp$ $-\frac{1}{ N_c }}{-2.62348}
\dataPab{6}{7}{\xi\dIleft(-\frac{ C_0 }{2}\dIbsp$ $+2\log(2)\dIbsp$ $+\frac{1}{2}\dIright)\dIbsp$ $-\frac{3 C_0 }{2}\dIbsp$ $+2\log(2)\dIbsp$ $+2}{-0.12957}
\dataPab{7}{6}{\xi\dIleft(2\log(2)\dIbsp$ $-\frac{ C_0 }{2}\dIright)\dIbsp$ $+2\log(2)\dIbsp$ $-3}{-1.61371}
\dataPab{7}{7}{\xi\dIleft(-\frac{ C_0  N_c }{2}\dIbsp$ $+\frac{ C_0 }{ N_c }\dIbsp$ $-\frac{2\log(2)}{ N_c }\dIbsp$ $+ N_c \dIbsp$ $-\frac{1}{ N_c }\dIright)\dIbsp$ $-\frac{3 C_0  N_c }{2}\dIbsp$ $+\frac{3 C_0 }{2 N_c }\dIbsp$ $-\frac{2\log(2)}{ N_c }\dIbsp$ $+4 N_c \dIbsp$ $-\frac{1}{ N_c }}{1.82894}
\\
      \hline
    \end{tabular}
    \caption{One-loop conversion matrix $\Delta
  r^{\ri\to\msbar}$ for the \reduced{} basis in the
  $\rismom(\gamma_\mu,\slashed{q})$ scheme.  We also give numerical
  values $\Delta r^{*}$ for $N_c=3$ and $\xi=0$.}
    \label{tab:gammaCQNEPQ}
  \end{table}

  \renewcommand{\dataINL}{} 
  \begin{table}[tp]
    \centering
    \begin{tabular}{lclcr}
      \hline
      $(i,j)$ &  &  $\Delta r^{\ri\to\msbar}_{ij}/\frac{\alpha_s}{4\pi}$ & & $\Delta r^{*}_{ij}/\frac{\alpha_s}{4\pi}$ \\
      \hline\hline
      \dataPab{1}{1}{\xi\dIleft(\frac{ C_0 }{2 N_c }\dIbsp$ $-\frac{ C_0 }{2}\dIbsp$ $-\frac{4\log(2)}{ N_c }\dIbsp$ $+4\log(2)\dIbsp$ $+\frac{1}{2 N_c }\dIbsp$ $-\frac{1}{2}\dIright)\dIbsp$ $-\frac{12\log(2)}{ N_c }\dIbsp$ $+12\log(2)\dIbsp$ $+\frac{8}{ N_c }\dIbsp$ $-8}{0.21184}
\dataPab{2}{2}{\xi\dIleft(\frac{ C_0 }{2 N_c }\dIbsp$ $-\frac{4\log(2)}{ N_c }\dIbsp$ $+\frac{1}{2 N_c }\dIright)\dIbsp$ $-\frac{12\log(2)}{ N_c }\dIbsp$ $+\frac{8}{ N_c }}{-0.10592}
\dataPab{2}{3}{\xi\dIleft(-\frac{ C_0 }{2}\dIbsp$ $+4\log(2)\dIbsp$ $-\frac{1}{2}\dIright)\dIbsp$ $+12\log(2)\dIbsp$ $-8}{0.31777}
\dataPab{3}{2}{\xi\dIleft(-\frac{ C_0 }{2}\dIbsp$ $+4\log(2)\dIbsp$ $-\frac{1}{2}\dIright)\dIbsp$ $+12\log(2)\dIbsp$ $+\frac{2}{3 N_c }\dIbsp$ $-\frac{76}{9}}{0.09554}
\dataPab{3}{3}{\xi\dIleft(\frac{ C_0 }{2 N_c }\dIbsp$ $-\frac{4\log(2)}{ N_c }\dIbsp$ $+\frac{1}{2 N_c }\dIright)\dIbsp$ $-\frac{12\log(2)}{ N_c }\dIbsp$ $+\frac{76}{9 N_c }\dIbsp$ $-\frac{2}{3}}{-0.62444}
\dataPab{3}{4}{\frac{2}{9 N_c }}{0.07407}
\dataPab{3}{5}{-\frac{2}{9}}{-0.22222}
\dataPab{4}{4}{\xi\dIleft(\frac{ C_0 }{2 N_c }\dIbsp$ $-\frac{2\log(2)}{ N_c }\dIbsp$ $-\frac{1}{2 N_c }\dIright)\dIbsp$ $+\frac{3 C_0 }{2 N_c }\dIbsp$ $-\frac{2\log(2)}{ N_c }\dIbsp$ $-\frac{2}{ N_c }}{0.04319}
\dataPab{4}{5}{\xi\dIleft(-\frac{ C_0 }{2}\dIbsp$ $+2\log(2)\dIbsp$ $+\frac{1}{2}\dIright)\dIbsp$ $-\frac{3 C_0 }{2}\dIbsp$ $+2\log(2)\dIbsp$ $+2}{-0.12957}
\dataPab{5}{2}{\frac{5}{ N_c }\dIbsp$ $-\frac{10}{3}}{-1.66667}
\dataPab{5}{3}{\frac{10}{3 N_c }\dIbsp$ $-5}{-3.88889}
\dataPab{5}{4}{\xi\dIleft(2\log(2)\dIbsp$ $-\frac{ C_0 }{2}\dIright)\dIbsp$ $+2\log(2)\dIbsp$ $+\frac{5}{3 N_c }\dIbsp$ $-3}{-1.05815}
\dataPab{5}{5}{\xi\dIleft(\frac{ C_0 }{2 N_c }\dIbsp$ $-\frac{2\log(2)}{ N_c }\dIbsp$ $+\frac{ N_c }{2}\dIbsp$ $-\frac{1}{2 N_c }\dIright)\dIbsp$ $-\frac{3 C_0  N_c }{2}\dIbsp$ $+\frac{3 C_0 }{2 N_c }\dIbsp$ $-\frac{2\log(2)}{ N_c }\dIbsp$ $+5 N_c \dIbsp$ $-\frac{2}{ N_c }\dIbsp$ $-\frac{5}{3}}{2.82894}
\dataPab{6}{6}{\xi\dIleft(\frac{ C_0 }{2 N_c }\dIbsp$ $-\frac{2\log(2)}{ N_c }\dIbsp$ $-\frac{1}{2 N_c }\dIright)\dIbsp$ $+\frac{3 C_0 }{2 N_c }\dIbsp$ $-\frac{2\log(2)}{ N_c }\dIbsp$ $-\frac{2}{ N_c }}{0.04319}
\dataPab{6}{7}{\xi\dIleft(-\frac{ C_0 }{2}\dIbsp$ $+2\log(2)\dIbsp$ $+\frac{1}{2}\dIright)\dIbsp$ $-\frac{3 C_0 }{2}\dIbsp$ $+2\log(2)\dIbsp$ $+2}{-0.12957}
\dataPab{7}{6}{\xi\dIleft(2\log(2)\dIbsp$ $-\frac{ C_0 }{2}\dIright)\dIbsp$ $+2\log(2)\dIbsp$ $-3}{-1.61371}
\dataPab{7}{7}{\xi\dIleft(\frac{ C_0 }{2 N_c }\dIbsp$ $-\frac{2\log(2)}{ N_c }\dIbsp$ $+\frac{ N_c }{2}\dIbsp$ $-\frac{1}{2 N_c }\dIright)\dIbsp$ $-\frac{3 C_0  N_c }{2}\dIbsp$ $+\frac{3 C_0 }{2 N_c }\dIbsp$ $-\frac{2\log(2)}{ N_c }\dIbsp$ $+5 N_c \dIbsp$ $-\frac{2}{ N_c }}{4.49561}
\\
      \hline
    \end{tabular}
    \caption{One-loop conversion matrix $\Delta
  r^{\ri\to\msbar}$ for the \reduced{} basis in the
  $\rismom(\gamma_\mu,\gamma_\mu)$ scheme.  We also give numerical
  values $\Delta r^{*}$ for $N_c=3$ and $\xi=0$.}
    \label{tab:gammaCGNEPQ}
  \end{table}

\subsection{\boldmath Results for the $\bfrismom(\slashed{q},y)$ schemes\label{sec:convq}}
In the following we discuss the non-exceptional momentum configuration
and $\rismom$ schemes defined by projectors of the class
$P^{\slashed{q}}$.  This allows for the definition of new and
independent $\ri$ schemes which we call $\rismom$($\slashed{q},y$)
schemes, where $y$ denotes the choice of the wave function
renormalization scheme as in the previous section.  We give results
only for the \reduced{} basis which does not contain linearly
dependent operators.  Since mixing only occurs within the blocks of
the $(27,1)$, $(8,1)$, and $(8,8)$ operators, we renormalize each
block separately.

We first note that for a scheme with projectors of the class
$P^{\slashed{q}}$ the conversion matrices cannot be simply read off from
Tabs.~\ref{tab:gammaEPQ}--\ref{tab:gammaEPQG} due to the nonzero
contribution of $\GAMMA^{\slashed{q}}$. This means that the schemes
defined in this section, unlike the schemes discussed in the previous
section \ref{sec:convg}, depend on the specific choice of the
projectors.  However, any non-degenerate linear transformation of a given
set of projectors leaves the conversion matrix of Eq.~\eqref{eq:co}
invariant.  We give the projectors used to define the
$\rismom(\slashed{q},y)$ schemes in the following.  The projectors will
be expressed in terms of
\begin{align}\label{eqn:defmasterproj}
  P^{VV \pm AA,\slashed{q}}_{(1),f'} &= \delta_{ff'}
  \delta_{ij}\delta_{kl} \notag\\&\quad\times
  [(\slashed{q})_{\beta\alpha}(\slashed{q})_{\delta\gamma} \pm
  (\slashed{q}\gamma_5)_{\beta\alpha}(\slashed{q}\gamma_5)_{\delta\gamma}] / q^2\,, \notag\\
  P^{VV \pm AA,\slashed{q}}_{(2),f'} &= \delta_{ff'}
  \delta_{il}\delta_{kj} \notag\\&\quad\times
  [(\slashed{q})_{\beta\alpha}(\slashed{q})_{\delta\gamma} \pm
  (\slashed{q}\gamma_5)_{\beta\alpha}(\slashed{q}\gamma_5)_{\delta\gamma}]/ q^2\,, \notag\\
  P^{VV \pm AA,\slashed{q}}_{(3),f'} &= \delta_{ff'}
  \delta_{il}\delta_{kj} \notag\\&\quad\times
  [(\slashed{q})_{\beta\gamma}(\slashed{q})_{\delta\alpha} \pm
  (\slashed{q}\gamma_5)_{\beta\gamma}(\slashed{q}\gamma_5)_{\delta\alpha}]/q^2\,, \notag\\
  P^{VV \pm AA,\slashed{q}}_{(4),f'} &= \delta_{ff'}
  \delta_{ij}\delta_{kl} \notag\\&\quad\times
  [(\slashed{q})_{\beta\gamma}(\slashed{q})_{\delta\alpha} \pm
  (\slashed{q}\gamma_5)_{\beta\gamma}(\slashed{q}\gamma_5)_{\delta\alpha}]/q^2\,,
\end{align}
where all indices are as shown in Fig.~\ref{fig:cctree}.

The $(27,1)$ operator $Q'_1$ only mixes with itself, so we only need
to provide a single projector $P^{\slashed{q}}_{(27,1)}$.  We adopt
the $\rismom(\slashed{q},\slashed{q})$ and
$\rismom(\slashed{q},\gamma_\mu)$ schemes defined in
Ref.~\cite{Aoki:2010pe} so that for the $(27,1)$ operator we use
\begin{align}
  P^{\slashed{q}}_{(27,1),1} &=
  \frac{P^{VV+AA,\slashed{q}}_{(1),u}}{64N_c(N_c+1)}\,.
\end{align}
The $(8,8)$ operators $Q'_7$ and $Q'_8$ also only mix with themselves 
under renormalization, and therefore we have to provide two
projectors.  We choose
\begin{align}
  P^{\slashed{q}}_{(8,8),7} &=
  \frac{N_c P^{VV-AA,\slashed{q}}_{(1),u}-P^{VV-AA,\slashed{q}}_{(2),u}}{32 N_c (N_c^2-1)}
\,, \notag\\
  P^{\slashed{q}}_{(8,8),8} &=
  \frac{-P^{VV-AA,\slashed{q}}_{(1),u}+N_c P^{VV-AA,\slashed{q}}_{(2),u}}{32 N_c (N_c^2-1)}
\end{align}
with the property
\begin{align}
  P^{\slashed{q}}_{(8,8),i} \Gamma^\tree_4({Q'}_j) = \delta_{ij}\,, \quad i,j=7,8\,.
\end{align}
The one-loop mixing coefficients for the $(8,1)$ operators shall be
determined using the projectors
\begin{align}
  P^{\slashed{q}}_{(8,1),2} &=
  \frac{(3N_c-2)P^{VV+AA,\slashed{q}}_{(1),u} + (2N_c-3)P^{VV+AA,\slashed{q}}_{(2),u}}{32 N_c (N_c^2-1)}
  \,, \notag\\
  P^{\slashed{q}}_{(8,1),3} &=
  \frac{(2N_c-3)P^{VV+AA,\slashed{q}}_{(1),u} + (3N_c-2)P^{VV+AA,\slashed{q}}_{(2),u}}{32 N_c (N_c^2-1)}
  \,,\notag\\
  P^{\slashed{q}}_{(8,1),5} &=
  P^{\slashed{q}}_{(8,8),7} \,, \qquad
  P^{\slashed{q}}_{(8,1),6} =
  P^{\slashed{q}}_{(8,8),8}
  \,,
\end{align}
where
\begin{align}
  P^{\slashed{q}}_{(8,1),i} \Gamma^\tree_4({Q'}_j) &= \delta_{ij}\,, & i,j &=2,3,5,6\,.
\end{align}
Other choices of projectors within class $P^{\slashed{q}}$, which are
not linear combinations of the projectors given above, are possible
and might lead to smaller conversion factors and a better convergence
of the perturbative expansion.  The reader can easily obtain the
respective projections using the results which we provide in
Tabs.~\ref{tab:gammaEPQ}--\ref{tab:gammaEPQG}.

In Tabs.~\ref{tab:gammaCGQNEPQ} and \ref{tab:gammaCQQNEPQ} we provide
the resulting one-loop conversion matrices $\Delta r^{\ri\to\msbar}$
for the $\rismom(\slashed{q},\gamma_\mu)$ and
$\rismom(\slashed{q},\slashed{q})$ schemes.  We note that the result
for the $(27,1)$ operator agrees with the result of
Ref.~\cite{Aoki:2010pe} for the conversion of the (VV+AA)--$(\Delta
S=2)$ operator.  In
Tabs.~\ref{tab:gammaCGQNEPQA}--\ref{tab:gammaCQQNEPQA} we give the
individual conversion matrices $\Delta a^{\ri\to\msbar}$ for the
$(8,1)$ operators in the same schemes.

  \renewcommand{\dataINL}{} 
  \begin{table}[tp]
    \centering
    \begin{tabular}{lclcr}
      \hline
      $(i,j)$ &  &  $\Delta r^{\ri\to\msbar}_{ij}/\frac{\alpha_s}{4\pi}$ & & $\Delta r^{*}_{ij}/\frac{\alpha_s}{4\pi}$ \\
      \hline\hline
      \dataPab{1}{1}{\xi\dIleft(\frac{ C_0  N_c }{2}\dIbsp$ $+\frac{ C_0 }{2 N_c }\dIbsp$ $- C_0 \dIbsp$ $-\frac{4\log(2)}{ N_c }\dIbsp$ $+4\log(2)\dIbsp$ $-\frac{ N_c }{2}\dIbsp$ $+\frac{1}{2 N_c }\dIright)\dIbsp$ $-\frac{12\log(2)}{ N_c }\dIbsp$ $+12\log(2)\dIbsp$ $+ N_c \dIbsp$ $+\frac{8}{ N_c }\dIbsp$ $-9}{2.21184}
\dataPab{2}{2}{\xi\dIleft(\frac{13 C_0  N_c }{10}\dIbsp$ $+\frac{ C_0 }{2 N_c }\dIbsp$ $-\frac{6 C_0 }{5}\dIbsp$ $-\frac{4\log(2)}{ N_c }\dIbsp$ $-\frac{13 N_c }{10}\dIbsp$ $+\frac{1}{2 N_c }\dIbsp$ $+\frac{6}{5}\dIright)\dIbsp$ $-\frac{12\log(2)}{ N_c }\dIbsp$ $+\frac{13 N_c }{5}\dIbsp$ $+\frac{8}{ N_c }\dIbsp$ $-\frac{12}{5}}{5.29408}
\dataPab{2}{3}{\xi\dIleft(\frac{6 C_0  N_c }{5}\dIbsp$ $-\frac{9 C_0 }{5}\dIbsp$ $+4\log(2)\dIbsp$ $-\frac{6 N_c }{5}\dIbsp$ $+\frac{4}{5}\dIright)\dIbsp$ $+12\log(2)\dIbsp$ $+\frac{12 N_c }{5}\dIbsp$ $-\frac{53}{5}}{4.91777}
\dataPab{3}{2}{\xi\dIleft(-\frac{6 C_0  N_c }{5}\dIbsp$ $+\frac{4 C_0 }{5}\dIbsp$ $+4\log(2)\dIbsp$ $+\frac{6 N_c }{5}\dIbsp$ $-\frac{9}{5}\dIright)\dIbsp$ $+12\log(2)\dIbsp$ $-\frac{12 N_c }{5}\dIbsp$ $+\frac{2}{3 N_c }\dIbsp$ $-\frac{263}{45}}{-4.50446}
\dataPab{3}{3}{\xi\dIleft(-\frac{13 C_0  N_c }{10}\dIbsp$ $+\frac{ C_0 }{2 N_c }\dIbsp$ $+\frac{6 C_0 }{5}\dIbsp$ $-\frac{4\log(2)}{ N_c }\dIbsp$ $+\frac{13 N_c }{10}\dIbsp$ $+\frac{1}{2 N_c }\dIbsp$ $-\frac{6}{5}\dIright)\dIbsp$ $-\frac{12\log(2)}{ N_c }\dIbsp$ $-\frac{13 N_c }{5}\dIbsp$ $+\frac{76}{9 N_c }\dIbsp$ $+\frac{26}{15}}{-6.02444}
\dataPab{3}{4}{\frac{2}{9 N_c }}{0.07407}
\dataPab{3}{5}{-\frac{2}{9}}{-0.22222}
\dataPab{4}{4}{\xi\dIleft(\frac{ C_0  N_c }{2}\dIbsp$ $-\frac{2\log(2)}{ N_c }\dIbsp$ $-\frac{ N_c }{2}\dIright)\dIbsp$ $+\frac{3 C_0 }{2 N_c }\dIbsp$ $-\frac{2\log(2)}{ N_c }\dIbsp$ $+ N_c \dIbsp$ $-\frac{3}{ N_c }}{2.70986}
\dataPab{4}{5}{\xi\dIleft(-\frac{ C_0 }{2}\dIbsp$ $+2\log(2)\dIbsp$ $+\frac{1}{2}\dIright)\dIbsp$ $-\frac{3 C_0 }{2}\dIbsp$ $+2\log(2)\dIbsp$ $+2}{-0.12957}
\dataPab{5}{2}{\frac{5}{ N_c }\dIbsp$ $-\frac{10}{3}}{-1.66667}
\dataPab{5}{3}{\frac{10}{3 N_c }\dIbsp$ $-5}{-3.88889}
\dataPab{5}{4}{\dIleft(2\log(2)\dIbsp$ $-\frac{1}{2}\dIright)\xi\dIbsp$ $+2\log(2)\dIbsp$ $+\frac{5}{3 N_c }\dIbsp$ $-2}{-0.05815}
\dataPab{5}{5}{-\frac{3 C_0  N_c }{2}\dIbsp$ $+\frac{3 C_0 }{2 N_c }\dIbsp$ $+\xi\dIleft(\frac{ N_c }{2}\dIbsp$ $-\frac{2\log(2)}{ N_c }\dIright)\dIbsp$ $-\frac{2\log(2)}{ N_c }\dIbsp$ $+5 N_c \dIbsp$ $-\frac{3}{ N_c }\dIbsp$ $-\frac{5}{3}}{2.49561}
\dataPab{6}{6}{\xi\dIleft(\frac{ C_0  N_c }{2}\dIbsp$ $-\frac{2\log(2)}{ N_c }\dIbsp$ $-\frac{ N_c }{2}\dIright)\dIbsp$ $+\frac{3 C_0 }{2 N_c }\dIbsp$ $-\frac{2\log(2)}{ N_c }\dIbsp$ $+ N_c \dIbsp$ $-\frac{3}{ N_c }}{2.70986}
\dataPab{6}{7}{\xi\dIleft(-\frac{ C_0 }{2}\dIbsp$ $+2\log(2)\dIbsp$ $+\frac{1}{2}\dIright)\dIbsp$ $-\frac{3 C_0 }{2}\dIbsp$ $+2\log(2)\dIbsp$ $+2}{-0.12957}
\dataPab{7}{6}{\dIleft(2\log(2)\dIbsp$ $-\frac{1}{2}\dIright)\xi\dIbsp$ $+2\log(2)\dIbsp$ $-2}{-0.61371}
\dataPab{7}{7}{-\frac{3 C_0  N_c }{2}\dIbsp$ $+\frac{3 C_0 }{2 N_c }\dIbsp$ $+\xi\dIleft(\frac{ N_c }{2}\dIbsp$ $-\frac{2\log(2)}{ N_c }\dIright)\dIbsp$ $-\frac{2\log(2)}{ N_c }\dIbsp$ $+5 N_c \dIbsp$ $-\frac{3}{ N_c }}{4.16227}
\\
      \hline
    \end{tabular}
    \caption{One-loop conversion matrix $\Delta
  r^{\ri\to\msbar}$ for the \reduced{} basis in the
  $\rismom(\slashed{q},\gamma_\mu)$ scheme.  We also give numerical
  values $\Delta r^{*}$ for $N_c=3$ and $\xi=0$.}
    \label{tab:gammaCGQNEPQ}
  \end{table}

  \renewcommand{\dataINL}{} 
  \begin{table}[tp]
    \centering
    \begin{tabular}{lclcr}
      \hline
      $(i,j)$ &  &  $\Delta r^{\ri\to\msbar}_{ij}/\frac{\alpha_s}{4\pi}$ & & $\Delta r^{*}_{ij}/\frac{\alpha_s}{4\pi}$ \\
      \hline\hline
      \dataPab{1}{1}{\xi\dIleft(\frac{ C_0 }{ N_c }\dIbsp$ $- C_0 \dIbsp$ $-\frac{4\log(2)}{ N_c }\dIbsp$ $+4\log(2)\dIright)\dIbsp$ $-\frac{12\log(2)}{ N_c }\dIbsp$ $+12\log(2)\dIbsp$ $+\frac{9}{ N_c }\dIbsp$ $-9}{-0.45482}
\dataPab{2}{2}{\xi\dIleft(\frac{4 C_0  N_c }{5}\dIbsp$ $+\frac{ C_0 }{ N_c }\dIbsp$ $-\frac{6 C_0 }{5}\dIbsp$ $-\frac{4\log(2)}{ N_c }\dIbsp$ $-\frac{4 N_c }{5}\dIbsp$ $+\frac{6}{5}\dIright)\dIbsp$ $-\frac{12\log(2)}{ N_c }\dIbsp$ $+\frac{8 N_c }{5}\dIbsp$ $+\frac{9}{ N_c }\dIbsp$ $-\frac{12}{5}}{2.62741}
\dataPab{2}{3}{\xi\dIleft(\frac{6 C_0  N_c }{5}\dIbsp$ $-\frac{9 C_0 }{5}\dIbsp$ $+4\log(2)\dIbsp$ $-\frac{6 N_c }{5}\dIbsp$ $+\frac{4}{5}\dIright)\dIbsp$ $+12\log(2)\dIbsp$ $+\frac{12 N_c }{5}\dIbsp$ $-\frac{53}{5}}{4.91777}
\dataPab{3}{2}{\xi\dIleft(-\frac{6 C_0  N_c }{5}\dIbsp$ $+\frac{4 C_0 }{5}\dIbsp$ $+4\log(2)\dIbsp$ $+\frac{6 N_c }{5}\dIbsp$ $-\frac{9}{5}\dIright)\dIbsp$ $+12\log(2)\dIbsp$ $-\frac{12 N_c }{5}\dIbsp$ $+\frac{2}{3 N_c }\dIbsp$ $-\frac{263}{45}}{-4.50446}
\dataPab{3}{3}{\xi\dIleft(-\frac{9 C_0  N_c }{5}\dIbsp$ $+\frac{ C_0 }{ N_c }\dIbsp$ $+\frac{6 C_0 }{5}\dIbsp$ $-\frac{4\log(2)}{ N_c }\dIbsp$ $+\frac{9 N_c }{5}\dIbsp$ $-\frac{6}{5}\dIright)\dIbsp$ $-\frac{12\log(2)}{ N_c }\dIbsp$ $-\frac{18 N_c }{5}\dIbsp$ $+\frac{85}{9 N_c }\dIbsp$ $+\frac{26}{15}}{-8.69111}
\dataPab{3}{4}{\frac{2}{9 N_c }}{0.07407}
\dataPab{3}{5}{-\frac{2}{9}}{-0.22222}
\dataPab{4}{4}{\xi\dIleft(\frac{ C_0 }{2 N_c }\dIbsp$ $-\frac{2\log(2)}{ N_c }\dIbsp$ $-\frac{1}{2 N_c }\dIright)\dIbsp$ $+\frac{3 C_0 }{2 N_c }\dIbsp$ $-\frac{2\log(2)}{ N_c }\dIbsp$ $-\frac{2}{ N_c }}{0.04319}
\dataPab{4}{5}{\xi\dIleft(-\frac{ C_0 }{2}\dIbsp$ $+2\log(2)\dIbsp$ $+\frac{1}{2}\dIright)\dIbsp$ $-\frac{3 C_0 }{2}\dIbsp$ $+2\log(2)\dIbsp$ $+2}{-0.12957}
\dataPab{5}{2}{\frac{5}{ N_c }\dIbsp$ $-\frac{10}{3}}{-1.66667}
\dataPab{5}{3}{\frac{10}{3 N_c }\dIbsp$ $-5}{-3.88889}
\dataPab{5}{4}{\dIleft(2\log(2)\dIbsp$ $-\frac{1}{2}\dIright)\xi\dIbsp$ $+2\log(2)\dIbsp$ $+\frac{5}{3 N_c }\dIbsp$ $-2}{-0.05815}
\dataPab{5}{5}{\xi\dIleft(-\frac{ C_0  N_c }{2}\dIbsp$ $+\frac{ C_0 }{2 N_c }\dIbsp$ $-\frac{2\log(2)}{ N_c }\dIbsp$ $+ N_c \dIbsp$ $-\frac{1}{2 N_c }\dIright)\dIbsp$ $-\frac{3 C_0  N_c }{2}\dIbsp$ $+\frac{3 C_0 }{2 N_c }\dIbsp$ $-\frac{2\log(2)}{ N_c }\dIbsp$ $+4 N_c \dIbsp$ $-\frac{2}{ N_c }\dIbsp$ $-\frac{5}{3}}{-0.17106}
\dataPab{6}{6}{\xi\dIleft(\frac{ C_0 }{2 N_c }\dIbsp$ $-\frac{2\log(2)}{ N_c }\dIbsp$ $-\frac{1}{2 N_c }\dIright)\dIbsp$ $+\frac{3 C_0 }{2 N_c }\dIbsp$ $-\frac{2\log(2)}{ N_c }\dIbsp$ $-\frac{2}{ N_c }}{0.04319}
\dataPab{6}{7}{\xi\dIleft(-\frac{ C_0 }{2}\dIbsp$ $+2\log(2)\dIbsp$ $+\frac{1}{2}\dIright)\dIbsp$ $-\frac{3 C_0 }{2}\dIbsp$ $+2\log(2)\dIbsp$ $+2}{-0.12957}
\dataPab{7}{6}{\dIleft(2\log(2)\dIbsp$ $-\frac{1}{2}\dIright)\xi\dIbsp$ $+2\log(2)\dIbsp$ $-2}{-0.61371}
\dataPab{7}{7}{\xi\dIleft(-\frac{ C_0  N_c }{2}\dIbsp$ $+\frac{ C_0 }{2 N_c }\dIbsp$ $-\frac{2\log(2)}{ N_c }\dIbsp$ $+ N_c \dIbsp$ $-\frac{1}{2 N_c }\dIright)\dIbsp$ $-\frac{3 C_0  N_c }{2}\dIbsp$ $+\frac{3 C_0 }{2 N_c }\dIbsp$ $-\frac{2\log(2)}{ N_c }\dIbsp$ $+4 N_c \dIbsp$ $-\frac{2}{ N_c }}{1.49561}
\\
      \hline
    \end{tabular}
    \caption{One-loop conversion matrix $\Delta
  r^{\ri\to\msbar}$ for the \reduced{} basis in the
  $\rismom(\slashed{q},\slashed{q})$ scheme.  We also give numerical
  values $\Delta r^{*}$ for $N_c=3$ and $\xi=0$.}
    \label{tab:gammaCQQNEPQ}
  \end{table}

  \begin{table}[tp]
    \centering
    \renewcommand{\dataINL}{}
    \begin{tabular}{lcl}
      \hline
      $(i,j)$ &  &  $\Delta a^{\ri\to\msbar}_{ij}/\frac{\alpha_s}{4\pi}$ \\
      \hline\hline
      \dataPa{2}{2}{\xi\dIleft(\frac{13 C_0  N_c }{10}\dIbsp$ $+\frac{ C_0 }{2 N_c }\dIbsp$ $-\frac{6 C_0 }{5}\dIbsp$ $-\frac{4\log(2)}{ N_c }\dIbsp$ $-\frac{13 N_c }{10}\dIbsp$ $+\frac{1}{2 N_c }\dIbsp$ $+\frac{6}{5}\dIright)\dIbsp$ $-\frac{12\log(2)}{ N_c }\dIbsp$ $+\frac{13 N_c }{5}\dIbsp$ $+\frac{8}{ N_c }\dIbsp$ $-\frac{12}{5}}
\dataPa{2}{3}{\xi\dIleft(\frac{6 C_0  N_c }{5}\dIbsp$ $-\frac{9 C_0 }{5}\dIbsp$ $+4\log(2)\dIbsp$ $-\frac{6 N_c }{5}\dIbsp$ $+\frac{4}{5}\dIright)\dIbsp$ $+12\log(2)\dIbsp$ $+\frac{12 N_c }{5}\dIbsp$ $-\frac{53}{5}}
\dataPa{3}{2}{\xi\dIleft(-\frac{6 C_0  N_c }{5}\dIbsp$ $+\frac{4 C_0 }{5}\dIbsp$ $+4\log(2)\dIbsp$ $+\frac{6 N_c }{5}\dIbsp$ $-\frac{9}{5}\dIright)\dIbsp$ $+12\log(2)\dIbsp$ $-\frac{12 N_c }{5}\dIbsp$ $-\frac{27}{5}}
\dataPa{3}{3}{\xi\dIleft(-\frac{13 C_0  N_c }{10}\dIbsp$ $+\frac{ C_0 }{2 N_c }\dIbsp$ $+\frac{6 C_0 }{5}\dIbsp$ $-\frac{4\log(2)}{ N_c }\dIbsp$ $+\frac{13 N_c }{10}\dIbsp$ $+\frac{1}{2 N_c }\dIbsp$ $-\frac{6}{5}\dIright)\dIbsp$ $-\frac{12\log(2)}{ N_c }\dIbsp$ $-\frac{13 N_c }{5}\dIbsp$ $+\frac{8}{ N_c }\dIbsp$ $+\frac{12}{5}}
\dataPa{4}{4}{\xi\dIleft(\frac{ C_0  N_c }{2}\dIbsp$ $-\frac{2\log(2)}{ N_c }\dIbsp$ $-\frac{ N_c }{2}\dIright)\dIbsp$ $+\frac{3 C_0 }{2 N_c }\dIbsp$ $-\frac{2\log(2)}{ N_c }\dIbsp$ $+ N_c \dIbsp$ $-\frac{3}{ N_c }}
\dataPa{4}{5}{\xi\dIleft(-\frac{ C_0 }{2}\dIbsp$ $+2\log(2)\dIbsp$ $+\frac{1}{2}\dIright)\dIbsp$ $-\frac{3 C_0 }{2}\dIbsp$ $+2\log(2)\dIbsp$ $+2}
\dataPa{5}{4}{\dIleft(2\log(2)\dIbsp$ $-\frac{1}{2}\dIright)\xi\dIbsp$ $+2\log(2)\dIbsp$ $-2}
\dataPa{5}{5}{-\frac{3 C_0  N_c }{2}\dIbsp$ $+\frac{3 C_0 }{2 N_c }\dIbsp$ $+\xi\dIleft(\frac{ N_c }{2}\dIbsp$ $-\frac{2\log(2)}{ N_c }\dIright)\dIbsp$ $-\frac{2\log(2)}{ N_c }\dIbsp$ $+5 N_c \dIbsp$ $-\frac{3}{ N_c }}
\\
      \hline
    \end{tabular}
    \caption{One-loop conversion matrix $\Delta
  a^{\ri\to\msbar}$ for the $(8,1)$ operators of the \reduced{} basis in
  the $\rismom(\slashed{q},\gamma_\mu)$ scheme.}
    \label{tab:gammaCGQNEPQA}
  \end{table}

  \begin{table}[tp]
    \centering
    \renewcommand{\dataINL}{}
    \begin{tabular}{lcl}
      \hline
      $(i,j)$ &  &  $\Delta a^{\ri\to\msbar}_{ij}/\frac{\alpha_s}{4\pi}$ \\
      \hline\hline
      \dataPa{2}{2}{\xi\dIleft(\frac{4 C_0  N_c }{5}\dIbsp$ $+\frac{ C_0 }{ N_c }\dIbsp$ $-\frac{6 C_0 }{5}\dIbsp$ $-\frac{4\log(2)}{ N_c }\dIbsp$ $-\frac{4 N_c }{5}\dIbsp$ $+\frac{6}{5}\dIright)\dIbsp$ $-\frac{12\log(2)}{ N_c }\dIbsp$ $+\frac{8 N_c }{5}\dIbsp$ $+\frac{9}{ N_c }\dIbsp$ $-\frac{12}{5}}
\dataPa{2}{3}{\xi\dIleft(\frac{6 C_0  N_c }{5}\dIbsp$ $-\frac{9 C_0 }{5}\dIbsp$ $+4\log(2)\dIbsp$ $-\frac{6 N_c }{5}\dIbsp$ $+\frac{4}{5}\dIright)\dIbsp$ $+12\log(2)\dIbsp$ $+\frac{12 N_c }{5}\dIbsp$ $-\frac{53}{5}}
\dataPa{3}{2}{\xi\dIleft(-\frac{6 C_0  N_c }{5}\dIbsp$ $+\frac{4 C_0 }{5}\dIbsp$ $+4\log(2)\dIbsp$ $+\frac{6 N_c }{5}\dIbsp$ $-\frac{9}{5}\dIright)\dIbsp$ $+12\log(2)\dIbsp$ $-\frac{12 N_c }{5}\dIbsp$ $-\frac{27}{5}}
\dataPa{3}{3}{\xi\dIleft(-\frac{9 C_0  N_c }{5}\dIbsp$ $+\frac{ C_0 }{ N_c }\dIbsp$ $+\frac{6 C_0 }{5}\dIbsp$ $-\frac{4\log(2)}{ N_c }\dIbsp$ $+\frac{9 N_c }{5}\dIbsp$ $-\frac{6}{5}\dIright)\dIbsp$ $-\frac{12\log(2)}{ N_c }\dIbsp$ $-\frac{18 N_c }{5}\dIbsp$ $+\frac{9}{ N_c }\dIbsp$ $+\frac{12}{5}}
\dataPa{4}{4}{\xi\dIleft(\frac{ C_0 }{2 N_c }\dIbsp$ $-\frac{2\log(2)}{ N_c }\dIbsp$ $-\frac{1}{2 N_c }\dIright)\dIbsp$ $+\frac{3 C_0 }{2 N_c }\dIbsp$ $-\frac{2\log(2)}{ N_c }\dIbsp$ $-\frac{2}{ N_c }}
\dataPa{4}{5}{\xi\dIleft(-\frac{ C_0 }{2}\dIbsp$ $+2\log(2)\dIbsp$ $+\frac{1}{2}\dIright)\dIbsp$ $-\frac{3 C_0 }{2}\dIbsp$ $+2\log(2)\dIbsp$ $+2}
\dataPa{5}{4}{\dIleft(2\log(2)\dIbsp$ $-\frac{1}{2}\dIright)\xi\dIbsp$ $+2\log(2)\dIbsp$ $-2}
\dataPa{5}{5}{\xi\dIleft(-\frac{ C_0  N_c }{2}\dIbsp$ $+\frac{ C_0 }{2 N_c }\dIbsp$ $-\frac{2\log(2)}{ N_c }\dIbsp$ $+ N_c \dIbsp$ $-\frac{1}{2 N_c }\dIright)\dIbsp$ $-\frac{3 C_0  N_c }{2}\dIbsp$ $+\frac{3 C_0 }{2 N_c }\dIbsp$ $-\frac{2\log(2)}{ N_c }\dIbsp$ $+4 N_c \dIbsp$ $-\frac{2}{ N_c }}
\\
      \hline
    \end{tabular}
    \caption{One-loop conversion matrix $\Delta
  a^{\ri\to\msbar}$ for the $(8,1)$ operators of the \reduced{} basis in the
  $\rismom(\slashed{q},\slashed{q})$ scheme.}
    \label{tab:gammaCQQNEPQA}
  \end{table}

\subsection{Discussion}
In the previous sections we provided results for the conversion from
different $\rimom$ and $\rismom$ schemes to the $\NDR$ scheme.  In
general the different schemes have a different magnitude of one-loop
corrections $\Delta r^{\ri\to\msbar}$ and a different rate of
convergence of the perturbative expansion of $\Delta r^{\ri\to\msbar}$
in $\alpha_s$.  In the context of lattice QCD and non-perturbative
renormalization it is thus useful to have several $\ri$ schemes to
choose from in order to estimate the effects of missing higher-order
terms in the perturbative expansion.  In this work we give four
different $\rismom$ schemes for each operator of the \reduced{}
$\Delta S=1$ operator basis.

Since we only give one-loop results, we cannot estimate the rate of
convergence of the different schemes presented in this paper. In some
cases where higher-order results in $\rismom$ schemes are known, such
as the conversion relation for light quark mass determinations, the
convergence in the $\rismom$ schemes is significantly faster than the
convergence in the traditional $\rimom$ schemes.

It is, however, interesting to compare the magnitude of the one-loop
corrections $\Delta r^{\ri\to\msbar}$ of different $\ri$ schemes.  In
Tab.~\ref{tab:matrixnorms} we give the spectral matrix norm and the
maximum norm for the conversion matrices $\Delta r^*$ at $N_c=3$ and
$\xi=0$ for the different blocks of operators in the \reduced{} basis.
The spectral norm of matrix $M$ is defined as the maximum singular
value of the matrix $M$ and the maximum norm of matrix $M$ is defined
as the maximum absolute value of the matrix elements.  Therefore the
spectral norm gives a good estimate of the general magnitude of the
one-loop corrections since it is invariant under a change of operator
basis.  The maximum norm gives a good estimate of especially large
individual mixing coefficients.

One observes that for the $(27,1)$ operator the one-loop coefficients
for the $\rimom$ scheme with $\ripmom$ wave function renormalization
are of order $1$, while the $\rismom$ results vary from order $1/10$
to order $1$.  The $\rismom(x,y)$ schemes with $x=y$ have especially
small one-loop coefficients.  In this case the spectral norm is, of
course, equal to the maximum norm.  For the $(8,1)$ operators the
spectral norm and the maximum norm for the $\rimom,\ripmom$ scheme and
the $\rismom(\slashed{q},y)$ schemes is approximately twice the size
of the $\rismom(\gamma_\mu,y)$ schemes.  In the case of the $(8,8)$
operators the maximum norm as well as the spectral norm in the
$\rimom,\ripmom$ scheme is significantly larger than the respective
norm in the $\rismom$ schemes.  The difference is especially
pronounced for the $\rismom(\slashed{q},\slashed{q})$ scheme.

We conclude that the magnitude of the one-loop contributions to
$\Delta r^{\ri\to\msbar}$ varies significantly between the different
schemes, and we expect that the same holds for the rate of convergence
of the perturbative expansion of $\Delta r^{\ri\to\msbar}$ in
$\alpha_s$.  Note that Tab.~\ref{tab:matrixnorms} only gives the
numerical values for Landau gauge and that the qualitative analysis of
this section may be different for other gauges.

\begin{table}[tp]
  \centering
  \begin{tabular}[t]{llrrrr}\hline
    Scheme & $(L,R)$ &&  $\norm{\Delta r^*/\frac{\alpha_s}{4\pi}}_s$\relax
    &&$\norm{\Delta r^*/\frac{\alpha_s}{4\pi}}_\infty$\\
    \hline\hline
    $\rimom,\ripmom$ & (27,1) && 0.87851
\\
    $\rismom(\gamma_\mu,\slashed{q})$ & (27,1) && 
    2.45482\\
    $\rismom(\gamma_\mu,\gamma_\mu)$ & (27,1) && 
    0.21184\\
    $\rismom(\slashed{q},\slashed{q})$ & (27,1) && 
    0.45482
\\
    $\rismom(\slashed{q},\gamma_\mu)$ & (27,1) && 
    2.21184
\\
    \hline
    $\rimom,\ripmom$ & (8,1) && \shownorm{10.93031}{10.61021}{9.20457}
\\
    $\rismom(\gamma_\mu,\slashed{q})$ & (8,1) && 
    \shownorm{}{5.34301}{3.88889}\\
    $\rismom(\gamma_\mu,\gamma_\mu)$ & (8,1) && 
    \shownorm{}{5.21189}{3.88889}\\
    $\rismom(\slashed{q},\slashed{q})$ & (8,1) && 
    \shownorm{12.03830}{12.03242}{8.69111}
\\
    $\rismom(\slashed{q},\gamma_\mu)$ & (8,1) && 
    \shownorm{11.84666}{11.17384}{6.02444}
\\
    \hline
    $\rimom,\ripmom$ & (8,8) && \shownorm{11.45869}{11.42390}{10.87124}
\\
    $\rismom(\gamma_\mu,\slashed{q})$ & (8,8) && 
    \shownorm{}{3.23249}{2.62348}\\
    $\rismom(\gamma_\mu,\gamma_\mu)$ & (8,8) && 
    \shownorm{}{4.77841}{4.49561}\\
    $\rismom(\slashed{q},\slashed{q})$ & (8,8) && 
    \shownorm{1.62238}{1.62236}{1.49561}
\\
    $\rismom(\slashed{q},\gamma_\mu)$ & (8,8) && 
    \shownorm{5.00612}{4.26036}{4.16227}
\\
    \hline
  \end{tabular}
  \caption{Magnitude of the one-loop conversion matrices 
    $\Delta r^*/\frac{\alpha_s}{4\pi}$ for operators in the \reduced{} basis and
    $N_c=3$, $\xi=0$. 
    We give the spectral norm $\norm{\Delta r^*/\frac{\alpha_s}{4\pi}}_s$ 
    as well as the maximum norm 
    $\norm{\Delta r^*/\frac{\alpha_s}{4\pi}}_\infty$ for operators in 
    the $(L,R)$ representation of $\SU(3)_L\otimes \SU(3)_R$.}
  \label{tab:matrixnorms}
\end{table}

\section{Summary and conclusion\label{sec:conclusion}}
Physical processes that change the strangeness by one unit play an
important role in the field of flavor phenomenology.  These processes
can be studied in lattice simulations using an effective $\Delta S=1$
Hamiltonian of electroweak interactions that is formulated in terms of
$\Delta S=1$ flavor-changing four-quark operators.  In order to
renormalize these operators non-perturbatively and to convert measured
matrix elements to the $\msbar$ scheme it is necessary to define
renormalization schemes that are independent of the specific
regulator.  In this work we define $\rismom$ renormalization schemes
for different $\Delta S=1$ operator bases and provide one-loop
matching factors to the $\NDR$ scheme.

Since the different $\ri$ schemes project out different components of
the perturbative series, the variety of $\ri$ schemes can be used to
estimate the effects of higher-order corrections to the conversion
matrices which are presently unknown.  In this work we define four
different $\rismom$ schemes for each $\Delta S=1$ operator.  We also
provide a compact expression for the finite one-loop amplitudes that
can be used by the reader to define further $\rismom$ schemes in a
straightforward way.

The numerical size of the one-loop contributions to the matching
factors is discussed briefly for different $\ri$ schemes in the Landau
gauge.  We find that their magnitude varies significantly between the
different $\rimom$ and $\rismom$ schemes.

In future work two-loop conversion factors will be calculated which
will further reduce the systematic uncertainties in lattice
calculations involving the effective $\Delta S=1$ Hamiltonian of
electroweak interactions.

\acknowledgments We would like to thank the RBC/UKQCD collaboration,
especially N.H.~Christ, N.~Garron, T.~Izubuchi, R.D.~Mawhinney,
C.T.~Sachrajda, and A.~Soni, for many interesting and inspiring
discussions.  C.L.~acknowledges support from the RIKEN FPR
program. C.S.~was partially supported by U.S.~DOE under Contract
No.~DE-AC02-98CH10886.

\appendix
\section{Flavor and isospin decomposition of four-quark operators}
\label{app:flavordecompose}
In this appendix we give the flavor and isospin decomposition of the
four-quark operators defined in
Eqs.~\eqref{eqn:q1FTdef}-\eqref{eqn:q10def}.  We proceed along the
lines of App.~B of Ref.~\cite{Blum:2001xb}, carefully avoiding the use
of Fierz transformations.  We can then convert operators $\tilde Q_i$
to operators $Q_i$ including the evanescent operators of
Eq.~\eqref{eqn:defevanqqtilde}.

\subsection{Left-left operators}
The left-left operators $Q_1,\ldots,Q_4,Q_9,Q_{10}$ can be written as
\begin{align}\label{eqn:opformll}
  Q_{LL} &= (T_{LL})^{ij}_{kl} ( \bar q_L^i \otimes q_L^k)
  (\bar q_L^j \otimes q_L^l)
\end{align}
with left-handed quark fields $q_L^i$ and flavor indices $i$, $j$,
$k$, $l$.  We identify $q_L^1$ with an up quark, $q_L^2$ with a down
quark, and $q_L^3$ with a strange quark.  The color and spinor
contractions are denoted by $\otimes$.  The individual quark fields
transform in the fundamental representation of $\SU(3)_L$, i.e., under
$V \in \SU(3)_L$ we find
\begin{align}
  Q'_{LL} &= (V^\dagger)_{ia}(V^\dagger)_{jb}(T_{LL})^{ab}_{cd}V_{ck} V_{dl} ( \bar q_L^i \otimes q_L^k)
  (\bar q_L^j \otimes q_L^l) \notag\\
&=(T'_{LL})^{ij}_{kl} ( \bar q_L^i \otimes q_L^k)
  (\bar q_L^j \otimes q_L^l)
\end{align}
with
\begin{align}\label{eqn:trafotl}
  (T'_{LL})^{ij}_{kl} =
  (V^\dagger)_{ia}(V^\dagger)_{jb}(T_{LL})^{ab}_{cd}V_{ck} V_{dl}\,.
\end{align}
This transformation corresponds to the 81-dimensional representation
$(\bar 3 \otimes 3) \otimes (\bar 3 \otimes 3)$ of $\SU(3)_L$.  We can
decompose the tensor of this $81$-dimensional representation as
\begin{align}
  (T_{LL})^{ij}_{kl} &= (T_{LL})^{\{i,j\}}_{\{k,l\}} + (T_{LL})^{\{i,j\}}_{[k,l]} \notag\\&\quad
 + (T_{LL})^{[i,j]}_{\{k,l\}} + (T_{LL})^{[i,j]}_{[k,l]}\,,
\end{align}
where for a general tensor $t_{ij}$ we have $2t_{\{i,j\}} = t_{ij} +
t_{ji}$ and $2t_{[i,j]} = t_{ij} - t_{ji}$.  It is straightforward to
check that the respective subspaces are invariant under
Eq.~\eqref{eqn:trafotl} and that their dimensionality is given by
\begin{align}
 \dim \left[  (T_{LL})^{\{i,j\}}_{\{k,l\}} \right] &= 36 \,, &
 \dim \left[  (T_{LL})^{[i,j]}_{\{k,l\}} \right] &= 18\,, \\
 \dim \left[  (T_{LL})^{\{i,j\}}_{[k,l]} \right] &= 18 \,, &
 \dim \left[  (T_{LL})^{[i,j]}_{[k,l]} \right] &= 9\,.
\end{align}
Since $Q_{LL}$ is symmetric under simultaneous exchange of $(i,k)
\leftrightarrow (j,l)$ we only need to consider the completely
symmetric case $(T_{LL})^{\{i,j\}}_{\{k,l\}}$ and the completely
antisymmetric case $(T_{LL})^{[i,j]}_{[k,l]}$.

We can further decompose the remaining subspaces by considering the
trace of a pair of upper and lower indices.  If such a trace vanishes,
it also vanishes after applying the transformation of
Eq.~\eqref{eqn:trafotl}, and therefore such a constraint defines an
invariant subspace.  For the completely symmetric case we distinguish
between
\begin{align}
 ({\rm i}) &&  (T_{LL})^{\{i,j\}}_{\{i,l\}} &= 0 & \Rightarrow &&  (T_{LL})^{\{i,j\}}_{\{i,j\}} &= 0 \,, \\
 ({\rm ii}) &&  (T_{LL})^{\{i,j\}}_{\{i,l\}} &\ne 0 & \land &&  (T_{LL})^{\{i,j\}}_{\{i,j\}} &= 0 \,, \\
 ({\rm iii}) &&  (T_{LL})^{\{i,j\}}_{\{i,l\}} &\ne 0 & \land &&  (T_{LL})^{\{i,j\}}_{\{i,j\}} &\ne 0 \,,
\end{align}
where we sum over repeated indices.  The subspace (i) has nine
constraints and is thus $27$-dimensional, the subspace (ii) is
orthogonal to (i) and has one constraint and is thus $8$-dimensional,
and the remaining subspace (iii) is $1$-dimensional.
Therefore the completely symmetric case can be decomposed as
\begin{align}\nonumber
  27 \oplus 8 \oplus 1\,.
\end{align}
For the completely antisymmetric case the analog definition of
subspaces leads to a zero-dimensional space (i), a $8$-dimensional
space (ii), and a $1$-dimensional space (iii), i.e., the completely
antisymmetric case can be decomposed as
\begin{align}\nonumber
  0 \oplus 8 \oplus 1\,.
\end{align}

This completes the classification of the left-left operators of the
form given in Eq.~\eqref{eqn:opformll} according to representations of
$\SU(3)_L$.  From now on we restrict the discussion to $Q_{LL}$ with
either (I) exactly one $\bar s$ field or (II) exactly one $u$ or $d$
field.  In the case (I) only the elements $(T_{LL})^{3j}_{kl}$ and
$(T_{LL})^{j3}_{kl}$ with $j,k,l=1,2$ are nonzero.  Therefore they live
in a $8$-dimensional representation of $\SU(2)$ isospin, and we can
classify them according to irreducible isospin representations in the
following.  In the completely symmetric case (which is especially
symmetric in $k \leftrightarrow l$) they live in a $6$-dimensional
isospin representation, and we distinguish between
\begin{align}
 ({\rm Ia}) &&  (T_{LL})^{\{3,j\}}_{\{k,j\}} &= 0 \,, & \\
 ({\rm Ib}) &&  (T_{LL})^{\{3,j\}}_{\{k,j\}} &\ne 0\,, &
\end{align}
where we sum over $j=1,2$.  The subspace (Ia) has two constraints and
is thus $4$-dimensional ($I=3/2$), the subspace (Ib) is orthogonal to
(Ia) and is thus $2$-dimensional ($I=1/2$).  In the completely
antisymmetric case $T_{LL}$ lives in a $2$-dimensional isospin
representation.  The respective subspace (Ia) has two constraints and
is thus zero-dimensional, the respective subspace (Ib) is orthogonal
to (Ia) and is thus two-dimensional ($I=1/2$).  In the case (II) the
tensor transforms in the fundamental isospin representation ($I=1/2$).

In the remainder of this subsection we give a list of left-left
operators with $\Delta S=1$ transforming in irreducible
representations of $\SU(3)_L$ and isospin in which the operator
\begin{align}
  (\bar s d)_{V-A} (\bar u u)_{V-A}
\end{align}
enters.  The completely symmetric operators are given by
\begin{align}\label{eqn:fclsLLS}
  Q_S^{(27,1),3/2} &= (\bar s
  d)_{V-A}(\bar u u)_{V-A} + (\bar s u)_{V-A}(\bar u d)_{V-A} \notag\\&\quad
  -(\bar s d)_{V-A}(\bar d d)_{V-A} \,, \notag\\
  Q_S^{(27,1),1/2} &= (\bar s
  d)_{V-A}(\bar u u)_{V-A} + (\bar s u)_{V-A}(\bar u d)_{V-A} \notag\\&\quad
  +2(\bar s d)_{V-A}(\bar d d)_{V-A}
  -3(\bar s d)_{V-A}(\bar s s)_{V-A} \,, \notag\\
  Q_S^{(8,1),1/2} &= (\bar s
  d)_{V-A}(\bar u u)_{V-A} + (\bar s u)_{V-A}(\bar u d)_{V-A} \notag\\&\quad
  +2(\bar s d)_{V-A}(\bar d d)_{V-A} \notag\\&\quad
  +2(\bar s d)_{V-A}(\bar s s)_{V-A} \,,
\end{align}
where $(L,R)$ indicates the representation of $\SU(3)_L \otimes
\SU(3)_R$, and $1/2$ ($3/2$) indicates the isospin.
The completely antisymmetric operator is given by
\begin{align}\label{eqn:fclsLLA}
  Q_A^{(8,1),1/2} &= (\bar s
  d)_{V-A}(\bar u u)_{V-A} - (\bar s u)_{V-A}(\bar u d)_{V-A}\,.
\end{align}

\subsection{Left-right operators}
The left-right operators $Q_5,\ldots,Q_8$ can be written as
\begin{align}\label{eqn:opformlr}
  Q_{LR} &= (T_{LR})^{ij}_{kl} (\bar q_L^i \otimes q_L^k) (\bar q_R^j
  \otimes q_R^l)
\end{align}
with right-handed quark fields $q_R^i$.  The individual quark fields
transform in the fundamental representation of $\SU(3)_{L/R}$, i.e.,
under $V_L \in \SU(3)_L$ and $V_R \in \SU(3)_R$ we find
\begin{align}
  Q'_{LR} &=
  (V_L^\dagger)_{ia}(V_R^\dagger)_{jb}(T_{LR})^{ab}_{cd}(V_L)_{ck}
  (V_R)_{dl} \notag\\&\quad\times(\bar q_L^i \otimes q_L^k) (\bar q_R^j
  \otimes q_R^l) \notag\\
  &=(T'_{LR})^{ij}_{kl} (\bar q_L^i \otimes q_L^k) (\bar q_R^j \otimes
  q_R^l)
\end{align}
with
\begin{align}\label{eqn:trafotlr}
  (T'_{LR})^{ij}_{kl} =
  (V_L^\dagger)_{ia}(V_R^\dagger)_{jb}(T_{LR})^{ab}_{cd}(V_L)_{ck}
  (V_R)_{dl}\,.
\end{align}
This transformation corresponds to the 9-dimensional $\bar 3 \otimes
3$ representations of $\SU(3)_{L}$ and $\SU(3)_R$.  These
$9$-dimensional representations can be decomposed in a $8$-dimensional
subspace with vanishing trace of left (right) indices and in a
$1$-dimensional subspace with nonzero trace, i.e.,
\begin{align}
 \bar 3 \otimes 3 = 8 \oplus 1\,.
\end{align}
This completes the classification of the left-right operators of the
form given in Eq.~\eqref{eqn:opformlr} according to representations of
$\SU(3)_L \otimes \SU(3)_R$.  From now on we restrict the discussion
to $Q_{LR}$ with $(T_{LR})^{ij}_{kl} \ne 0$ only for (I) $i=3$ and
$j,k,l=1,2$ or (II) $i=j=l=3$, $k=1,2$.  The classification according
to irreducible representations of isospin now follows from the analog
discussion of left-left operators.

In the remainder of this subsection we give a list of left-right
operators with $\Delta S=1$ transforming in irreducible
representations of $\SU(3)_L \otimes \SU(3)_R$ and isospin in which
the operator
\begin{align}
 (\bar s d)_{V-A} (\bar u u)_{V+A}
\end{align}
enters.  It is straightforward to see that $Q_5$ and $Q_6$ transform
in the $(8,1)$ representation of $\SU(3)_L \otimes \SU(3)_R$ and in
isospin representation $I=1/2$, i.e., we define
\begin{align}
  Q_{5,6}^{(8,1),1/2}=Q_{5,6}\,.
\end{align}
Operators symmetric under $k \leftrightarrow l$ are given by
\begin{align}
  Q_S^{(8,8),3/2} &= (\bar s d)_{V-A}(\bar u u)_{V+A} + (\bar s
  u)_{V-A}(\bar u d)_{V+A} \notag\\&\quad
  -(\bar s d)_{V-A}(\bar d d)_{V+A} \,, \notag\\
  Q_S^{(8,8),1/2} &= (\bar s d)_{V-A}(\bar u u)_{V+A} + (\bar s
  u)_{V-A}(\bar u d)_{V+A} \notag\\&\quad +2(\bar s d)_{V-A}(\bar d
  d)_{V+A} \notag\\&\quad -3 (\bar s d)_{V-A}(\bar s s)_{V+A}\,.
\end{align}
One can also construct an operator that is antisymmetric under $k
\leftrightarrow l$ for $k,l=1,2$ with well-defined transformation
under flavor and isospin:
\begin{align}
  Q_A^{(8,8),1/2} &= (\bar s d)_{V-A}(\bar u u)_{V+A} - (\bar s
  u)_{V-A}(\bar u d)_{V+A} \notag\\ &\quad - (\bar s d)_{V-A}(\bar s
  s)_{V+A}\,.
\end{align}

\subsection{Change of basis}
Operators of the physical basis I and II with spinor structure
$(V-A)$--$(V-A)$ can be decomposed in the color-diagonal operators of
Eqs.~\eqref{eqn:fclsLLS} and \eqref{eqn:fclsLLA}.  We find
\begin{align}
 \tilde Q_1 &= \frac1{10} Q_S^{(8,1),1/2} + \frac1{15} Q_S^{(27,1),1/2}
  \notag\\& \quad+ \frac13 Q_S^{(27,1),3/2} + \frac12 Q_A^{(8,1),1/2} \,, \\
  Q_3 &= \frac12 Q_S^{(8,1),1/2} + \frac12 Q_A^{(8,1),1/2} \,, \\
  Q_9 &= -\frac1{10} Q_S^{(8,1),1/2} + \frac1{10} Q_S^{(27,1),1/2}
  \notag\\& \quad+ \frac12 Q_S^{(27,1),3/2} + \frac12 Q_A^{(8,1),1/2}
\end{align}
for the color-diagonal operators
and
\begin{align}
  Q_2 &= \frac1{10} Q_S^{(8,1),1/2} + \frac1{15} Q_S^{(27,1),1/2}
  \notag\\& \quad+ \frac13 Q_S^{(27,1),3/2} - \frac12 Q_A^{(8,1),1/2} \,, \\
  \tilde Q_4 &= \frac12 Q_S^{(8,1),1/2} - \frac12 Q_A^{(8,1),1/2} \,, \\
  \tilde Q_{10} &= -\frac1{10} Q_S^{(8,1),1/2} + \frac1{10} Q_S^{(27,1),1/2}
  \notag\\& \quad+ \frac12 Q_S^{(27,1),3/2} - \frac12 Q_A^{(8,1),1/2}
\end{align}
for the color-mixed operators
with
\begin{align}
  \tilde Q_4 &= \sum_{q=u,d,s} (\bar s_a q_a)_{V-A} (\bar
  q_b
  d_b)_{V-A} \notag\,,\\
  \tilde Q_{10} &= \frac32\sum_{q=u,d,s} e_q (\bar s_a
  q_a)_{V-A} (\bar q_b d_b)_{V-A}\,.
\end{align}

Operators with spinor structure $(V-A)$--$(V+A)$ are already in
explicit $(8,1)$ and $(8,8)$ representations, i.e.,
\begin{align}
  Q_{5,6} &= Q_{5,6}^{(8,1),1/2}\,, \\
  Q_7 &= \frac12 Q_S^{(8,8),3/2} + \frac12 Q_A^{(8,8),1/2}\,, \\
  Q_8 &= \frac12 {Q'}_S^{(8,8),3/2} + \frac12 {Q'}_A^{(8,8),1/2}
\end{align}
with
\begin{align}
  {Q'}_S^{(8,8),3/2} &= (\bar s_a d_b)_{V-A}(\bar u_b
  u_a)_{V+A} \notag\\&\quad + (\bar s_a u_b)_{V-A}(\bar
  u_b d_a)_{V+A} \notag\\&\quad
  -(\bar s_a d_b)_{V-A}(\bar d_b d_a)_{V+A} \,, \\
  {Q'}_A^{(8,8),1/2} &= (\bar s_a d_b)_{V-A}(\bar u_b
  u_a)_{V+A} \notag\\&\quad - (\bar s_a u_b)_{V-A}(\bar
  u_b d_a)_{V+A} \notag\\ &\quad - (\bar s_a
  d_b)_{V-A}(\bar s_b s_a)_{V+A}\,.
\end{align}

\section{\boldmath Alternative determination of $\Delta T$}\label{app:rhut}
\newcommand{\meI}{P^{VV+AA,\slashed{q}}_{(1),d}}
\newcommand{\meII}{P^{VV+AA,\slashed{q}}_{(2),d}}
\newcommand{\meIII}{P^{VV-AA,\slashed{q}}_{(1),d}}
\newcommand{\meIV}{P^{VV-AA,\slashed{q}}_{(2),d}}
\newcommand{\meV}{P^{VV+AA,\slashed{q}}_{(1),u}}
\newcommand{\meVI}{P^{VV+AA,\slashed{q}}_{(2),u}}
\newcommand{\meVII}{P^{VV-AA,\slashed{q}}_{(1),u}}
\newcommand{\meVIII}{P^{VV-AA,\slashed{q}}_{(2),u}}
\newcommand{\meIX}{P^{VV+AA,\slashed{q}}_{(1),s}}
\newcommand{\meX}{P^{VV+AA,\slashed{q}}_{(2),s}}
\newcommand{\meXI}{P^{VV-AA,\slashed{q}}_{(1),s}}
\newcommand{\meXII}{P^{VV-AA,\slashed{q}}_{(2),s}}
\newcommand{\meXIII}{P^{VV+AA,\slashed{q}}_{(4),u}}
\newcommand{\meXIV}{P^{VV+AA,\slashed{q}}_{(3),u}}
\newcommand{\meXV}{P^{VV-AA,\slashed{q}}_{(4),u}}
\newcommand{\meXVI}{P^{VV-AA,\slashed{q}}_{(3),u}} 

In this appendix we provide an alternative method to determine $\Delta
T$ defined in Eq.~\eqref{eqn:defdt} relating the Wilson coefficients of
the $\msbar$-renormalized operator basis I, II to the respective
coefficients of the \reduced{} basis.

To this end we consider an arbitrary $\ri$ scheme and express on-shell
matrix elements of the effective Hamiltonian in terms of operators
${Q'}^\ri_j$ and first in terms of Wilson coefficients
${C'}^\msbar_i$,
\begin{align}
  \braket{\mathcal{H}^{\Delta S=1}_{\rm eff}}
&={G_{F}\over\sqrt{2}}\sum_{i,j}{C'}^{\msbar}_{i}(\mu) {R'}^{\ri\to\msbar}_{ij}(\mu)\notag\\
&\qquad\qquad\times\braket{{Q'}^{\ri}_{j}(\mu)}\,,
\end{align}
and then in terms of Wilson coefficients ${C}^\msbar_i$,
\begin{align}
  \braket{\mathcal{H}^{\Delta S=1}_{\rm eff}}
&={G_{F}\over\sqrt{2}}\sum_{i,j,l}{C}^{\msbar}_{i}(\mu) {R}^{\ri\to\msbar}_{ij}(\mu)\notag\\
&\qquad\qquad\times T_{jl}\braket{{Q'}^{\ri}_{l}(\mu)}\,.
\end{align}
Since these equations hold for an arbitrary matrix element, we find
\begin{align}\label{eqn:defcprime}
 ({C'}^\msbar)^T &=  (C^\msbar)^T {R}^{\ri\to\msbar} T ({R'}^{\ri\to\msbar})^{-1}
\end{align}
where $T$ is defined in Eq.~\eqref{eqn:deft}.  A comparison with
Eq.~\eqref{eqn:defdt} yields
\begin{align}\label{eqn:defdtb}
  T + \Delta T = {R}^{\ri\to\msbar} T ({R'}^{\ri\to\msbar})^{-1}\,.
\end{align}
Note that the left-hand side is just a conversion factor of Wilson
coefficients in the $\msbar$ scheme, and thus the right-hand side must
also be independent of the $\ri$ scheme used to calculate ${
  R}^{\ri\to\msbar}$ and ${R'}^{\ri\to\msbar}$.

We calculate $\Delta T$ using Eq.~\eqref{eqn:defdtb} for the
non-exceptional momentum configuration with projectors
\begin{align}
 P_{1} &= \frac{\meV - \meXIV -\meI}{160 N_c (N_c+1)}\,, \\
  P_2 &= \bigl(3\meV+2\meI +2\meXIV \bigr) \notag\\&\quad \times \frac{3N_c-2}{160N_c(N_c^2-1)} \notag\\&\quad
  +\bigl(3\meVI +2\meII+2\meXIII\bigr) \notag\\&\quad\times \frac{2N_c-3}{160N_c(N_c^2-1)}\,, \\
  P_3 &= \bigl(3\meV+2\meI +2\meXIV \bigr) \notag\\&\quad \times \frac{2N_c-3}{160N_c(N_c^2-1)} \notag\\&\quad
  +\bigl(3\meVI +2\meII+2\meXIII\bigr) \notag\\&\quad\times \frac{3N_c-2}{160N_c(N_c^2-1)}\,, \\
  P_5 &= \frac{\meVII + 2\meIII}{96(N_c^2-1)} \notag\\&\quad
  - \frac{\meVIII + 2\meIV}{96 N_c (N_c^2-1)}\,, \\
  P_6 &=  \frac{\meVIII + 2\meIV}{96(N_c^2-1)} \notag\\&\quad
  -\frac{\meVII + 2\meIII}{96N_c(N_c^2-1)}\,,\\
  P_7 &= \frac{\meVII - \meIII}{48(N_c^2-1)}\notag\\&\quad
  -\frac{\meVIII - \meIV}{48N_c(N_c^2-1)} \,,\\
  P_8 &= \frac{\meVIII-\meIV}{48(N_c^2-1)} \notag\\&\quad
  - \frac{\meVII - \meIII}{48N_c(N_c^2-1)}
\end{align}
with
\begin{align}
  P_i \Gamma^\tree_4({Q'}_j) &= \delta_{ij}\,, & i,j &=
  1,2,3,5,6,7,8\,.
\end{align}
This choice of projectors recovers the conversion matrices of the
$\rismom(\slashed{q},y)$ schemes for the non-exceptional momentum
configuration.  It makes use of projectors with a more complex flavor
structure which allows for a definition of this scheme without
considering each irreducible representation of operators separately.
We use the basis of projectors $\{P^{VV \pm AA}_{(i),f'}\}$ with
$i=1,2,3,4$ defined in Eqs.~\eqref{eqn:defmasterproj}.  The details of
the wave function renormalization given in $\Delta^y_q$ drop out in
Eq.~\eqref{eqn:defdtb} and hence we do not need to specify the wave
function renormalization scheme $y$.  The result for $\Delta T$ using
this method agrees with $\Delta T$ given in Sec.~\ref{sec:wilson}.

\end{document}